\documentclass[11pt,a4paper]{article}

\usepackage[utf8]{inputenc}
\usepackage[T1]{fontenc}
\usepackage{lmodern}
\usepackage[margin=2.6cm]{geometry}
\usepackage{amsmath,amssymb,amsthm,mathtools}
\usepackage{booktabs,array}
\usepackage[shortlabels]{enumitem}
\usepackage{xcolor}
\usepackage{tikz}
\usetikzlibrary{arrows.meta,calc,patterns,shapes.geometric}
\usepackage{pgfplots}
\pgfplotsset{compat=1.18}
\usepgfplotslibrary{groupplots}
\usepackage{graphicx}
\usepackage[font=small,labelfont=bf]{caption}
\usepackage{titlesec}
\usepackage{fancyhdr}
\usepackage{url}
\usepackage[hidelinks]{hyperref}
\usepackage[capitalise,noabbrev]{cleveref}
\usepackage{microtype}
\usepackage{placeins}

\titleformat{\section}{\normalfont\normalsize\bfseries}{\thesection}{0.7em}{\MakeUppercase}
\titleformat{\subsection}{\normalfont\normalsize\bfseries}{\thesubsection}{0.7em}{}
\titlespacing*{\section}{0pt}{3.0ex plus 1ex minus 0.4ex}{1.4ex plus 0.3ex}
\titlespacing*{\subsection}{0pt}{2.4ex plus 0.8ex minus 0.3ex}{1.0ex plus 0.2ex}

\fancypagestyle{plain}{\fancyhf{}\fancyfoot[C]{\small\thepage}}

\newtheorem{theorem}{Theorem}[section]
\newtheorem{lemma}[theorem]{Lemma}
\newtheorem{proposition}[theorem]{Proposition}
\newtheorem{corollary}[theorem]{Corollary}
\theoremstyle{definition}
\newtheorem{definition}[theorem]{Definition}
\theoremstyle{remark}
\newtheorem{remark}[theorem]{Remark}

\newcommand{\Lq}{\mathsf L}
\newcommand{\Gq}{\mathsf G}
\newcommand{\Aq}{\mathsf A}
\newcommand{\Bq}{\mathsf B}
\newcommand{\Cq}{\mathsf C}
\newcommand{\Fq}{\mathsf F}
\newcommand{\G}{\mathcal G}
\newcommand{\PUMP}{\mathrm{PUMP}}
\newcommand{\MT}{\mathrm{MT}}

\definecolor{stB}{HTML}{C6D5E6}
\definecolor{stA}{HTML}{7C9DC2}
\definecolor{stC}{HTML}{3F6894}
\definecolor{stG}{HTML}{1A2638}
\definecolor{stF}{HTML}{E8650E}
\definecolor{rgAgree}{HTML}{E9EFF6}
\definecolor{rgInput}{HTML}{B4C7DC}
\definecolor{rgCone}{HTML}{4A75A3}
\definecolor{rgReturn}{HTML}{6B3519}
\definecolor{rgRefl}{HTML}{F5E6D7}
\definecolor{rgRefl2}{HTML}{E4C8AE}
\definecolor{ink}{HTML}{111111}
\definecolor{grid}{HTML}{8A8A8A}
\definecolor{diffSame}{HTML}{DCE6F1}
\definecolor{diffDiff}{HTML}{8C3B12}
\definecolor{gFull}{HTML}{4A75A3}
\definecolor{gEarly}{HTML}{B4C7DC}
\definecolor{gReturn}{HTML}{6B3519}
\definecolor{gCone}{HTML}{F2C48D}
\definecolor{gNone}{HTML}{F1F1F1}
\definecolor{stUnknown}{HTML}{D9D9D9}
\definecolor{coneR}{HTML}{F2C48D}
\definecolor{coneM}{HTML}{A9C1DA}
\tikzset{
  coneRight/.style={fill=coneR},
  coneMid/.style={fill=coneM,postaction={pattern=north east lines,pattern color=ink!55}},
  coneBoth/.style={fill=coneR,postaction={pattern=north east lines,pattern color=ink!55}},
  coneNone/.style={fill=gNone}}

\newcommand{\staircase}[2]{(0,-#1) \foreach \i in {1,...,#2} { -- (\i,-#1+\i-1) -- (\i,-#1+\i) }}
\newcommand{\swatch}[1]{\tikz[baseline=-0.6ex,x=1cm,y=1cm]\filldraw[fill=#1,draw=grid,line width=0.3pt] (0,-0.09) rectangle (0.24,0.15);}

\title{Two barriers for minimal-time firing squads,\\ with applications to four and five states}
\author{Kia Keng Giam\\[0.4ex]
  \small Temasek Junior College, Singapore\\
  \small \texttt{giamkiakeng@gmail.com}}
\date{}

\begin{document}

\maketitle
\thispagestyle{plain}

\begin{abstract}
\noindent A minimal-time solution of the firing squad synchronization problem
synchronizes every line of $n$ cells at time $2n-2$. Six states suffice; whether
five do has been open since 1987. Every line
agrees with the half-line until its right end's echo returns. For minimal-time
solutions with any number of states, we prove that cells at distance $j$ behind
the half-line's front become time-periodic only after time $(3/2-\varepsilon)j$,
for every $\varepsilon>0$ and infinitely many $j$, and that the half-line is
never periodic in time and space on a wedge at its left border, hence not
eventually regular in our sense. With certified
satisfiability checks, we exclude four states using the lines of lengths $2$ to
$5$ and parts of the next five, and show that every five-state minimal-time
solution uses at least $15$ non-quiescent neighbourhoods on its half-line's
first $78$ anti-diagonals. We give five-state rules synchronizing
exactly the lengths $2$ to $14$.
\end{abstract}


%% ---------------------------------------------------------------------------
\section{Introduction}\label{sec:intro}

The firing squad synchronization problem, posed by Myhill in 1957 and made
popular by Moore~\cite{Moore64}, asks for a finite automaton with the following
property. Copies of the automaton are placed in the cells of a line of
arbitrary length $n$. At time $0$ the leftmost cell, the \emph{general}, is in a
special state and all other cells are \emph{quiescent}. At each step every cell
updates its state from its own state and the states of its two neighbours. All
cells must then enter a \emph{firing} state at the same time, and none of them
earlier. The difficulty is that the cells are identical, that each sees only
its two neighbours, and that a cell with $k$ states cannot count up to $n$:
the same finite rule has to synchronize lines of every length.

No line can be synchronized before time $2n-2$: the order of the general needs
$n-1$ steps to reach the far end of the line, and the news that the far end
exists needs $n-1$ more steps to come back. Goto~\cite{Goto62} and
Waksman~\cite{Waksman66} showed that time $2n-2$ can be achieved, and a
solution that synchronizes every line of length $n\ge2$ at time $2n-2$ is
called a \emph{minimal-time solution}. Since then, the least number of states
of a minimal-time solution has been a benchmark question of cellular automata
theory~\cite{Mazoyer86,Umeo09,Correa17}. Waksman's solution has $16$ states,
Balzer's $8$~\cite{Balzer67}, Gerken's $7$~\cite{Gerken87} and Mazoyer's
$6$~\cite{Mazoyer87}. In the other direction, Balzer~\cite{Balzer67} reported
a computer search showing that no four-state minimal-time solution exists;
Sanders~\cite{Sanders94} confirmed this with a search of his own, and
Yun\`es~\cite{Yunes93} repeated the search and observed that the lines of
lengths $2,\dots,9$ suffice. Many
six-state minimal-time solutions are known today~\cite{Clergue18,NguyenMaignan20};
solutions with at most five states are known for restricted families of
lengths, such as the powers of two, some of them in non-minimal
time~\cite{UmeoYanagihara07,Yunes08}, and four-state protocols synchronize the
rings whose lengths are powers of two~\cite{UmeoKamikawaYunes09}. Whether a
five-state minimal-time solution exists has been open since Mazoyer's
construction~\cite{Umeo09,Correa17}.

\paragraph{Why lower bounds are hard}
A five-state rule is a table of $96$ entries (\cref{sec:lines}), two of which
are fixed by quiescence, so there are $5^{94}>10^{65}$ rules. A rule can be
refuted by a single line that it fails to synchronize, but a minimal-time
solution has to synchronize every line, so a proof that five states do not
suffice must exclude all rules. As there are finitely many rules, it suffices
to find one length $N$ such that no five-state rule synchronizes the lines of
lengths $2,\dots,N$; a negative answer can thus consist of finitely many
conditions. The difficulty is to find conditions that are strong enough and
cheap enough to check. This paper proves such conditions for every number of
states and measures how far they reach for four and five states.

\begin{figure}[t]
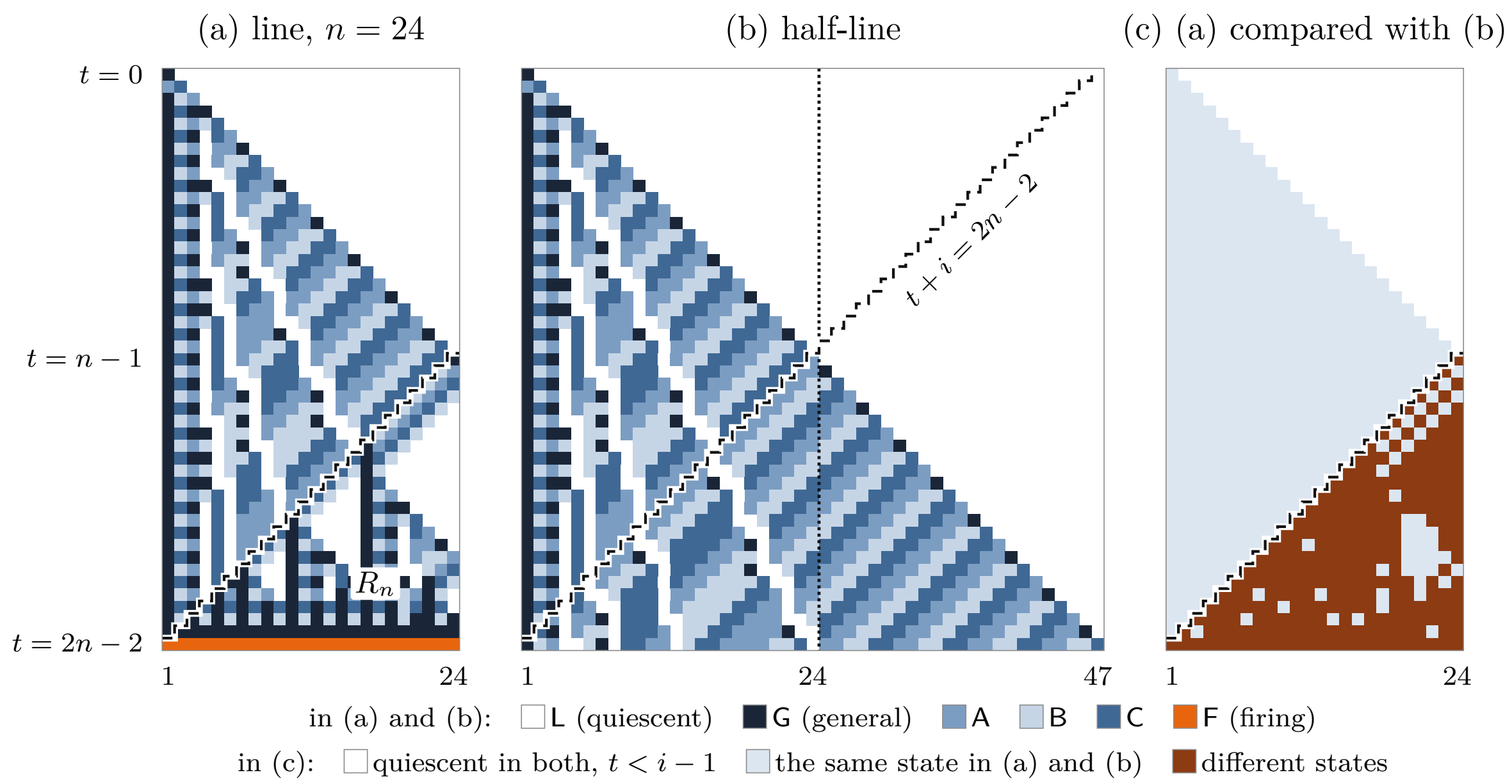

\centering
%% Figure: Mazoyer's solution on the line of length 24 and on the half-line, and where they differ
%% (cells from figures.py).
\begin{tikzpicture}[x=1.28mm,y=1.28mm]
  % (a) the line of length n = 24
  \begin{scope}
    \input{figures/cells_mazoyer_line24}
    \draw[white,line width=2.0pt] \staircase{46}{23} -- (24,-23);
    \draw[ink,densely dashed,line width=0.8pt] \staircase{46}{23} -- (24,-23);
    \draw[grid,line width=0.3pt] (0,0) rectangle (24,-47);
    \node[font=\footnotesize,fill=white,inner sep=1.2pt,rounded corners=1pt] at (17.2,-41.6) {$R_n$};
    \node[font=\small,anchor=south] at (12,0.6) {(a) line, $n=24$};
    \foreach \y/\l in {0/{t=0},23/{t=n-1},46/{t=2n-2}}
      \node[font=\scriptsize,anchor=east] at (-0.5,-\y-0.5) {$\l$};
    \foreach \x/\l in {1/1,24/24}
      \node[font=\scriptsize,anchor=north] at (\x-0.5,-47.3) {$\l$};
  \end{scope}
  % (b) the half-line, cells 1..47
  \begin{scope}[shift={(29,0)}]
    \input{figures/cells_mazoyer_half46}
    \draw[ink,densely dotted,line width=0.9pt] (24,0) -- (24,-47);
    \draw[white,line width=2.0pt] \staircase{46}{46};
    \draw[ink,densely dashed,line width=0.8pt] \staircase{46}{46};
    \draw[grid,line width=0.3pt] (0,0) rectangle (47,-47);
    \node[font=\scriptsize,fill=white,inner sep=1pt,rotate=45] at (36.3,-13.9) {$t+i=2n-2$};
    \node[font=\small,anchor=south] at (23.5,0.6) {(b) half-line};
    \foreach \x/\l in {1/1,24/24,47/47}
      \node[font=\scriptsize,anchor=north] at (\x-0.5,-47.3) {$\l$};
  \end{scope}
  % (c) where the line of length 24 differs from the half-line
  \begin{scope}[shift={(81,0)}]
    %% generated by figures.py; do not edit
\fill[diffSame] (0,0) rectangle (1,-1);
\fill[diffSame] (0,-1) rectangle (2,-2);
\fill[diffSame] (0,-2) rectangle (3,-3);
\fill[diffSame] (0,-3) rectangle (4,-4);
\fill[diffSame] (0,-4) rectangle (5,-5);
\fill[diffSame] (0,-5) rectangle (6,-6);
\fill[diffSame] (0,-6) rectangle (7,-7);
\fill[diffSame] (0,-7) rectangle (8,-8);
\fill[diffSame] (0,-8) rectangle (9,-9);
\fill[diffSame] (0,-9) rectangle (10,-10);
\fill[diffSame] (0,-10) rectangle (11,-11);
\fill[diffSame] (0,-11) rectangle (12,-12);
\fill[diffSame] (0,-12) rectangle (13,-13);
\fill[diffSame] (0,-13) rectangle (14,-14);
\fill[diffSame] (0,-14) rectangle (15,-15);
\fill[diffSame] (0,-15) rectangle (16,-16);
\fill[diffSame] (0,-16) rectangle (17,-17);
\fill[diffSame] (0,-17) rectangle (18,-18);
\fill[diffSame] (0,-18) rectangle (19,-19);
\fill[diffSame] (0,-19) rectangle (20,-20);
\fill[diffSame] (0,-20) rectangle (21,-21);
\fill[diffSame] (0,-21) rectangle (22,-22);
\fill[diffSame] (0,-22) rectangle (23,-23);
\fill[diffSame] (0,-23) rectangle (23,-24);
\fill[diffDiff] (23,-23) rectangle (24,-24);
\fill[diffSame] (0,-24) rectangle (22,-25);
\fill[diffDiff] (22,-24) rectangle (23,-25);
\fill[diffSame] (23,-24) rectangle (24,-25);
\fill[diffSame] (0,-25) rectangle (21,-26);
\fill[diffDiff] (21,-25) rectangle (22,-26);
\fill[diffSame] (22,-25) rectangle (23,-26);
\fill[diffDiff] (23,-25) rectangle (24,-26);
\fill[diffSame] (0,-26) rectangle (20,-27);
\fill[diffDiff] (20,-26) rectangle (21,-27);
\fill[diffSame] (21,-26) rectangle (22,-27);
\fill[diffDiff] (22,-26) rectangle (23,-27);
\fill[diffSame] (23,-26) rectangle (24,-27);
\fill[diffSame] (0,-27) rectangle (19,-28);
\fill[diffDiff] (19,-27) rectangle (20,-28);
\fill[diffSame] (20,-27) rectangle (21,-28);
\fill[diffDiff] (21,-27) rectangle (22,-28);
\fill[diffSame] (22,-27) rectangle (23,-28);
\fill[diffDiff] (23,-27) rectangle (24,-28);
\fill[diffSame] (0,-28) rectangle (18,-29);
\fill[diffDiff] (18,-28) rectangle (19,-29);
\fill[diffSame] (19,-28) rectangle (20,-29);
\fill[diffDiff] (20,-28) rectangle (21,-29);
\fill[diffSame] (21,-28) rectangle (22,-29);
\fill[diffDiff] (22,-28) rectangle (24,-29);
\fill[diffSame] (0,-29) rectangle (17,-30);
\fill[diffDiff] (17,-29) rectangle (18,-30);
\fill[diffSame] (18,-29) rectangle (19,-30);
\fill[diffDiff] (19,-29) rectangle (20,-30);
\fill[diffSame] (20,-29) rectangle (21,-30);
\fill[diffDiff] (21,-29) rectangle (24,-30);
\fill[diffSame] (0,-30) rectangle (16,-31);
\fill[diffDiff] (16,-30) rectangle (17,-31);
\fill[diffSame] (17,-30) rectangle (18,-31);
\fill[diffDiff] (18,-30) rectangle (19,-31);
\fill[diffSame] (19,-30) rectangle (20,-31);
\fill[diffDiff] (20,-30) rectangle (24,-31);
\fill[diffSame] (0,-31) rectangle (15,-32);
\fill[diffDiff] (15,-31) rectangle (18,-32);
\fill[diffSame] (18,-31) rectangle (19,-32);
\fill[diffDiff] (19,-31) rectangle (24,-32);
\fill[diffSame] (0,-32) rectangle (14,-33);
\fill[diffDiff] (14,-32) rectangle (17,-33);
\fill[diffSame] (17,-32) rectangle (18,-33);
\fill[diffDiff] (18,-32) rectangle (24,-33);
\fill[diffSame] (0,-33) rectangle (13,-34);
\fill[diffDiff] (13,-33) rectangle (24,-34);
\fill[diffSame] (0,-34) rectangle (12,-35);
\fill[diffDiff] (12,-34) rectangle (18,-35);
\fill[diffSame] (18,-34) rectangle (19,-35);
\fill[diffDiff] (19,-34) rectangle (24,-35);
\fill[diffSame] (0,-35) rectangle (11,-36);
\fill[diffDiff] (11,-35) rectangle (24,-36);
\fill[diffSame] (0,-36) rectangle (10,-37);
\fill[diffDiff] (10,-36) rectangle (19,-37);
\fill[diffSame] (19,-36) rectangle (21,-37);
\fill[diffDiff] (21,-36) rectangle (24,-37);
\fill[diffSame] (0,-37) rectangle (9,-38);
\fill[diffDiff] (9,-37) rectangle (19,-38);
\fill[diffSame] (19,-37) rectangle (22,-38);
\fill[diffDiff] (22,-37) rectangle (24,-38);
\fill[diffSame] (0,-38) rectangle (8,-39);
\fill[diffDiff] (8,-38) rectangle (11,-39);
\fill[diffSame] (11,-38) rectangle (12,-39);
\fill[diffDiff] (12,-38) rectangle (19,-39);
\fill[diffSame] (19,-38) rectangle (22,-39);
\fill[diffDiff] (22,-38) rectangle (24,-39);
\fill[diffSame] (0,-39) rectangle (7,-40);
\fill[diffDiff] (7,-39) rectangle (19,-40);
\fill[diffSame] (19,-39) rectangle (23,-40);
\fill[diffDiff] (23,-39) rectangle (24,-40);
\fill[diffSame] (0,-40) rectangle (6,-41);
\fill[diffDiff] (6,-40) rectangle (17,-41);
\fill[diffSame] (17,-40) rectangle (18,-41);
\fill[diffDiff] (18,-40) rectangle (19,-41);
\fill[diffSame] (19,-40) rectangle (22,-41);
\fill[diffDiff] (22,-40) rectangle (23,-41);
\fill[diffSame] (23,-40) rectangle (24,-41);
\fill[diffSame] (0,-41) rectangle (5,-42);
\fill[diffDiff] (5,-41) rectangle (20,-42);
\fill[diffSame] (20,-41) rectangle (21,-42);
\fill[diffDiff] (21,-41) rectangle (22,-42);
\fill[diffSame] (22,-41) rectangle (23,-42);
\fill[diffDiff] (23,-41) rectangle (24,-42);
\fill[diffSame] (0,-42) rectangle (4,-43);
\fill[diffDiff] (4,-42) rectangle (7,-43);
\fill[diffSame] (7,-42) rectangle (8,-43);
\fill[diffDiff] (8,-42) rectangle (12,-43);
\fill[diffSame] (12,-42) rectangle (13,-43);
\fill[diffDiff] (13,-42) rectangle (17,-43);
\fill[diffSame] (17,-42) rectangle (18,-43);
\fill[diffDiff] (18,-42) rectangle (20,-43);
\fill[diffSame] (20,-42) rectangle (21,-43);
\fill[diffDiff] (21,-42) rectangle (24,-43);
\fill[diffSame] (0,-43) rectangle (3,-44);
\fill[diffDiff] (3,-43) rectangle (9,-44);
\fill[diffSame] (9,-43) rectangle (10,-44);
\fill[diffDiff] (10,-43) rectangle (17,-44);
\fill[diffSame] (17,-43) rectangle (18,-44);
\fill[diffDiff] (18,-43) rectangle (24,-44);
\fill[diffSame] (0,-44) rectangle (2,-45);
\fill[diffDiff] (2,-44) rectangle (6,-45);
\fill[diffSame] (6,-44) rectangle (7,-45);
\fill[diffDiff] (7,-44) rectangle (13,-45);
\fill[diffSame] (13,-44) rectangle (14,-45);
\fill[diffDiff] (14,-44) rectangle (15,-45);
\fill[diffSame] (15,-44) rectangle (16,-45);
\fill[diffDiff] (16,-44) rectangle (24,-45);
\fill[diffSame] (0,-45) rectangle (1,-46);
\fill[diffDiff] (1,-45) rectangle (2,-46);
\fill[diffSame] (2,-45) rectangle (3,-46);
\fill[diffDiff] (3,-45) rectangle (21,-46);
\fill[diffSame] (21,-45) rectangle (22,-46);
\fill[diffDiff] (22,-45) rectangle (24,-46);
\fill[diffDiff] (0,-46) rectangle (24,-47);
    \draw[white,line width=2.0pt] \staircase{46}{23} -- (24,-23);
    \draw[ink,densely dashed,line width=0.8pt] \staircase{46}{23} -- (24,-23);
    \draw[grid,line width=0.3pt] (0,0) rectangle (24,-47);
    \node[font=\small,anchor=south] at (12,0.6) {(c) (a) compared with (b)};
    \foreach \x/\l in {1/1,24/24}
      \node[font=\scriptsize,anchor=north] at (\x-0.5,-47.3) {$\l$};
  \end{scope}
  % legends
  \node[anchor=north,font=\scriptsize] at (52.5,-50.3) {in (a) and (b):\quad \swatch{white}\,$\Lq$ (quiescent)\quad
    \swatch{stG}\,$\Gq$ (general)\quad \swatch{stA}\,$\Aq$\quad \swatch{stB}\,$\Bq$\quad
    \swatch{stC}\,$\Cq$\quad \swatch{stF}\,$\Fq$ (firing)};
  \node[anchor=north,font=\scriptsize] at (52.5,-53.9) {in (c):\quad \swatch{white}\,quiescent in both,
    $t<i-1$\quad \swatch{diffSame}\,the same state in (a) and (b)\quad
    \swatch{diffDiff}\,different states};
\end{tikzpicture}
\caption{Mazoyer's six-state minimal-time solution. (a) The line of length
$n=24$; time runs downwards from $t=0$ to the firing time $t=2n-2=46$, and the
cells $i=1,\dots,24$ run from left to right. (b) The half-line, which is
started in the same way but has no right end, for the same times; the dotted
line marks the right end of the line in (a). (c) Each cell of (a) compared with
the same cell of (b). Above the front $t=i-1$, both are quiescent
(\cref{lem:lightcone}). The dashed staircase separates the cells with $t+i\le
2n-2$, on which (a) and (b) coincide (\cref{lem:agree}), from the reflected
triangle $R_n$, in which the line reacts to its right end. On the first
anti-diagonal of $R_n$, the \emph{return chain} $t+i=2n-1$, every cell of the
line differs from the same cell of the half-line (\cref{lem:return}); further
down, the two diagrams coincide on some cells and differ on others.}
\label{fig:line}
\end{figure}

\paragraph{The idea}
Write $(t,i)$ for the cell $i$ at time $t$. The \emph{half-line} is started like
a line, with the general in cell $1$, but has no right end. Until the echo of
its right end comes back, a line cannot tell that it is finite: the line of
length $n$ and the half-line agree on every cell $(t,i)$ with $t+i\le2n-2$
(\cref{lem:agree}; \cref{fig:line} shows this for Mazoyer's solution). The line
can therefore differ from the half-line only on the cells $(t,i)$ with
$t+i\ge2n-1$ and $t\le2n-2$, which form a triangle next to the right end, the
\emph{reflected triangle}. So a minimal-time solution consists of one
space-time diagram of the half-line, shared by all lengths, and, for each
length $n$, a reflected triangle in which the line reacts to its right end.
Everything that the reflected triangle of the line of length $n$ learns from
the half-line is contained in the cells of the half-line with $t+i=2n-3$ or
$t+i=2n-2$ (\cref{lem:determinacy}). Conditions on the half-line alone are
therefore cheap to check, and they apply to all lengths at once.

\paragraph{Results}
Our main results concern minimal-time solutions with any number of states.
\begin{enumerate}[(1)]
\item \emph{A late onset of periodicity} (\cref{thm:third}). The front of the
  half-line is the cell $(t,t+1)$ reached by the general's signal at time $t$.
  For $j\ge0$, the cells $(t,t+1-j)$ with $t\ge j$, which stay at distance $j$
  behind the front, form the \emph{depth row} $j$. In a minimal-time solution
  every depth row is eventually periodic in time; let $\theta(j)$ be the time
  from which depth row $j$ is periodic. We prove that
  $\limsup_{j\to\infty}\theta(j)/j\ge3/2$: for every $\varepsilon>0$ there are
  infinitely many $j$ with $\theta(j)\ge(3/2-\varepsilon)j$. The proof rests on
  a pumping lemma (\cref{lem:pumping}): if the reflected triangles of two lines
  of different lengths received the same values on the cells that determine
  their right ends, these ends would fire at the same number of steps after the
  front reaches them, which is impossible (\cref{fig:pumping}). At the time
  $\theta(j)$, depth row $j$ is at the cell $\theta(j)+1-j$, which is close to
  the cell $\theta(j)/3$ or beyond when $\theta(j)\ge(3/2-\varepsilon)j$; for this
  reason we also call the theorem the \emph{speed-$1/3$ barrier}. The name
  describes this bound only: the theorem concerns infinitely many depth rows, not
  all of them, and it does not assert that a boundary or a signal moves at speed
  $1/3$. If the depth rows become periodic where they cross a signal of the
  general, as they do approximately in the two rules that we simulated, the
  bound says that the first division point of the line is asymptotically not in
  the half next to the general; this reading is a heuristic
  (\cref{rem:division}). In simulations up to time $1400$, the depth rows
  $10\le j\le300$ of the five-state rule $\delta_{12}$ of \cref{sec:five}, which
  synchronizes the lines of lengths $2,\dots,12$, become periodic at the time
  $\lceil 3j/2\rceil+2$, on the bound (\cref{fig:theta}).
\item \emph{The left-border barrier} (\cref{thm:left,cor:regular}). The
  half-line is never periodic in time and space on a wedge
  $\{(t,i):s\le i\le\alpha t,\ t\ge t_0\}$ at its left border. In particular it
  is not \emph{eventually regular} in the sense of \cref{sec:left}, that is, it
  is not true that from some time on the half-line consists of finitely many
  periodic blocks, separated by signals of rational speeds, whose lengths grow
  linearly. Other kinds of regular
  behaviour, and periodic regions away from the left border, are not excluded.
  The reason is that the cells near the left border are fired by a finite
  automaton that reads the half-line along the cells with $t+i=2n-3$ and
  $t+i=2n-2$, and a long periodic input would make it fire too early
  (\cref{lem:left,fig:leftproof}). For example, no five-state rule whose
  half-line agrees with that of $\delta_{12}$ on the cells with $t+i\le78$
  synchronizes the line of length $518$ in minimal time
  (\cref{cor:germ12,fig:delta12}).
\item \emph{Finitely many early rows} (\cref{cor:reduce}). The front reaches
  the right end of the line of length $n$ at time $n-1$. For a fixed $T$, the
  cells of the reflected triangles at the times $n-1,\dots,n-1+T$, which we call
  the first $T+1$ rows, take only finitely many forms over all lengths $n$, and a
  certificate computed from the half-line determines them. This turns the
  condition that no line fires during these $T+1$ steps into finitely many
  checks (\cref{lem:G,fig:early}).
\end{enumerate}
With these conditions and satisfiability solvers whose negative answers are
certified by independently checked proofs, we obtain the following results.
Four states are excluded by the lines of lengths $2,\dots,5$ together with
three parts of the lines of lengths $6,\dots,10$: the first five rows, the
return chains in these rows, and the dependency cones of the right ends at the
firing time (\cref{thm:four}); this is sharp in several directions
(\cref{prop:sharp}). Five rules with five states synchronize exactly
the lines of lengths $2,\dots,14$ (\cref{prop:nearmiss}), so these lines alone
cannot exclude five states; for four states, the longest such range is $2,\dots,8$
(\cref{prop:sharp}). On
the cells of its half-line with $t+i\le78$, every five-state minimal-time
solution uses at least $15$ distinct neighbourhoods besides the quiescent one
(\cref{prop:germs}). Finally, we confirm a conditional five-state result
reported by Balzer in 1967, even under a weaker reading of his conditions: no
five-state rule that satisfies them synchronizes all lines of lengths
$2,\dots,12$ (\cref{prop:balzer}). His symmetry condition is needed at the
borders of the line as well: without it there, some rule that satisfies the
remaining conditions in the weak reading synchronizes all these lines
(\cref{sec:balzer}).
\Cref{tab:results} lists the results with the kind of evidence behind each of
them, and separately the observations from simulations.

\begin{table}[tbp]
\centering\small
\caption{The results of this paper, the evidence behind them, and the figures
and tables that illustrate them. ``Proof'' is a proof in the text. A
``certified computation'' is the unsatisfiability of a propositional formula,
with a proof in the LRAT format that two independent checkers verify, one of
them formally verified. An
``exhaustive enumeration'' or an ``exact finite simulation'' decides a finite
statement completely, and the proof in the text extends it where needed. The
last row lists observations from finite simulations, which the paper does not
claim beyond the simulated ranges.}
\label{tab:results}
\begin{tabular}{@{}>{\raggedright\arraybackslash}p{0.36\textwidth}l>{\raggedright\arraybackslash}p{0.25\textwidth}>{\raggedright\arraybackslash}p{0.14\textwidth}@{}}
\toprule
Result & Where & Evidence & Shown in\\
\midrule
Light cone; agreement with the half-line; return chain & \ref{lem:lightcone}, \ref{lem:agree}, \ref{lem:return} & proof & Figs.~\ref{fig:line}--\ref{fig:return}\\
Determinacy; cone of the right end & \ref{lem:determinacy}, \ref{lem:cone} & proof & Figs.~\ref{fig:anatomy}, \ref{fig:pumping}\\
Pumping lemma; late onset of periodicity (speed-$1/3$ barrier) & \ref{lem:pumping}, \ref{thm:third} & proof & Figs.~\ref{fig:pumping}, \ref{fig:theta}\\
Right band & \ref{lem:rightband} & proof & Fig.~\ref{fig:rightband}\\
Left-border lemma and barrier; no eventually regular half-line (\cref{sec:left}) & \ref{lem:left}, \ref{thm:left}, \ref{cor:regular} & proof & Figs.~\ref{fig:chains}--\ref{fig:wedge}\\
Finitely many early rows; the formulas $\G_k(T,N_1,C)$ & \ref{cor:reduce}, \ref{lem:G} & proof & Figs.~\ref{fig:early}, \ref{fig:regions}\\
Four states excluded by the lines of lengths $2,\dots,5$ and parts of those of lengths $6,\dots,10$ & \ref{thm:four} & proof and certified computation & Fig.~\ref{fig:regions}, Table~\ref{tab:four}\\
Sharpness of \cref{thm:four} & \ref{prop:sharp} & exhaustive enumeration; models checked by simulation & Table~\ref{tab:four}\\
Five rules for exactly the lengths $2,\dots,14$ & \ref{prop:nearmiss} & proof and exact finite simulation & Figs.~\ref{fig:delta14}, \ref{fig:firstfire}, Tables~\ref{tab:delta14}, \ref{tab:rules}\\
At least $15$ neighbourhoods on the half-line of a five-state solution & \ref{prop:germs} & proof, exhaustive enumeration and certified computation & Fig.~\ref{fig:germs}, Table~\ref{tab:rules}\\
No five-state minimal-time solution has the half-line of $\delta_{12}$ and $\delta_{13}$ & \ref{cor:germ12} & proof and exact finite simulation & Fig.~\ref{fig:delta12}\\
Balzer's conditional five-state result; (B2) at the borders is needed & \ref{prop:balzer}, \ref{sec:balzer} & certified computation; models checked by simulation & Table~\ref{tab:balzer}\\
\midrule
\emph{Observations:} the onset times $\theta_H(j)$ for Mazoyer's solution and $\delta_{12}$; the patterns of the half-lines of $\delta_{12}$ and $\delta_{14}$; $\PUMP$ and \cref{lem:left} in finite ranges & \ref{rem:division}, \ref{sec:left}, \ref{sec:five} & finite simulation only & Figs.~\ref{fig:theta}, \ref{fig:halflines}, \ref{fig:delta14}\\
\bottomrule
\end{tabular}
\end{table}

\paragraph{Related work}
For the structure of minimal-time solutions see Mazoyer~\cite{Mazoyer96}.
Nonexistence results for minimal-time solutions are known for variants of the
problem, such as the variant with sub-generals~\cite{YamashitaEtAl14} and some
geometric configurations~\cite{Kobayashi19}. Our pumping lemma and our
left-border lemma are indistinguishability arguments for the original problem:
repetitions forced by the pigeonhole principle would make a solution unable to
tell two situations apart, so that it would fire too early. The depth rows of
the half-line are eventually periodic by a simple finite-state argument
(\cref{thm:third}), and periodic regions separated by signals are a classical
way to describe the behaviour of cellular automata~\cite{MazoyerTerrier99};
our contribution is a lower bound on where this periodicity can start. Lower bounds on the number of
states are known for restricted classes and for variants such as the
ring~\cite{BerthiaumeEtAl04}; see the survey~\cite{Correa17}. Satisfiability
solvers have been applied to cellular automata before~\cite{DAntonioDelzanno04},
and clausal proofs checked by programs independent of the
solvers~\cite{WetzlerHeuleHunt14,CruzFilipeEtAl17} have certified large
computer-assisted results in combinatorics~\cite{HeuleKullmannMarek16,Heule18}.
We are not aware
of statements equivalent to
\cref{thm:third,thm:left,cor:regular,cor:reduce} in the literature.

\paragraph{Organization}
\Cref{sec:lines} introduces lines, the half-line and their agreement.
\Cref{sec:right} studies the right end of the reflected triangles and proves
the late onset of periodicity, \cref{sec:left} studies their left end and proves the
left-border barrier, and \cref{sec:early} reduces their first rows to finitely
many cases. \Cref{sec:applications} applies these results to four and five
states, and \cref{sec:discussion} discusses open questions.
\Cref{app:encoding} gives the clauses of the encodings, and
\cref{app:artefacts} lists the computational claims with their programs,
formulas and proofs. A list of symbols is given at the end. All data behind the
computational statements, with the scripts that rerun them, are archived at
Zenodo, as described in the data availability statement at the end.

%% ---------------------------------------------------------------------------
\section{Lines and the Half-Line}\label{sec:lines}

\subsection{Rules, Lines and Synchronization}

Let $S$ be a finite set of \emph{states} that contains three distinct states:
the \emph{quiescent} state $\Lq$, the \emph{general} $\Gq$ and the
\emph{firing} state $\Fq$. The other states are called \emph{auxiliary}; the
rules of this paper use $\Aq$, $\Bq$ and, for Mazoyer's solution, $\Cq$. The
\emph{working} states are $W:=S\setminus\{\Fq\}$, and $\ast\notin S$ denotes the
border. A \emph{neighbourhood} is a triple $(x,y,z)\in(W\cup\{\ast\})\times
W\times(W\cup\{\ast\})$ with $(x,z)\ne(\ast,\ast)$, and a \emph{rule} is a map
$\delta$ from the set of neighbourhoods to $S$ such that
\begin{equation}\label{eq:quiescence}
\delta(\Lq,\Lq,\Lq)=\delta(\Lq,\Lq,\ast)=\Lq .
\end{equation}
If $|S|=k$, there are $k(k-1)k-(k-1)=(k-1)^2(k+1)$ neighbourhoods: $45$ for
four states, $96$ for five and $175$ for six.

The \emph{line of length $n\ge2$} is the partial map
$C_n:\mathbb N\times\{1,\dots,n\}\to S$ given by $C_n(0,1)=\Gq$, $C_n(0,i)=\Lq$
for $2\le i\le n$, and
\[
C_n(t+1,i)=\delta\bigl(C_n(t,i-1),\,C_n(t,i),\,C_n(t,i+1)\bigr)
\]
if the three arguments are defined and none of them is $\Fq$; otherwise
$C_n(t+1,i)$ is undefined. Here $C_n(t,0)=C_n(t,n+1)=\ast$ stand for the
borders. Thus a cell is defined as long as nothing in its neighbourhood has
fired. The rule \emph{synchronizes} the line of length $n$ at time $T$ if
$C_n(T,i)=\Fq$ for $1\le i\le n$ and $C_n(t,i)\in W$ for all $t<T$ and $1\le
i\le n$. A \emph{minimal-time solution} is a rule that synchronizes the line of
length $n$ at time $2n-2$ for every $n\ge2$; it has $k$ states if $|S|=k$.

The \emph{half-line} $C_\infty:\mathbb N\times\{1,2,\dots\}\to S$ is defined in
the same way, with $C_\infty(0,1)=\Gq$, $C_\infty(0,i)=\Lq$ for $i\ge2$, and a
border on the left only. We draw space-time diagrams with time running
downwards and the cells from left to right, as in \cref{fig:line}. The
\emph{front} is the diagonal $t=i-1$, and the \emph{anti-diagonal} $t+i=m$ is
the set of cells $(t,i)$ with $t+i=m$.

\subsection{Agreement}

Information travels at most one cell per step, so the cells that the general
cannot yet have influenced stay quiescent.

\begin{lemma}[light cone]\label{lem:lightcone}
For every rule, $C_n(t,i)=C_\infty(t,i)=\Lq$ whenever $t<i-1$ (and $i\le n$
for $C_n$).
\end{lemma}

\begin{proof}
By induction on $t$. For $t=0$ this is the initial condition. For $t\ge1$, the
cells $(t-1,i')$ with $i'\in\{i-1,i,i+1\}$ satisfy $t-1<i'-1$, so each of them
is $\Lq$ or the right border, and \eqref{eq:quiescence} gives
$C_n(t,i)=C_\infty(t,i)=\Lq$.
\end{proof}

The right border of the line of length $n$ is first felt by the cell $n$ at
time $n-1$, when the front reaches it, and its influence travels back to the
left at speed at most one. Before this echo arrives, the line cannot be told
apart from the half-line.

\begin{lemma}[agreement]\label{lem:agree}
For every rule, every $n\ge2$, $1\le i\le n$ and $t\ge0$ with $t+i\le 2n-2$,
the values $C_n(t,i)$ and $C_\infty(t,i)$ are both undefined or both defined
and equal.
\end{lemma}

\begin{proof}
By induction on $t$; the case $t=0$ is clear. Let $t\ge1$. If $i=n$, then
$t\le n-2<i-1$, and both values are $\Lq$ by \cref{lem:lightcone}. If $i<n$,
the cells $(t-1,j)$, $j\in\{i-1,i,i+1\}$, satisfy $(t-1)+j\le 2n-2$ and $j\le
n$, so by induction they are both undefined or equal in $C_n$ and $C_\infty$
(for $j=0$ both are the border), and the recursion gives the claim.
\end{proof}

So the border can only matter in the \emph{reflected triangle}
\[
R_n:=\{(t,i):\ 1\le i\le n,\ t+i\ge 2n-1,\ t\le 2n-2\},
\]
which is bounded by the anti-diagonal $t+i=2n-1$, the right border and the
firing row $t=2n-2$ (\cref{fig:line}). The agreement lemma already shows why
time $2n-2$ is optimal.

\begin{figure}[t]
\centering
%% Figure: why no line is synchronized before time 2n-2 (Remark 2.3). Schematic, with n = 8, a
%% hypothetical firing time T = 11 <= 2n-3 and n' = 13 > T+1; cell (t,i) is [i-1,i] x [-t-1,-t].
\begin{tikzpicture}[x=3.1mm,y=3.1mm]
  % (a) the line of length n = 8, synchronized at time T = 11
  \begin{scope}
    \foreach \t in {0,...,11} {\foreach \i in {1,...,8} {%
      \pgfmathtruncatemacro{\s}{\t+\i}\pgfmathtruncatemacro{\lc}{\i-1}%
      \def\cc{rgRefl}%
      \ifnum\s<15 \def\cc{rgAgree}\fi
      \ifnum\t=11 \def\cc{stF}\fi
      \ifnum\t<\lc \def\cc{white}\fi
      \filldraw[fill=\cc,draw=grid!45,line width=0.15pt] (\i-1,-\t) rectangle (\i,-\t-1);}}
    \begin{scope}\clip (0,0) rectangle (8,-12);
      \draw[white,line width=1.8pt] \staircase{14}{7} -- (8,-7);
      \draw[ink,densely dashed,line width=0.8pt] \staircase{14}{7} -- (8,-7);
    \end{scope}
    \draw[ink,line width=0.6pt] (0,0) rectangle (8,-12);
    \node[font=\small,anchor=south] at (4,0.4) {(a) line of length $n$};
    \node[font=\scriptsize,anchor=east] at (-0.3,-0.5) {$t=0$};
    \node[font=\scriptsize,anchor=east] at (-0.3,-11.5) {$t=T$};
    \node[font=\scriptsize,anchor=north,align=center] at (4,-12.4) {synchronized at a\\ time $T\le 2n-3$};
  \end{scope}
  % (b) the half-line up to time T
  \begin{scope}[shift={(12,0)}]
    \foreach \t in {0,...,11} {\foreach \i in {1,...,14} {%
      \pgfmathtruncatemacro{\s}{\t+\i}\pgfmathtruncatemacro{\lc}{\i-1}%
      \def\cc{gNone}%
      \ifnum\s<15 \def\cc{rgAgree}\ifnum\t=11 \def\cc{stF}\fi\fi
      \ifnum\t<\lc \def\cc{white}\fi
      \filldraw[fill=\cc,draw=grid!45,line width=0.15pt] (\i-1,-\t) rectangle (\i,-\t-1);}}
    \begin{scope}\clip (0,0) rectangle (14,-12);
      \draw[white,line width=1.8pt] \staircase{14}{14};
      \draw[ink,densely dashed,line width=0.8pt] \staircase{14}{14};
    \end{scope}
    \draw[ink,line width=0.6pt] (0,0) -- (0,-12) -- (14,-12) (0,0) -- (14,0);
    \node[font=\scriptsize,fill=white,inner sep=1pt,rotate=45] at (9.6,-4.2) {$t+i=2n-2$};
    \node[font=\small,anchor=south] at (7,0.4) {(b) half-line};
    \node[font=\scriptsize,anchor=north,align=center] at (7,-12.4) {$C_\infty(T,1)=C_n(T,1)=\Fq$\\
      (\cref{lem:agree}, as $T+1\le 2n-2$)};
  \end{scope}
  % (c) the line of length n' = 13 > T+1 up to time T
  \begin{scope}[shift={(30,0)}]
    \foreach \t in {0,...,11} {\foreach \i in {1,...,13} {%
      \pgfmathtruncatemacro{\s}{\t+\i}\pgfmathtruncatemacro{\lc}{\i-1}%
      \def\cc{gNone}%
      \ifnum\s<15 \def\cc{rgAgree}\ifnum\t=11 \def\cc{stF}\fi\fi
      \ifnum\t<\lc \def\cc{white}\fi
      \filldraw[fill=\cc,draw=grid!45,line width=0.15pt] (\i-1,-\t) rectangle (\i,-\t-1);}}
    \begin{scope}\clip (0,0) rectangle (13,-12);
      \draw[white,line width=1.8pt] \staircase{14}{13};
      \draw[ink,densely dashed,line width=0.8pt] \staircase{14}{13};
    \end{scope}
    \draw[ink,line width=0.6pt] (0,0) rectangle (13,-12);
    \draw[ink,line width=1.1pt] (12,-11) rectangle (13,-12);
    \node[font=\small,anchor=south] at (6.5,0.4) {(c) line of length $n'>T+1$};
    \node[font=\scriptsize,anchor=north,align=center] at (6.5,-12.4) {$C_{n'}(T,1)=\Fq$, but the cell $n'$\\
      is still quiescent (\cref{lem:lightcone})};
  \end{scope}
  % legend
  \node[anchor=north,font=\scriptsize,align=left] at (21.5,-15.6) {%
    \swatch{white}\,quiescent, $t<i-1$\qquad
    \swatch{rgAgree}\,$t+i\le 2n-2$: the same in (a), (b) and (c) (\cref{lem:agree})\qquad
    \swatch{stF}\,$\Fq$\\[0.4ex]
    \swatch{rgRefl}\,the reflected triangle $R_n$\qquad
    \swatch{gNone}\,the same in (b) and (c) (\cref{lem:agree} for the length $n'$)};
\end{tikzpicture}
\caption{Why no line can be synchronized before time $2n-2$
(\cref{rem:lowerbound}), drawn for $n=8$, $T=11$ and $n'=13$. (a)~If the line
of length $n$ were synchronized at a time $T\le 2n-3$, its cell $(T,1)$ would
satisfy $T+1\le2n-2$ and lie above the dashed staircase, where the line agrees
with the half-line. (b)~Hence the cell $(T,1)$ of the half-line would be in
state $\Fq$. (c)~A line of length $n'>T+1$ agrees with the half-line on every
cell with $t\le T$, because then $t+i\le T+n'\le 2n'-2$. Its cell $1$ would be
in state $\Fq$ at time $T$ while its cell $n'$ is still quiescent, so this line
could never be synchronized.}
\label{fig:lowerbound}
\end{figure}
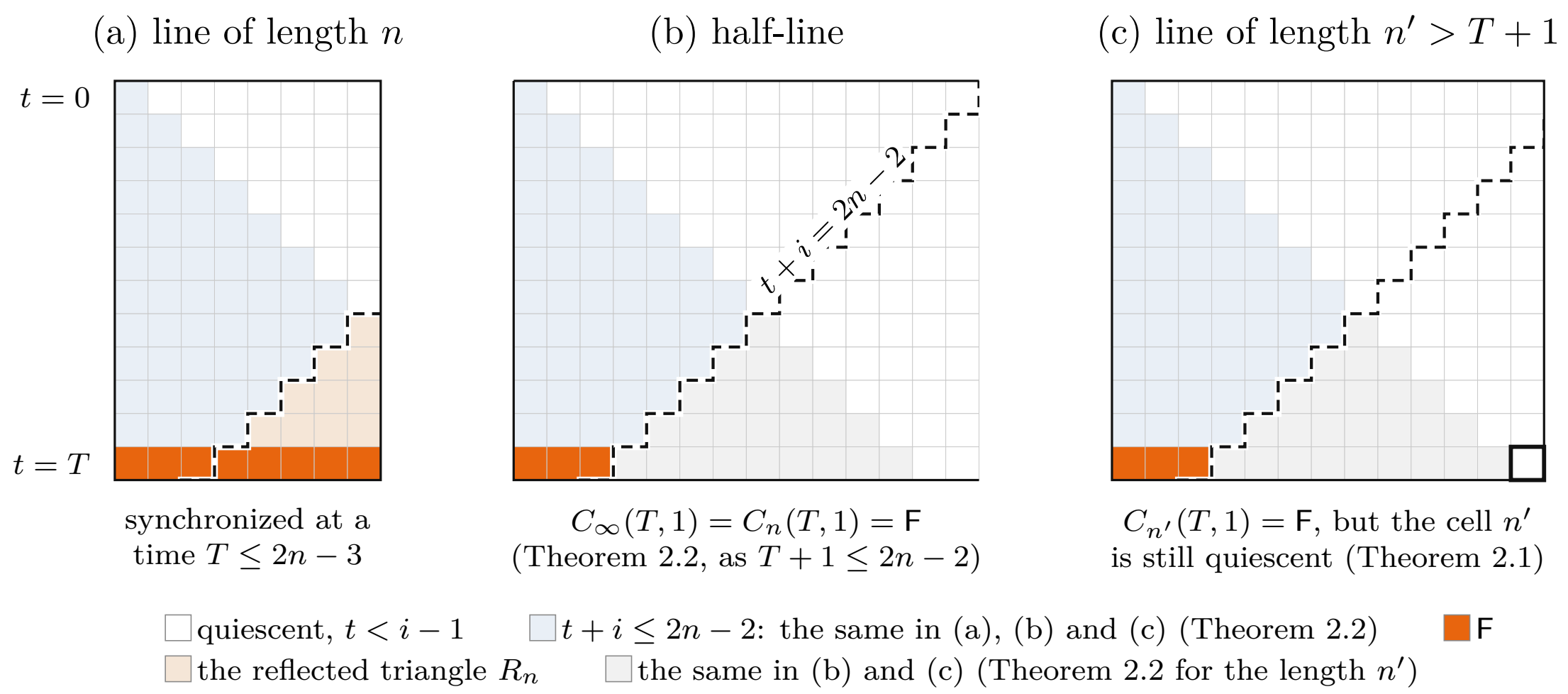

\begin{remark}\label{rem:lowerbound}
Suppose that a rule synchronizes every line, the line of length $n$ at some
time $T\le 2n-3$. Since $T+1\le 2n-2$, \cref{lem:agree} gives
$C_\infty(T,1)=C_n(T,1)=\Fq$, and then $C_{n'}(T,1)=\Fq$ for every $n'\ge n$,
by \cref{lem:agree} again. For $n'>T+1$ the cell $n'$ is still quiescent at
time $T$ by \cref{lem:lightcone}, so the line of length $n'$ has a cell in
state $\Fq$ at time $T$ although not all of its cells are in state $\Fq$; it
can then never be synchronized, a contradiction (\cref{fig:lowerbound}).
\end{remark}

\subsection{The Return Chain}

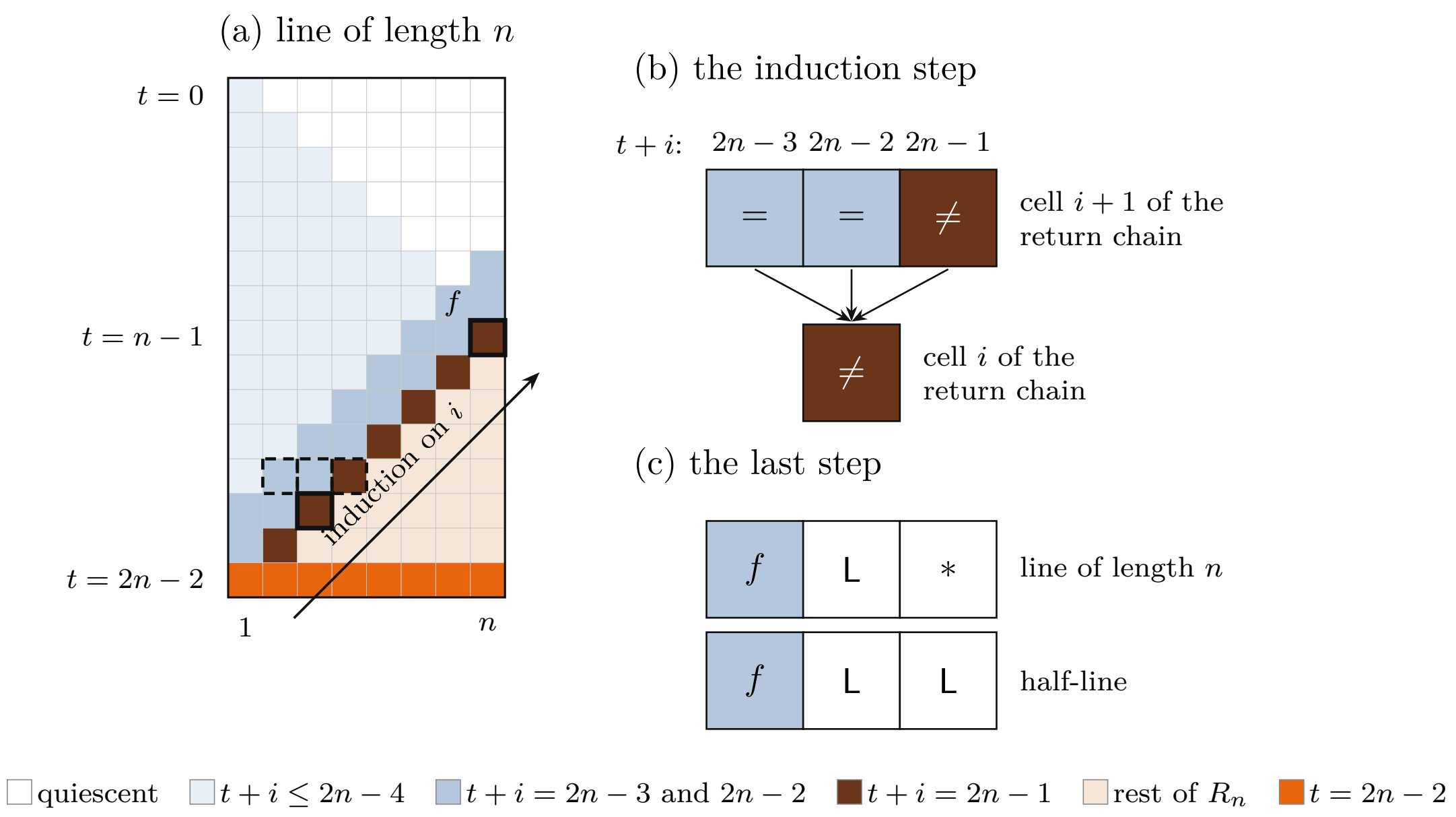
\begin{figure}[t]
\centering
%% Figure: the induction of Lemma 2.4 along the return chain. Schematic, n = 8; cell (t,i) is
%% [i-1,i] x [-t-1,-t] in panel (a).
\begin{tikzpicture}
  % (a) the line of length n = 8
  \begin{scope}[x=3.4mm,y=3.4mm]
    \foreach \t in {0,...,14} {\foreach \i in {1,...,8} {%
      \pgfmathtruncatemacro{\s}{\t+\i}\pgfmathtruncatemacro{\lc}{\i-1}%
      \def\cc{rgRefl}%
      \ifnum\s<16 \def\cc{rgReturn}\fi
      \ifnum\s<15 \def\cc{rgInput}\fi
      \ifnum\s<13 \def\cc{rgAgree}\ifnum\t<\lc \def\cc{white}\fi\fi
      \ifnum\t=14 \def\cc{stF}\fi
      \filldraw[fill=\cc,draw=grid!45,line width=0.15pt] (\i-1,-\t) rectangle (\i,-\t-1);}}
    \draw[ink,line width=0.6pt] (0,0) rectangle (8,-15);
    % the step of panel (b), for i = 3
    \foreach \x/\y in {1/11,2/11,3/11} \draw[ink,line width=0.9pt,densely dashed] (\x,-\y) rectangle (\x+1,-\y-1);
    \draw[ink,line width=1.3pt] (2,-12) rectangle (3,-13);
    % the front cell f_{n-1} = (n-2, n-1) and the last cell (n-1, n) of the return chain
    \node[font=\scriptsize] at (6.5,-6.5) {$f$};
    \draw[ink,line width=1.3pt] (7,-7) rectangle (8,-8);
    % direction of the induction
    \draw[-{Stealth[length=1.8mm]},ink,line width=0.7pt] (1.9,-15.6) -- (9.0,-8.5);
    \node[font=\scriptsize,rotate=45,anchor=south] at (5.2,-11.9) {induction on $i$};
    \node[font=\small,anchor=south] at (4,0.4) {(a) line of length $n$};
    \foreach \y/\l in {0/{t=0},7/{t=n-1},14/{t=2n-2}}
      \node[font=\scriptsize,anchor=east] at (-0.3,-\y-0.5) {$\l$};
    \foreach \x/\l in {1/1,8/n}
      \node[font=\scriptsize,anchor=north] at (\x-0.5,-15.2) {$\l$};
  \end{scope}
  % (b) the induction step
  \begin{scope}[shift={(4.7,-0.9)},x=9.5mm,y=9.5mm]
    \filldraw[fill=rgInput,draw=ink,line width=0.5pt] (0,0) rectangle (1,-1);
    \filldraw[fill=rgInput,draw=ink,line width=0.5pt] (1,0) rectangle (2,-1);
    \filldraw[fill=rgReturn,draw=ink,line width=0.5pt] (2,0) rectangle (3,-1);
    \filldraw[fill=rgReturn,draw=ink,line width=0.5pt] (1,-1.6) rectangle (2,-2.6);
    \node[font=\small] at (0.5,-0.5) {$=$};
    \node[font=\small] at (1.5,-0.5) {$=$};
    \node[font=\small,text=white] at (2.5,-0.5) {$\neq$};
    \node[font=\small,text=white] at (1.5,-2.1) {$\neq$};
    \foreach \x in {0.5,1.5,2.5} \draw[-{Stealth[length=1.5mm]},ink,line width=0.5pt] (\x,-1.03) -- (1.5,-1.57);
    \node[font=\scriptsize,anchor=south] at (0.5,0.02) {$2n-3$};
    \node[font=\scriptsize,anchor=south] at (1.5,0.02) {$2n-2$};
    \node[font=\scriptsize,anchor=south] at (2.5,0.02) {$2n-1$};
    \node[font=\scriptsize,anchor=east] at (-0.1,0.25) {$t+i$:};
    \node[font=\scriptsize,anchor=west,align=left] at (3.1,-0.5) {cell $i+1$ of the\\ return chain};
    \node[font=\scriptsize,anchor=west,align=left] at (2.1,-2.1) {cell $i$ of the\\ return chain};
    \node[font=\small,anchor=south west] at (-0.9,0.7) {(b) the induction step};
  \end{scope}
  % (c) the last cell of the return chain
  \begin{scope}[shift={(4.7,-4.35)},x=9.5mm,y=9.5mm]
    \foreach \row/\z/\lab in {0/{\ast}/{line of length $n$}, 1.15/{\Lq}/{half-line}} {
      \filldraw[fill=rgInput,draw=ink,line width=0.5pt] (0,-\row) rectangle (1,-\row-1);
      \filldraw[fill=white,draw=ink,line width=0.5pt] (1,-\row) rectangle (2,-\row-1);
      \filldraw[fill=white,draw=ink,line width=0.5pt] (2,-\row) rectangle (3,-\row-1);
      \node[font=\small] at (0.5,-\row-0.5) {$f$};
      \node[font=\small] at (1.5,-\row-0.5) {$\Lq$};
      \node[font=\small] at (2.5,-\row-0.5) {$\z$};
      \node[font=\scriptsize,anchor=west] at (3.1,-\row-0.5) {\lab};
    }
    \node[font=\small,anchor=south west] at (-0.9,0.25) {(c) the last step};
  \end{scope}
  % legend
  \node[anchor=north,font=\scriptsize,align=left] at (4.9,-6.75) {%
    \swatch{white}\,quiescent\quad \swatch{rgAgree}\,$t+i\le 2n-4$\quad
    \swatch{rgInput}\,$t+i=2n-3$ and $2n-2$\quad \swatch{rgReturn}\,$t+i=2n-1$\quad
    \swatch{rgRefl}\,rest of $R_n$\quad \swatch{stF}\,$t=2n-2$};
\end{tikzpicture}
\caption{The proof of \cref{lem:return}, drawn for $n=8$. (a)~The line of
length $n$. On the cells with $t+i\le2n-2$, including the two anti-diagonals
$t+i=2n-3$ and $t+i=2n-2$, it agrees with the half-line (\cref{lem:agree}). The
return chain $t+i=2n-1$ starts at the cell $(2n-2,1)$, which is $\Fq$ in the
line but not in the half-line, and the induction climbs the chain to the cell
$(n-1,n)$. (b)~The induction step, outlined in (a) for $i=3$: the left and
middle predecessors of the cell $i$ of the return chain have the same state in
the line and in the half-line ($=$); since the cell differs ($\neq$), so does
its right predecessor, the cell $i+1$ of the return chain. (c)~The last cell
$(n-1,n)$ of the return chain, outlined in (a), is computed from $f$, $\Lq$,
$\ast$ in the line and from $f$, $\Lq$, $\Lq$ in the half-line, where
$f=f_{n-1}$ is the front cell; by \eqref{eq:quiescence} the results can differ
only if $f\neq\Lq$.}
\label{fig:return}
\end{figure}

The first anti-diagonal of $R_n$ is where the line starts to differ from the
half-line, and in a minimal-time solution it differs there everywhere
(\cref{fig:line}(c) and \cref{fig:return}).

\begin{lemma}[return chain]\label{lem:return}
Let $n\ge2$. If $\delta$ synchronizes the lines of lengths $n$ and $n+1$ at
times $2n-2$ and $2n$, then $C_n(2n-1-i,i)\ne C_\infty(2n-1-i,i)$ for $1\le i\le
n$, and the front cell $f_{n-1}:=C_\infty(n-2,n-1)$ is not $\Lq$.
\end{lemma}

\begin{proof}
For $i=1$ we have $C_n(2n-2,1)=\Fq$, whereas
$C_\infty(2n-2,1)=C_{n+1}(2n-2,1)\ne\Fq$ by \cref{lem:agree} (applied to the
length $n+1$) and because $2n-2<2n$. Suppose that the claim holds for some
$i<n$. The left and middle predecessors of the cell $(2n-1-i,i)$, namely
$(2n-2-i,i-1)$ and $(2n-2-i,i)$, lie on the anti-diagonals $t+i=2n-3$ and
$2n-2$ (for $i=1$ the left one is the border), so they agree in $C_n$ and
$C_\infty$ by \cref{lem:agree}. As the cell itself differs, so do the right
predecessors $(2n-2-i,i+1)$. All these cells are defined: those of $C_n$
because no cell of the line of length $n$ fires before time $2n-2$, and those of
$C_\infty$ because they agree with $C_{n+1}$ (\cref{lem:agree}), which does not
fire before time $2n$. This is the claim for $i+1$. For
$i=n$ the claim concerns the cell $(n-1,n)$. Its left predecessor satisfies
$C_n(n-2,n-1)=C_\infty(n-2,n-1)=f_{n-1}$ by \cref{lem:agree}, its middle
predecessor $(n-2,n)$ is quiescent in both diagrams, and its right predecessor
is the border in $C_n$ and the quiescent cell $(n-2,n+1)$ in $C_\infty$
(\cref{lem:lightcone}). So the claim reads
$\delta(f_{n-1},\Lq,\ast)\ne\delta(f_{n-1},\Lq,\Lq)$. This is impossible
if $f_{n-1}=\Lq$, by \eqref{eq:quiescence}.
\end{proof}

We call the anti-diagonal $t+i=2n-1$ the \emph{return chain} of the line of
length $n$. Finally, a minimal-time solution never fires on the half-line:
given a cell $(t,i)$, take $n\ge t+i+1$; then $C_\infty(t,i)=C_n(t,i)$ by
\cref{lem:agree}, and $t<2n-2$, so
\begin{equation}\label{eq:neverfire}
C_\infty(t,i)\in W\qquad\text{for all }t\ge0\text{ and }i\ge1 .
\end{equation}

%% ---------------------------------------------------------------------------
\section{The Right End}\label{sec:right}

The reflected triangle $R_n$ is computed from the half-line, and its right end
fires at time $2n-2$. We show that the right end at the firing time depends
only on about $n/2$ cells of each of two anti-diagonals of the half-line. If
two lines of different lengths received the same values on these cells, their
right ends would fire at the same relative time, which is impossible (\cref{lem:pumping}). This
pumping lemma is a condition on the half-line alone. It implies a lower bound
on the times from which the cells behind the front of the half-line are
periodic (\cref{thm:third}), which we also call the speed-$1/3$ barrier, for a
reason explained after its proof.

\subsection{Relative Coordinates and Determinacy}

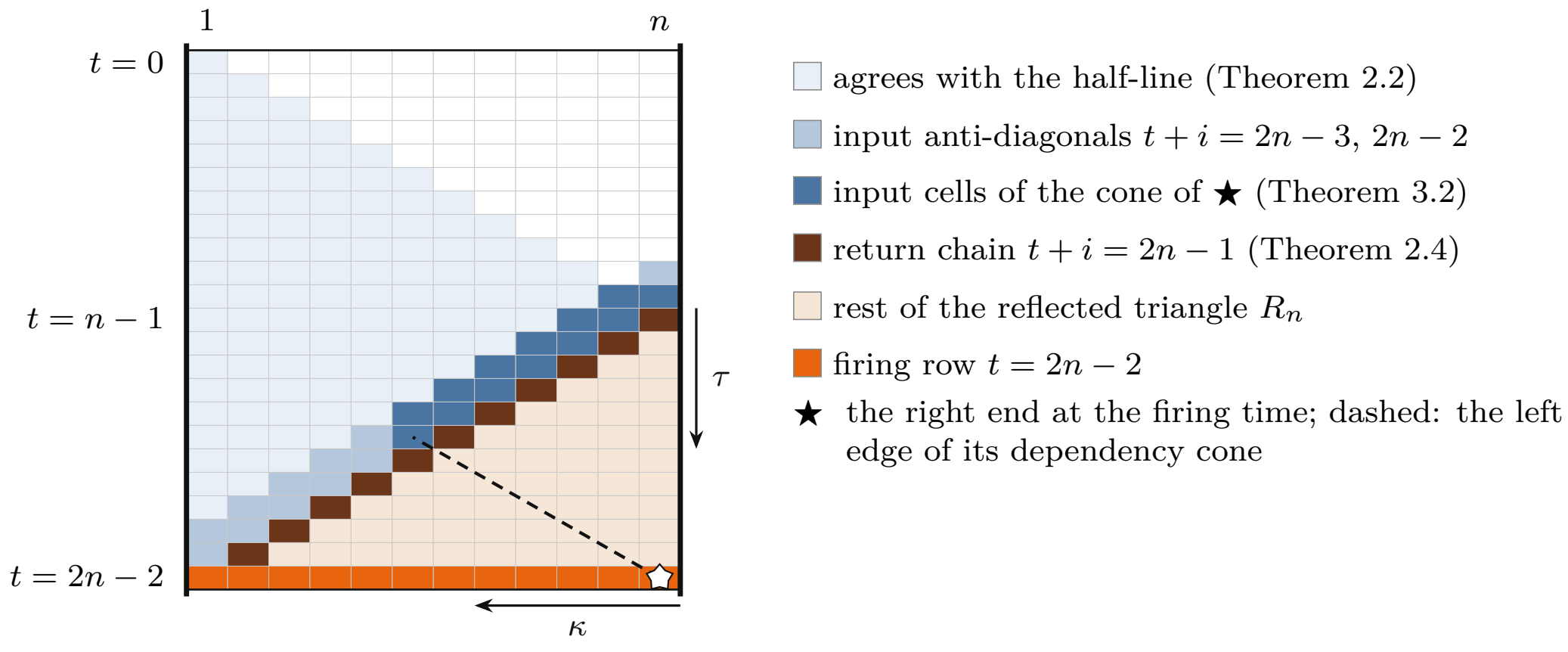
\begin{figure}[t]
\centering
%% Figure: anatomy of the line of length n = 12, coloured by region (Lemmas 2.2, 2.4, 3.1, 3.2).
\begin{tikzpicture}
  \def\u{0.36}\def\v{0.205}\def\n{12}
  \pgfmathtruncatemacro{\T}{2*\n-2}
  \foreach \t in {0,...,\T} {%
    \foreach \i in {1,...,\n} {%
      \pgfmathtruncatemacro{\s}{\t+\i}
      \pgfmathtruncatemacro{\lc}{\i-1}
      \pgfmathtruncatemacro{\kap}{\n-\i}
      \def\cc{white}%
      \ifnum\t<\lc\relax\else\def\cc{rgAgree}\fi
      \ifnum\s>20
        \ifnum\s<23
          \def\cc{rgInput}%
          \ifnum\kap<7
            \ifnum\s=22 \def\cc{rgCone}\fi
            \ifnum\s=21 \ifnum\kap>0 \def\cc{rgCone}\fi\fi
          \fi
        \fi
      \fi
      \ifnum\s=23 \def\cc{rgReturn}\fi
      \ifnum\s>23 \def\cc{rgRefl}\fi
      \ifnum\t=\T \def\cc{stF}\fi
      \filldraw[fill=\cc,draw=grid!45,line width=0.15pt]
        (\i*\u-\u,-\t*\v) rectangle (\i*\u,-\t*\v-\v);
    }}
  \draw[ink,line width=0.5pt] (0,0) rectangle (\n*\u,-\T*\v-\v);
  \draw[ink,line width=1.3pt] (0,0.06) -- (0,-\T*\v-\v-0.06);
  \draw[ink,line width=1.3pt] (\n*\u,0.06) -- (\n*\u,-\T*\v-\v-0.06);
  % left edge of the dependency cone of the right end at the firing time
  \draw[ink,densely dashed,line width=0.8pt]
    (\n*\u-0.5*\u,-\T*\v-0.5*\v) -- (5.5*\u,-16.5*\v);
  \node[star,star points=5,fill=white,draw=ink,line width=0.4pt,inner sep=0pt,
        minimum size=0.24cm] at (\n*\u-0.5*\u,-\T*\v-0.5*\v) {};
  % relative coordinates at the right end
  \draw[-{Stealth[length=1.6mm]},ink,line width=0.6pt]
    (\n*\u+0.14,-11*\v) -- node[right,font=\scriptsize] {$\tau$} (\n*\u+0.14,-17*\v);
  \draw[-{Stealth[length=1.6mm]},ink,line width=0.6pt]
    (\n*\u,-\T*\v-\v-0.14) -- node[below,font=\scriptsize] {$\kappa$} (\n*\u-5*\u,-\T*\v-\v-0.14);
  \node[font=\scriptsize,anchor=east] at (-0.06,-0.5*\v) {$t=0$};
  \node[font=\scriptsize,anchor=east] at (-0.06,-11.5*\v) {$t=n-1$};
  \node[font=\scriptsize,anchor=east] at (-0.06,-22.5*\v) {$t=2n-2$};
  \node[font=\scriptsize,anchor=south] at (0.5*\u,0.04) {$1$};
  \node[font=\scriptsize,anchor=south] at (\n*\u-0.5*\u,0.04) {$n$};
  % legend
  \begin{scope}[font=\scriptsize]
    \foreach \c/\l [count=\k from 0] in {
        rgAgree/{agrees with the half-line (\cref{lem:agree})},
        rgInput/{input anti-diagonals $t+i=2n-3,\,2n-2$},
        rgCone/{input cells of the cone of $\bigstar$ (\cref{lem:cone})},
        rgReturn/{return chain $t+i=2n-1$ (\cref{lem:return})},
        rgRefl/{rest of the reflected triangle $R_n$},
        stF/{firing row $t=2n-2$}}
      \node[anchor=west,inner sep=0pt] at (5.3,-0.25-0.5*\k) {\swatch{\c}\ \l};
    \node[anchor=west,inner sep=0pt,align=left] at (5.3,-3.35)
      {$\bigstar$\ \ the right end at the firing time; dashed: the left\\
       \phantom{$\bigstar$\ \ }edge of its dependency cone};
  \end{scope}
\end{tikzpicture}
\caption{Anatomy of the line of length $n=12$, coloured by region rather than
by state. The upper part agrees with the half-line (\cref{lem:agree}). The
reflected triangle $R_n$ (\cref{lem:determinacy}) is computed from the two
input anti-diagonals and the borders; its first anti-diagonal is the return
chain (\cref{lem:return}). The right end at the firing time ($\bigstar$)
depends only on the dark input cells (\cref{lem:cone}); the dashed line is the
left edge of its dependency cone. The arrows show the relative coordinates
$\tau$ and $\kappa$ of \cref{sec:right}.}
\label{fig:anatomy}
\end{figure}

Fix $n\ge2$ and use the relative coordinates $\tau:=t-(n-1)$ and $\kappa:=n-i$:
the relative time since the front reached the right end, and the distance from
the right end. Write $R_n(\tau,\kappa):=C_n(n-1+\tau,n-\kappa)$. Then
\[
R_n=\{(\tau,\kappa):\ 0\le\kappa\le\tau\le n-1\},
\]
the right border is at $\kappa=-1$, the left border at $\kappa=n$, the return
chain is the diagonal $\tau=\kappa$, and the firing row is $\tau=n-1$
(\cref{fig:anatomy}). The predecessors of a cell of $R_n$ lie in $R_n$, on a
border, or on one of the anti-diagonals $t+i=2n-3$ and $t+i=2n-2$, where the
line agrees with the half-line (\cref{lem:agree}). The \emph{input word} of
$R_n$ is the partial word
\[
J_n(\kappa):=\bigl(C_\infty(n-3+\kappa,\,n-\kappa),\;C_\infty(n-2+\kappa,\,n-\kappa)\bigr),
\qquad 0\le\kappa\le n-1,
\]
whose components are the \emph{lower} and \emph{upper input cells}, on the
anti-diagonals $t+i=2n-3$ and $t+i=2n-2$. (For $n=2$ and $\kappa=0$ the lower
cell lies at time $-1$; it is a predecessor of no cell of $R_n$, and we set it
to $\Lq$.)

\begin{lemma}[determinacy]\label{lem:determinacy}
There is a partial map $\Psi$, depending on $\delta$ only, such that for all
$n\ge2$ and $(\tau,\kappa)\in R_n$ the values $R_n(\tau,\kappa)$ and
$\Psi(J_n,n;\tau,\kappa)$ are both undefined or both defined and equal.
Moreover, $\Psi(J,m;\tau,\kappa)$ depends only on the input cells and the
borders in the backward dependency cone of $(\tau,\kappa)$.
\end{lemma}

In words: the reflected triangle is computed row by row by the local rule,
starting from the input word and the two borders, and each of its cells depends
only on the input cells from which information can reach it.

\begin{proof}
Define $\Psi(J,m;\tau,\kappa)$ for $0\le\kappa\le\tau\le m-1$ by induction on
$\tau$ as $\delta(x_{\kappa+1},x_\kappa,x_{\kappa-1})$, where each $x_{\kappa'}$
is given by the first applicable case: $\ast$ if $\kappa'\in\{-1,m\}$; the lower
or the upper component of $J(\kappa')$ if $\tau-1=\kappa'-2$ or
$\tau-1=\kappa'-1$, respectively; and $\Psi(J,m;\tau-1,\kappa')$ otherwise (then
$0\le\kappa'\le\tau-1$). The value $\Psi$ is undefined if one of the arguments
is $\Fq$ or undefined. For $m=n$ the three arguments are the left, middle and
right predecessors of the cell $(n-1+\tau,n-\kappa)$, and on the input cells
$C_n$ and $C_\infty$ are both undefined or equal by \cref{lem:agree}, so
induction on $\tau$ gives the first claim. Unfolding the definition gives the
second.
\end{proof}

For $(\tau,\kappa)\in R_n$ let $\mathrm{cone}_n(\tau,\kappa)$ be the set of input
cells that are reached from $(\tau,\kappa)$ by repeatedly passing to
predecessors, stopping at input cells and borders. For the right end at the
firing time, $(\tau,\kappa)=(n-1,0)$, this cone covers about the right half of
the two input anti-diagonals.

\begin{lemma}[cone of the right end]\label{lem:cone}
For $n\ge2$, the cone $\mathrm{cone}_n(n-1,0)$ avoids the left border. It
consists of the upper input cell at $\kappa'=0$, both input cells at
$1\le\kappa'\le\lfloor n/2\rfloor$ and, if $n$ is odd, the lower input cell at
$\kappa'=(n+1)/2$.
\end{lemma}

\begin{proof}
A step to a predecessor decreases $\tau$ by one and changes $\kappa$ by at most
one, and the input cells satisfy $\tau'\in\{\kappa'-1,\kappa'-2\}$. A path from
$(n-1,0)$ that reaches an input cell at $\kappa'$ after $r$ steps therefore has
$\kappa'\le r$ and $n-1-r\in\{\kappa'-1,\kappa'-2\}$, so $2\kappa'\le n+1$, and
$2\kappa'\le n$ for upper cells. Conversely, the upper cell at $\kappa'\le n/2$
is reached by $\kappa'$ steps to the left followed by vertical steps, and the
lower cell at $1\le\kappa'\le(n+1)/2$ by $\kappa'-1$ steps to the left,
vertical steps down to $(\kappa'-1,\kappa'-1)$ and one more step to the left,
through cells of $R_n$; the lower cell at $\kappa'=0$ is a predecessor of no
cell of $R_n$. Every cell of $R_n$ on such a path satisfies $\kappa\le\tau\le
n-1-\kappa$, hence $\kappa\le(n-1)/2$, so no such cell is adjacent to the left
border $\kappa=n$.
\end{proof}

\subsection{The Pumping Lemma}

\begin{lemma}[pumping]\label{lem:pumping}
Let $2\le n<n'$, and let $\delta$ synchronize the lines of lengths $n$ and
$n'$ at times $2n-2$ and $2n'-2$. Then $J_n$ and $J_{n'}$ differ on
$\mathrm{cone}_n(n-1,0)$.
\end{lemma}

\begin{figure}[t]
\centering
%% Figure: the pumping lemma (Lemma 3.3) and the proof of the speed-1/3 barrier (Theorem 3.4).
%% Schematic. In (a) the relative cell (tau,kappa) of R_m is [m-kappa-1,m-kappa] x [-tau-3,-tau-2].
\begin{tikzpicture}
  % (a) R_n and R_n' with the same values on the cone of the right end of R_n (n = 8, n' = 12)
  \begin{scope}[x=2.7mm,y=2.7mm]
    \foreach \m/\dx in {8/0,12/12} {
      \begin{scope}[shift={(\dx,0)}]
        \foreach \tau in {-2,...,11} {\foreach \kap in {0,...,11} {%
          \pgfmathtruncatemacro{\cls}{((\kap<=\tau)&&(\tau<=\m-1)) ? ((\tau==\m-1) ? 1 : ((\tau+\kap<=7) ? 2 : 3))
            : ((((\tau==\kap-1)||((\tau==\kap-2)&&(\kap>=1)))&&(\kap<=\m-1)) ? ((\kap<=4) ? 4 : 5) : 0)}%
          \ifnum\cls>0
            \def\cc{\ifcase\cls\or stF\or rgRefl2\or rgRefl\or rgCone\or rgInput\fi}%
            \filldraw[fill=\cc,draw=grid!45,line width=0.15pt] (\m-\kap-1,-\tau-2) rectangle (\m-\kap,-\tau-3);
          \fi}}
        \draw[ink,line width=1.2pt] (\m,0.1) -- (\m,-\m-2.1);
        \draw[ink,line width=1.2pt] (0,-\m+0.9) -- (0,-\m-2.1);
        \node[star,star points=5,fill=white,draw=ink,line width=0.4pt,inner sep=0pt,minimum size=0.24cm]
          at (\m-0.5,-9.5) {};
      \end{scope}}
    \node[font=\small,anchor=south] at (4,0.4) {$R_n$, $n=8$};
    \node[font=\small,anchor=south] at (20.5,0.4) {$R_{n'}$, $n'=12$};
    \draw[-{Stealth[length=1.8mm]},ink,line width=0.6pt] (7.6,-0.2) .. controls (9.6,2.7) and (14.4,2.7) .. (16.4,-0.2);
    \node[font=\scriptsize,anchor=south] at (12.0,2.25) {equal on the cone};
    \node[font=\scriptsize,anchor=north,align=center] at (4,-10.5) {$R_n(n-1,0)=\Fq$};
    \node[font=\scriptsize,anchor=north,align=center] at (18,-14.5) {then also $R_{n'}(n-1,0)=\Fq$,\\
      although $n-1<n'-1$};
  \end{scope}
  % (b) the half-line: the cones of the lengths n and n+P on the depth rows
  \begin{scope}[shift={(9.2cm,0cm)},x=1.6mm,y=1.6mm]
    \fill[gNone] (0,0) -- (36,-36) -- (0,-36) -- cycle;
    \foreach \j in {4,8,12} \draw[grid,line width=0.45pt] (0,-\j) -- (36-\j,-36);
    \draw[ink,densely dashed,line width=0.7pt] (0,0) -- (12,-36);
    \draw[rgCone,line width=2.4pt] (14,-14) -- (7,-21);
    \draw[rgCone,line width=2.4pt] (20,-20) -- (10,-30);
    \foreach \j in {4,8,12} {
      \fill[ink] (14-\j/2,-14-\j/2) circle (0.45);
      \fill[ink] (20-\j/2,-20-\j/2) circle (0.45);}
    \draw[-{Stealth[length=1.6mm]},stF,line width=0.8pt] (10.8,-18.8) -- (15.3,-23.3);
    \node[font=\scriptsize,anchor=south west,inner sep=1pt] at (13.0,-20.6) {$+P$};
    \draw[ink,line width=1.2pt] (0,0) -- (0,-36);
    \draw[ink,line width=0.7pt] (0,0) -- (36,-36);
    \node[font=\scriptsize,rotate=-45,anchor=south] at (6.5,-6.5) {front};
    \node[font=\scriptsize,anchor=north] at (12,-36.2) {$i=t/3$};
    \node[font=\scriptsize,anchor=north] at (29,-36.2) {depth rows};
    \node[font=\scriptsize,anchor=east,align=right] at (-0.8,-15.0) {cone of\\ the length $n$};
    \draw[-{Stealth[length=1.4mm]},ink,line width=0.45pt] (-0.7,-15.6) -- (8.6,-19.2);
    \node[font=\scriptsize,anchor=west,align=left] at (26.0,-19.0) {cone of the\\ length $n+P$};
    \draw[-{Stealth[length=1.4mm]},ink,line width=0.45pt] (25.9,-19.6) -- (17.2,-22.6);
  \end{scope}
  \node[font=\small,anchor=south west] at (-0.3,1.25) {(a) the pumping lemma};
  \node[font=\small,anchor=south west] at (8.2,1.25) {(b) the proof of \cref{thm:third}};
\end{tikzpicture}
\caption{(a)~The pumping lemma (\cref{lem:pumping}), drawn for $n=8$ and
$n'=12$ in relative coordinates: $\tau$ runs downwards and $\kappa$ to the left
of the right border (thick line). Above each triangle lie its input cells, on
the diagonals $\tau=\kappa-2$ and $\tau=\kappa-1$. The right end of $R_n$ at the
firing time ($\bigstar$) depends only on the dark input cells
(\cref{lem:cone}), through the darker cells of the triangle. If $R_{n'}$
received the same values at the same input positions, its relative cell
$(n-1,0)$ would also be in state $\Fq$, although $n-1<n'-1$. (b)~The proof of
\cref{thm:third}, in the half-line. Apart from the quiescent cell at $\kappa=0$,
the input cells of the cone of the right end of the line of length $n$ (thick)
lie on the depth rows $0,\dots,n-1$
(grey, parallel to the front), between the front and roughly the line $i=t/3$
(dashed). The cone of the length $n+P$ meets the same depth rows $P$ steps
later along each row (arrow). If each of these depth rows were periodic, with a
period dividing $P$, from its point on the cone of $n$ (dots) on, the two cones
would carry the same values, which (a) excludes.}
\label{fig:pumping}
\end{figure}
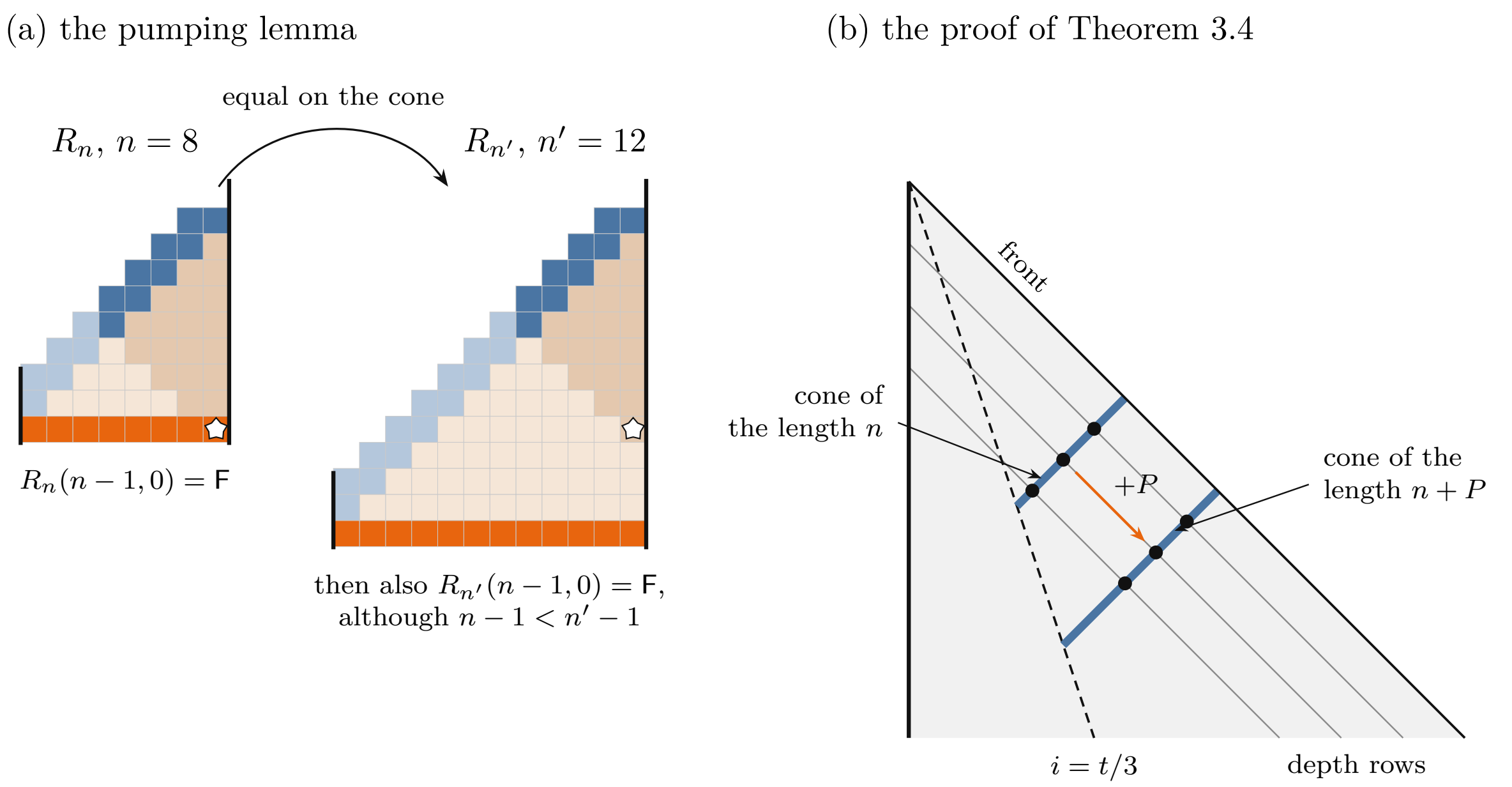

\begin{proof}
Suppose that they agree there. The relative cell $(n-1,0)$ lies in $R_{n'}$,
and its cone in $R_{n'}$ consists of the same positions, which are away from
the left border $\kappa=n'$. By \cref{lem:determinacy,lem:cone}, the relative
cell $(n-1,0)$ of $R_{n'}$ has the value of the relative cell $(n-1,0)$ of
$R_n$, namely $C_n(2n-2,n)=\Fq$. Thus $C_{n'}(n'+n-2,n')=\Fq$, although
$n'+n-2<2n'-2$ (\cref{fig:pumping}(a)).
\end{proof}

We write $\PUMP(n,n')$ for the conclusion of \cref{lem:pumping}. It is a
condition on the half-line alone: the half-line must keep supplying new
information to the right ends of longer and longer lines. If the depth rows
that carry these cones (defined below) were periodic, the input words of two
lengths that differ by a common period would agree on the cone; the next
theorem turns this observation into a bound on where the periodicity can start.

\subsection{Depth Rows and the Onset of Periodicity}

It is convenient to locate the cells of the half-line by their distance behind
the front, the cell $(t,t+1)$. Let
\[
\Phi_t(j):=C_\infty(t,\,t+1-j)\qquad(0\le j\le t)
\]
be the cell at \emph{depth} $j$ behind the front; the front cell has depth $0$.
The local rule becomes
\begin{equation}\label{eq:comoving}
\Phi_{t+1}(j)=\delta\bigl(\Phi_t(j),\,\Phi_t(j-1),\,\Phi_t(j-2)\bigr)
\qquad(0\le j\le t+1),
\end{equation}
with $\Phi_t(-1)=\Phi_t(-2)=\Lq$ (the quiescent cells ahead of the front) and
$\Phi_t(t+1)=\ast$ (the left border). So information only travels to larger
depths. The \emph{depth row} $j$ is the sequence $t\mapsto\Phi_t(j)$, $t\ge j$:
it consists of the cells $(t,t+1-j)$, so it starts in cell $1$ at time $j$ and
then advances by one cell per step, at distance $j$ behind the front. In these
coordinates,
\[
J_n(\kappa)=\bigl(\Phi_{n-3+\kappa}(2\kappa-2),\ \Phi_{n-2+\kappa}(2\kappa-1)\bigr):
\]
the input cells at distance $\kappa$ from the right end lie on the depth rows
$2\kappa-2$ and $2\kappa-1$.

\begin{theorem}[late onset of periodicity]\label{thm:third}
Let $\delta$ be a minimal-time solution. Every depth row $t\mapsto\Phi_t(j)$ is
eventually periodic; let $\theta(j)$ be the least $t_0\ge j$ such that it is
periodic on $[t_0,\infty)$. For every $n\ge4$ there is a depth $j\le n$ with
$\theta(j)>n+j/2-5/2$. Hence
$\limsup_{j\to\infty}\theta(j)/j\ge 3/2$.
\end{theorem}

\begin{proof}
Row $0$ satisfies $\Phi_{t+1}(0)=\delta(\Phi_t(0),\Lq,\Lq)$, so it is
eventually periodic. By \eqref{eq:comoving}, row $j$ is a finite-state system
driven by the rows $j-1$ and $j-2$; if these are eventually periodic, so is row
$j$. By \eqref{eq:neverfire} all values lie in $W$. Let $P_j$ be a period of
row $j$, fix $n\ge4$ and let $P$ be a common multiple of $P_0,\dots,P_{n-1}$.

We compare the input words of $R_n$ and $R_{n+P}$ on $\mathrm{cone}_n(n-1,0)$.
The input cell at $\kappa=0$ in this cone is the upper one, and it is quiescent
in both triangles by \cref{lem:lightcone}. The other input cells of the cone
have depth $2\kappa-2$ (lower cells) or $2\kappa-1$ (upper cells), at most
$n-1$ by \cref{lem:cone}. Suppose that each of them lies in the periodic part of
its row (\cref{fig:pumping}(b)), that is, $n-3+\kappa\ge\theta(2\kappa-2)$ for the lower cells and
$n-2+\kappa\ge\theta(2\kappa-1)$ for the upper cells. Replacing $n$ by $n+P$
moves each input cell along its depth row by $P$ steps, so $J_{n+P}=J_n$ on the
cone, which contradicts \cref{lem:pumping}. Hence some input cell of the cone,
of depth $j\le n$, satisfies $\theta(j)>n-3+\kappa$. For a lower cell
$\kappa=(j+2)/2$ and for an upper cell $\kappa=(j+1)/2$, so in both cases
$\theta(j)>n+j/2-5/2$.

Suppose now that $\limsup_{j\to\infty}\theta(j)/j<3/2$. Then there are
$\alpha<3/2$ and $\beta$ with $\theta(j)\le\alpha j+\beta$ for all $j$;
since $\theta(j)\ge j$, necessarily $\alpha\ge1$. For all $j\le n$,
\[
\theta(j)-\frac{j}{2}\le\Bigl(\alpha-\frac12\Bigr)j+\beta\le\Bigl(\alpha-\frac12\Bigr)n+\beta,
\]
and since $\alpha-1/2<1$, the right-hand side is at most $n-5/2$ for $n$ large.
Then $\theta(j)\le n+j/2-5/2$ for all $j\le n$, which contradicts the first
part.
\end{proof}

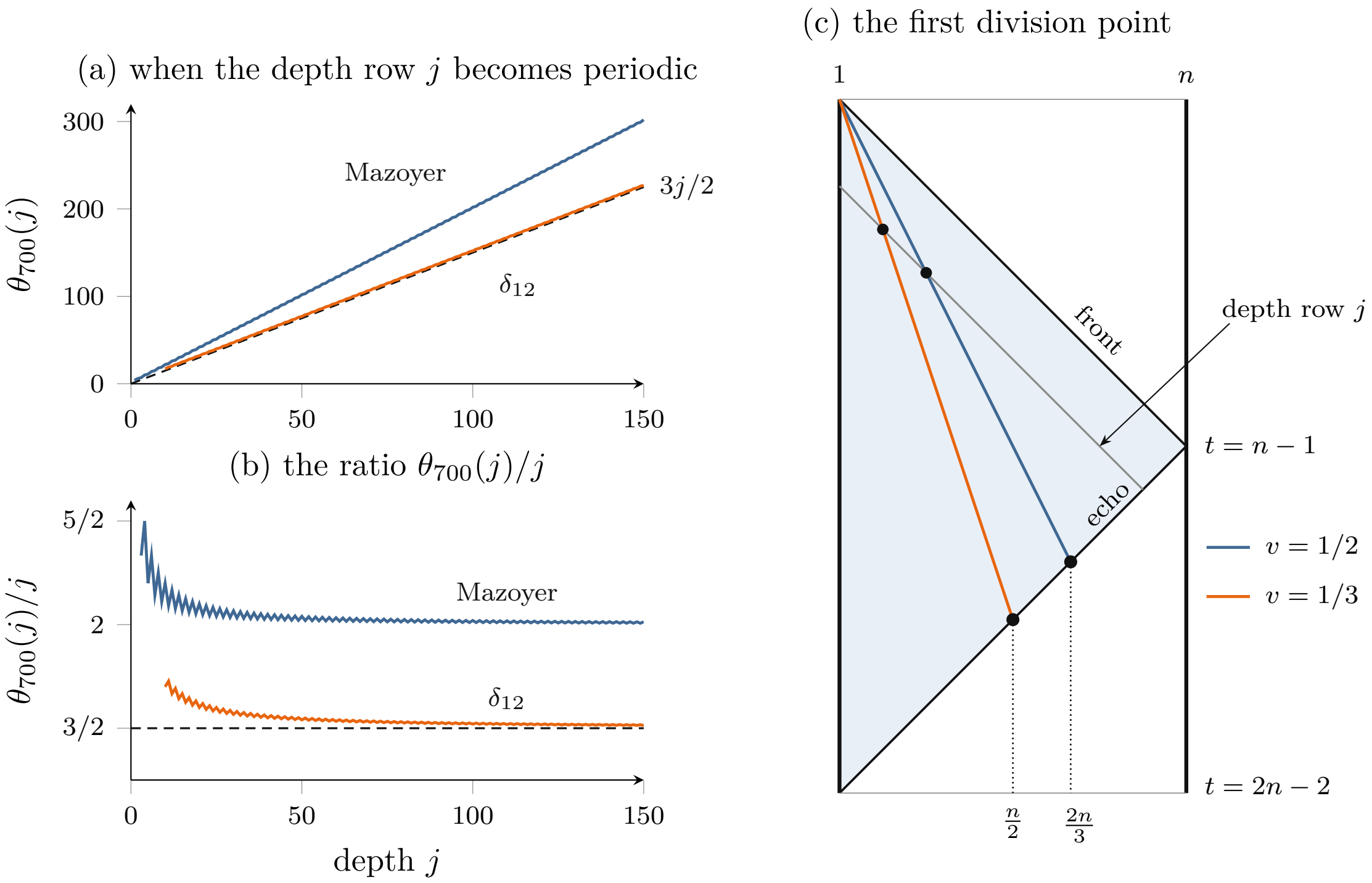
\begin{figure}[t]
\centering
%% Figure: theta(j) on the half-lines of Mazoyer's solution and of delta12 (data from figures.py),
%% and the division point of Remark 3.5 (schematic).
\begin{tikzpicture}
  \pgfplotsset{every axis/.append style={scale only axis,width=5.5cm,height=3.0cm,
    axis lines=left,tick label style={font=\scriptsize},label style={font=\small},
    xmin=0,xmax=150,xtick={0,50,100,150},tick align=outside,
    every axis plot/.append style={line width=0.9pt}}}
  % (a) theta(j)
  \begin{axis}[name=pa,ymin=0,ymax=320,ytick={0,100,200,300},ylabel={$\theta_{700}(j)$},clip=false,
      title={(a) when the depth row $j$ becomes periodic},title style={font=\small,yshift=-1ex}]
    \addplot[ink,densely dashed,line width=0.6pt,domain=0:150,samples=2] {1.5*x};
    \addplot[stC] table[x=j,y=theta] {figures/theta_mazoyer.dat};
    \addplot[stF] table[x=j,y=theta] {figures/theta_delta12.dat};
    \node[font=\scriptsize,anchor=south east,text=ink] at (axis cs:96,212) {Mazoyer};
    \node[font=\scriptsize,anchor=north west,text=ink] at (axis cs:104,140) {$\delta_{12}$};
    \node[font=\scriptsize,anchor=west,text=ink] at (axis cs:151,225) {$3j/2$};
  \end{axis}
  % (b) theta(j)/j
  \begin{axis}[name=pb,at={(pa.south west)},anchor=north west,yshift=-1.25cm,
      ymin=1.25,ymax=2.6,ytick={1.5,2,2.5},yticklabels={$3/2$,$2$,$5/2$},
      ylabel={$\theta_{700}(j)/j$},xlabel={depth $j$},
      title={(b) the ratio $\theta_{700}(j)/j$},title style={font=\small,yshift=-1ex}]
    \addplot[ink,densely dashed,line width=0.6pt,domain=0:150,samples=2] {1.5};
    \addplot[stC] table[x=j,y expr=\thisrow{theta}/\thisrow{j},restrict x to domain=3:150] {figures/theta_mazoyer.dat};
    \addplot[stF] table[x=j,y expr=\thisrow{theta}/\thisrow{j}] {figures/theta_delta12.dat};
    \node[font=\scriptsize,anchor=south,text=ink] at (axis cs:110,2.03) {Mazoyer};
    \node[font=\scriptsize,anchor=south,text=ink] at (axis cs:110,1.53) {$\delta_{12}$};
  \end{axis}
  % (c) the division point of Remark 3.5, for a line of length n = 24 (continuous coordinates)
  \begin{scope}[shift={(7.6cm,3.05cm)},x=1.55mm,y=1.55mm]
    \fill[rgAgree] (0,0) -- (24,-24) -- (0,-48) -- cycle;
    \draw[ink,line width=0.7pt] (0,0) -- (24,-24) -- (0,-48);
    \draw[ink,line width=1.2pt] (0,0) -- (0,-48);
    \draw[ink,line width=1.2pt] (24,0) -- (24,-48);
    \draw[grid,line width=0.3pt] (0,0) -- (24,0) (0,-48) -- (24,-48);
    \draw[stC,line width=0.9pt] (0,0) -- (16,-32);
    \draw[stF,line width=0.9pt] (0,0) -- (12,-36);
    \draw[grid,line width=0.6pt] (0,-6) -- (21,-27);
    \fill[ink] (16,-32) circle (0.45) (12,-36) circle (0.45);
    \fill[ink] (6,-12) circle (0.4) (3,-9) circle (0.4);
    \draw[ink,densely dotted,line width=0.5pt] (16,-32) -- (16,-48) (12,-36) -- (12,-48);
    \node[font=\scriptsize,anchor=north] at (12,-48.2) {$\frac n2$};
    \node[font=\scriptsize,anchor=north] at (16.6,-48.2) {$\frac{2n}3$};
    \node[font=\scriptsize,anchor=south] at (0,0.3) {$1$};
    \node[font=\scriptsize,anchor=south] at (24,0.3) {$n$};
    \node[font=\scriptsize,anchor=west] at (24.4,-24) {$t=n-1$};
    \node[font=\scriptsize,anchor=west] at (24.4,-47.6) {$t=2n-2$};
    \node[font=\scriptsize,rotate=-45,anchor=south] at (17,-17) {front};
    \node[font=\scriptsize,rotate=45,anchor=south] at (19.5,-29.0) {echo};
    \draw[-{Stealth[length=1.3mm]},ink,line width=0.4pt] (27.0,-15.5) -- (18.0,-24.0);
    \node[font=\scriptsize,anchor=west] at (25.6,-14.6) {depth row $j$};
    \draw[stC,line width=0.9pt] (25.4,-31.0) -- (28.4,-31.0);
    \node[font=\scriptsize,anchor=west] at (28.6,-31.0) {$v=1/2$};
    \draw[stF,line width=0.9pt] (25.4,-34.4) -- (28.4,-34.4);
    \node[font=\scriptsize,anchor=west] at (28.6,-34.4) {$v=1/3$};
    \node[font=\small,anchor=south west] at (-3.5,3.2) {(c) the first division point};
  \end{scope}
\end{tikzpicture}
\caption{(a)~The observed onset $\theta_{700}(j)$ of periodicity of depth row
$j$ (\cref{rem:division}) on the half-lines of Mazoyer's six-state solution
($1\le j\le150$) and of the five-state rule $\delta_{12}$ ($10\le j\le150$),
simulated up to time $700$; the dashed line is $3j/2$. In this range,
$\theta_{700}(j)\in\{2j+1,2j+2\}$ for Mazoyer's solution and
$\theta_{700}(j)=\lceil 3j/2\rceil+2$ for $\delta_{12}$. (b)~The same data as
the ratio $\theta_{700}(j)/j$ (from $j=3$ on for Mazoyer's solution).
\Cref{thm:third} says that every minimal-time solution has
$\theta(j)/j\ge3/2-\varepsilon$ for infinitely many $j$; the values of
$\delta_{12}$, which synchronizes only the lines of lengths $2,\dots,12$, lie at
this bound in the simulated range.
(c)~The heuristic reading of \cref{rem:division} on the line of length $n$: a signal of
speed $v$ sent by the general meets the echo of the right end at the cell
$v(2n-1)/(1+v)$, about $n/2$ for $v=1/3$ and $2n/3$ for $v=1/2$, and the depth
row $j$ (grey) crosses the signal at about the time $j/(1-v)$ (dots). Shaded:
the cells reached by the front but not yet by the echo, where the line agrees
with the half-line.}
\label{fig:theta}
\end{figure}

\Cref{thm:third} is a statement about the numbers $\theta(j)$, and only about
infinitely many of them: it says that $\theta(j)\ge(3/2-\varepsilon)j$ for
infinitely many $j$, and nothing about the other depths. Its geometric reading
explains its second name. Depth row $j$ becomes periodic at the cell
$(\theta(j),\theta(j)+1-j)$. If $\theta(j)\ge(3/2-\varepsilon)j$, then
$j\le\theta(j)/(3/2-\varepsilon)$, and the position of this cell satisfies
\[
\theta(j)+1-j\ \ge\ \frac{1-2\varepsilon}{3-2\varepsilon}\,\theta(j)+1,
\]
where the factor $(1-2\varepsilon)/(3-2\varepsilon)$ tends to $1/3$ as
$\varepsilon\to0$ (\cref{fig:theta}). So, for infinitely many $j$, depth row $j$
becomes periodic only at a time $t$ at which it is close to the cell $t/3$ or
beyond. Equivalently, there is no $c<1/3$ such that, for all large $j$, depth
row $j$ is periodic from the time at which it reaches the line $i=ct$; this is
the sense in which we speak of a speed-$1/3$ barrier. The theorem does not say
that the cells that are not yet periodic form a region bounded by a signal, nor
that such a boundary moves at speed $1/3$, nor that it exists at all.
\Cref{rem:division} discusses the constructions in which it does.

\begin{figure}[t]
\centering
%% Figure: the half-lines of Mazoyer's solution and of delta12 for 0 <= t <= 90 (cells from figures.py).
\begin{tikzpicture}[x=0.74mm,y=0.74mm]
  \foreach \file/\title/\dx in {cells_mazoyer_half90/{Mazoyer's solution (six states)}/0,
                                cells_delta12_half90/{$\delta_{12}$ (five states)}/99} {
    \begin{scope}[shift={(\dx,0)}]
      \input{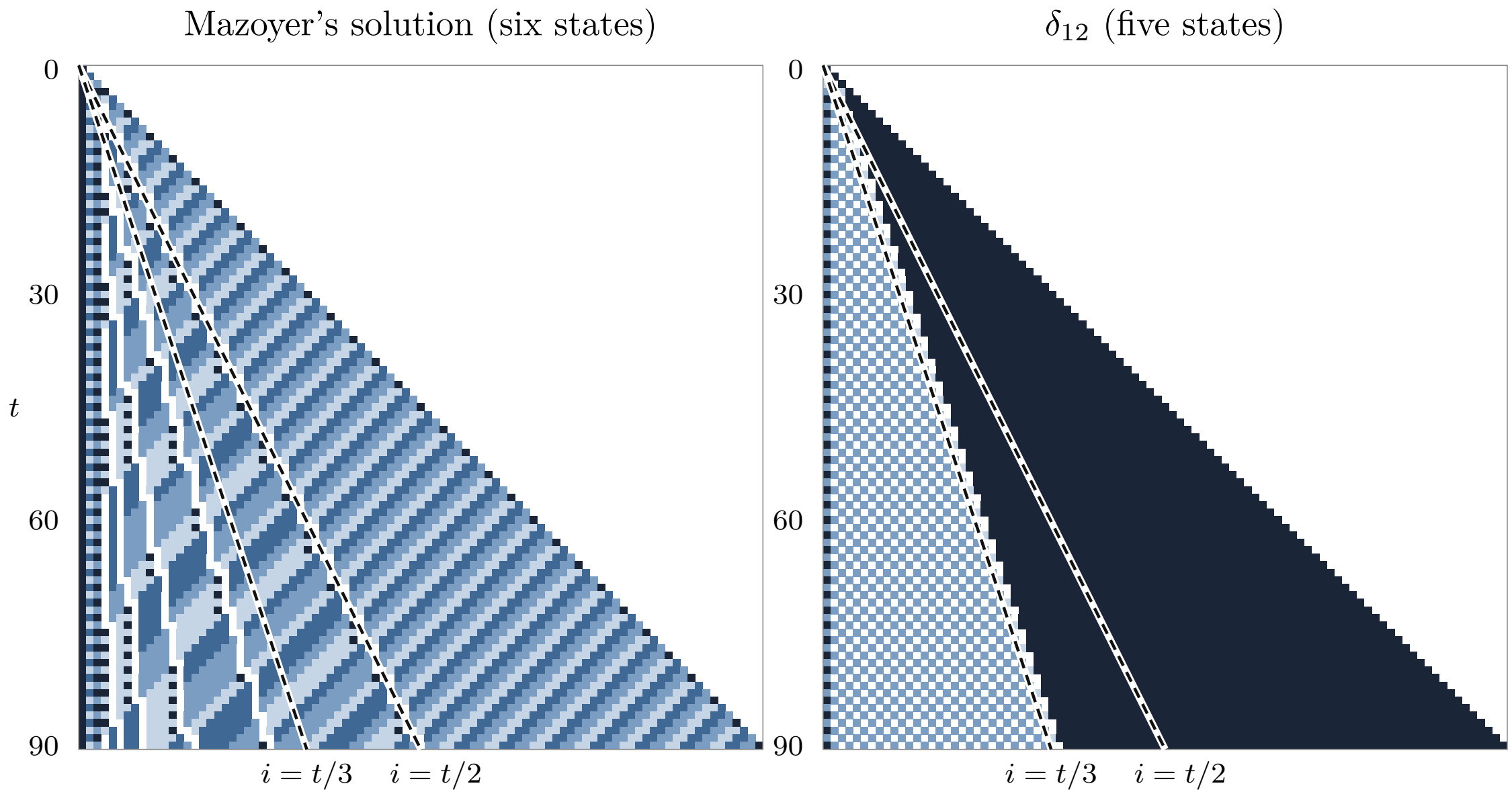}
      \draw[grid,line width=0.3pt] (0,0) rectangle (91,-91);
      \foreach \sl/\st in {2/densely dashed,3/densely dashed} {
        \draw[white,line width=2.0pt] (0,0) -- (91/\sl,-91);
        \draw[ink,\st,line width=0.85pt] (0,0) -- (91/\sl,-91);
      }
      \node[font=\scriptsize,anchor=north,fill=white,inner sep=1pt] at (91/3,-92) {$i=t/3$};
      \node[font=\scriptsize,anchor=north,fill=white,inner sep=1pt] at (91/2+2,-92) {$i=t/2$};
      \node[font=\small,anchor=south] at (45.5,1) {\title};
      \foreach \y in {0,30,60,90}
        \node[font=\scriptsize,anchor=east] at (-0.8,-\y-0.5) {$\y$};
    \end{scope}
  }
  \node[font=\scriptsize,anchor=east] at (-6,-45.5) {$t$};
\end{tikzpicture}
\caption{The half-lines of Mazoyer's six-state solution (left) and of the
five-state rule $\delta_{12}$ (right) for $0\le t\le90$. The dashed lines are
$i=t/2$ and $i=t/3$. In this range, the depth rows, that is, the cells at a
fixed distance behind the front, become periodic near the line $i=t/2$ for
Mazoyer's solution and near the line $i=t/3$ for $\delta_{12}$
(\cref{rem:division}). By \cref{thm:third}, in a minimal-time solution the
depth rows cannot all become periodic before they reach a line $i=ct$ with
$c<1/3$. In the range shown, the half-line of $\delta_{12}$ alternates between
$\Aq$ and $\Lq$ to the left of the line $i=t/3$, periodically in time and
space. \Cref{thm:left} excludes such a wedge at the left border if the pattern
continues forever, and \cref{cor:germ12} excludes it for the line of length
$518$ without this assumption.}
\label{fig:halflines}
\end{figure}

\begin{remark}[a heuristic reading]\label{rem:division}
In the classical constructions, the line is divided recursively, and the first
division point is where a signal sent by the general meets the echo of the right
end (\cref{fig:theta}(c)); a \emph{signal} of speed $v$ is a pattern that moves by
$v$ cells per step. A signal of speed $v$ that leaves the general at time $0$
meets the echo, which leaves the right end at time $n-1$ at speed one, at the
cell $v(2n-1)/(1+v)$. Suppose that the depth rows become periodic where they
cross this signal, as they do approximately for the two rules simulated below;
this is an assumption, not a consequence of minimal time. Then depth row $j$ meets the
signal at the time $(j-1)/(1-v)$, so $\theta(j)\approx j/(1-v)$, and
\cref{thm:third} gives $v\ge1/3$ asymptotically: the first division point, at the
cell $v(2n-1)/(1+v)\ge(2n-1)/4$, is then asymptotically never in the half of the
line next to the general. Waksman's and Balzer's solutions divide the line in
the middle, with $v=1/3$, and Mazoyer's solution divides it at two thirds of its
length from the general, with $v=1/2$.

Simulations can only approximate $\theta(j)$. For a horizon $H$, let
$\theta_H(j)$ be the least $t_0\ge j$ such that $\Phi_t(j)=\Phi_{t+p}(j)$ for
$t_0\le t\le H-p$, where $p\le12$ is the least period of depth row $j$ on its
last $200$ simulated steps; if $p$ is a period of the row from $\theta(j)$ on,
then $\theta_H(j)\le\theta(j)$. For Mazoyer's solution and $H=2600$,
$\theta_H(j)\in\{2j+1,2j+2\}$ for $1\le j<700$. For the five-state rule
$\delta_{12}$ of \cref{sec:five}, which synchronizes the lines of lengths
$2,\dots,12$, and $H=1400$, $\theta_H(j)=\lceil 3j/2\rceil+2$ for $10\le
j\le300$, and $\PUMP(n,n')$ holds for $4\le n<n'\le650$
(\cref{fig:theta,fig:halflines}). These are observations on finite horizons,
and the reading above need not apply to other rules.
\end{remark}

\subsection{The Right Band}

The pigeonhole argument behind the pumping lemma also works along the right
border, in the direction of time. The $d$ cells next to the right border form a
finite automaton driven by the column $\kappa=d$, and this automaton has to
reach its all-firing state exactly at the firing time.

\begin{lemma}[right band]\label{lem:rightband}
Let $\delta$ have $k$ states and synchronize the line of length $n$ at time
$2n-2$. Let $1\le d\le\tau_0\le n-2$ and $\ell\ge1$, and suppose that the
column $\tau\mapsto R_n(\tau,d)$ is $\ell$-periodic on $[\tau_0,n-2]$, that is,
$R_n(\tau+\ell,d)=R_n(\tau,d)$ whenever $\tau,\tau+\ell\in[\tau_0,n-2]$. Then
$n-1-\tau_0\le(k-1)^d\ell$.
\end{lemma}

\begin{figure}[t]
\centering
%% Figure: the proof of the right-band lemma (Lemma 3.6). Schematic, n = 22, d = 3, l = 3,
%% tau_0 = 5, tau_1 = 8, tau_2 = 14, h = 6. The relative cell (tau,kappa) is [n-kappa-1,n-kappa] x [-tau-1,-tau].
\begin{tikzpicture}[x=2.3mm,y=2.3mm]
  \foreach \tau in {0,...,21} {\foreach \kap in {0,...,\tau} {%
    \pgfmathtruncatemacro{\cls}{(\tau==21) ? 1 : ((\kap<3) ? ((\tau==15) ? 6 : 2)
      : ((\kap==3) ? ((\tau>=5) ? 3+mod(\tau-5,3) : 2) : 0))}%
    \def\cc{\ifcase\cls rgRefl\or stF\or rgRefl2\or stB\or stA\or stC\or stF\fi}%
    \filldraw[fill=\cc,draw=grid!45,line width=0.15pt] (21-\kap,-\tau) rectangle (22-\kap,-\tau-1);
    \ifnum\cls=6 \fill[pattern=north east lines,pattern color=white] (21-\kap,-\tau) rectangle (22-\kap,-\tau-1);\fi}}
  \draw[ink,line width=1.2pt] (22,0.2) -- (22,-22.2);
  \draw[ink,line width=0.5pt] (22,0) -- (0,-22) -- (22,-22);
  % the band states b(tau_1) = b(tau_2)
  \draw[ink,line width=1.2pt] (19,-8) rectangle (22,-9);
  \draw[ink,line width=1.2pt] (19,-14) rectangle (22,-15);
  % brackets: the band evolves in the same way from tau_1 and from tau_2
  \draw[ink,line width=0.6pt] (22.9,-8.1) -- (23.4,-8.1) -- (23.4,-15.9) -- (22.9,-15.9);
  \draw[ink,line width=0.6pt] (25.4,-14.1) -- (25.9,-14.1) -- (25.9,-21.9) -- (25.4,-21.9);
  \node[font=\scriptsize,anchor=west,align=left] at (26.4,-17.0) {the same evolution,\\ $h$ steps apart};
  % row labels
  \foreach \y/\l in {5/{\tau_0},8/{\tau_1},14/{\tau_2},15/{n-1-h},21/{n-1}}
    \node[font=\scriptsize,anchor=east] at (20.6-\y,-\y-0.5) {$\l$};
  \node[font=\scriptsize,anchor=north] at (21.5,-22.3) {$0$};
  \node[font=\scriptsize,anchor=north] at (18.5,-22.3) {$d$};
  \node[font=\scriptsize,anchor=north] at (11,-22.3) {$\kappa$};
  % annotations
  \draw[-{Stealth[length=1.5mm]},ink,line width=0.5pt] (30.5,-3.5) -- (19.3,-6.4);
  \node[font=\scriptsize,anchor=west,align=left] at (30.5,-3.5) {column $\kappa=d$: $\ell$-periodic\\ on $[\tau_0,n-2]$};
  \draw[-{Stealth[length=1.5mm]},ink,line width=0.5pt] (30.5,-9.6) -- (22.4,-8.6);
  \node[font=\scriptsize,anchor=west,align=left] at (30.5,-10.0) {$\mathbf b(\tau_1)=\mathbf b(\tau_2)$,\\ $\ell\mid h=\tau_2-\tau_1$};
  \draw[-{Stealth[length=1.5mm]},ink,line width=0.5pt] (30.5,-13.3) -- (22.4,-14.5);
  \draw[-{Stealth[length=1.5mm]},ink,line width=0.5pt] (31.0,-20.6) -- (22.4,-15.6);
  \node[font=\scriptsize,anchor=west,align=left] at (31.0,-21.0) {then $\mathbf b(n-1-h)=\Fq^d$:\\ $h$ steps too early};
  % legend
  \node[anchor=north west,font=\scriptsize,align=left] at (0,-24.2) {%
    \swatch{rgRefl2}\,the band $\kappa<d$\quad \swatch{stB}\,\swatch{stA}\,\swatch{stC}\,the column $\kappa=d$,
    shaded by $\tau\bmod\ell$\quad \swatch{stF}\,$\Fq$};
\end{tikzpicture}
\caption{The proof of \cref{lem:rightband}, drawn for $n=22$, $d=3$,
$\ell=3$ and $\tau_0=5$ in the relative coordinates of $R_n$. The band of the
$d$ cells next to the right border is a finite automaton driven by the column
$\kappa=d$, which is $\ell$-periodic on $[\tau_0,n-2]$. If the band takes the
same value $\mathbf b(\tau_1)=\mathbf b(\tau_2)$ (outlined) at two times whose
difference $h$ is a multiple of $\ell$, then it evolves in the same way from
$\tau_1$ and from $\tau_2$, so it reaches its all-firing value $\Fq^d$ at the
time $n-1-h$ (hatched), $h$ steps before the firing row. As the pairs
$(\mathbf b(\tau),\tau\bmod\ell)$ take at most $(k-1)^d\ell$ values, this
forces $n-1-\tau_0\le(k-1)^d\ell$.}
\label{fig:rightband}
\end{figure}
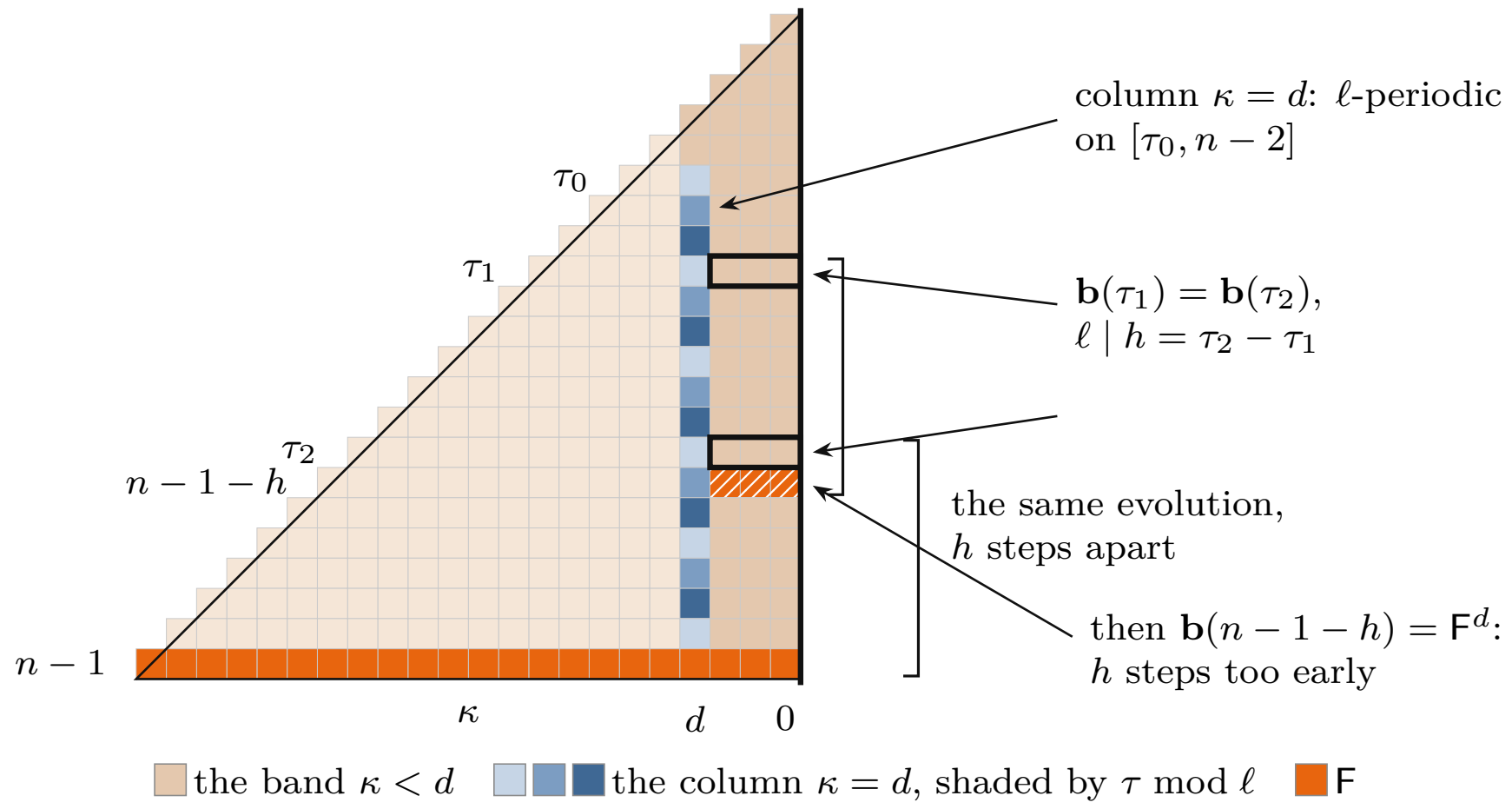

\begin{proof}
For $d-1\le\tau\le n-1$ put $\mathbf b(\tau):=(R_n(\tau,0),\dots,R_n(\tau,d-1))$.
Since $R_n(\tau+1,\kappa)=\delta(R_n(\tau,\kappa+1),R_n(\tau,\kappa),
R_n(\tau,\kappa-1))$ with $R_n(\tau,-1)=\ast$, there is a map $\Gamma$,
depending on $\delta$ and $d$ only, with $\mathbf b(\tau+1)=\Gamma(\mathbf
b(\tau),R_n(\tau,d))$ for $d\le\tau\le n-2$; its arguments lie in $W^d\times W$
because no cell fires before time $2n-2$. At $\tau=n-1$ the band is $\Fq^d$.
Suppose that $n-1-\tau_0>(k-1)^d\ell$. The pairs $(\mathbf
b(\tau),\tau\bmod\ell)$ with $\tau_0\le\tau\le n-2$ take at most $(k-1)^d\ell$
values, so there are $\tau_0\le\tau_1<\tau_2\le n-2$ with $\mathbf
b(\tau_1)=\mathbf b(\tau_2)$ and $\ell\mid h:=\tau_2-\tau_1$. For $0\le s\le
n-2-\tau_2$ the driving cells $R_n(\tau_1+s,d)$ and $R_n(\tau_2+s,d)$ lie in
$[\tau_0,n-2]$ and their times differ by a multiple of the period, so they are
equal, and induction on $s$ gives $\mathbf b(\tau_1+s)=\mathbf b(\tau_2+s)$ for
$0\le s\le n-1-\tau_2$. For $s=n-1-\tau_2$ this reads $\mathbf
b(n-1-h)=\mathbf b(n-1)=\Fq^d$, although $n-1-h<n-1$, a contradiction
(\cref{fig:rightband}).
\end{proof}

%% ---------------------------------------------------------------------------
\section{The Left End}\label{sec:left}

We cut the reflected triangle into the anti-diagonals $t+i=2n-1$, $t+i=2n$,
$\dots$, which follow the echo of the right end; we call them \emph{chains}.
Near the left border, the first few chains form a finite automaton that reads
the two input anti-diagonals $t+i=2n-3$ and $t+i=2n-2$ and must fire the cells
there exactly when the echo reaches them. If the input anti-diagonals were
periodic over a long stretch, this automaton would repeat a state and fire too
early (\cref{lem:left}). Applied to all lengths at once, this excludes periodic
wedges at the left border of the half-line (\cref{thm:left}).

\subsection{Chains and Slices}

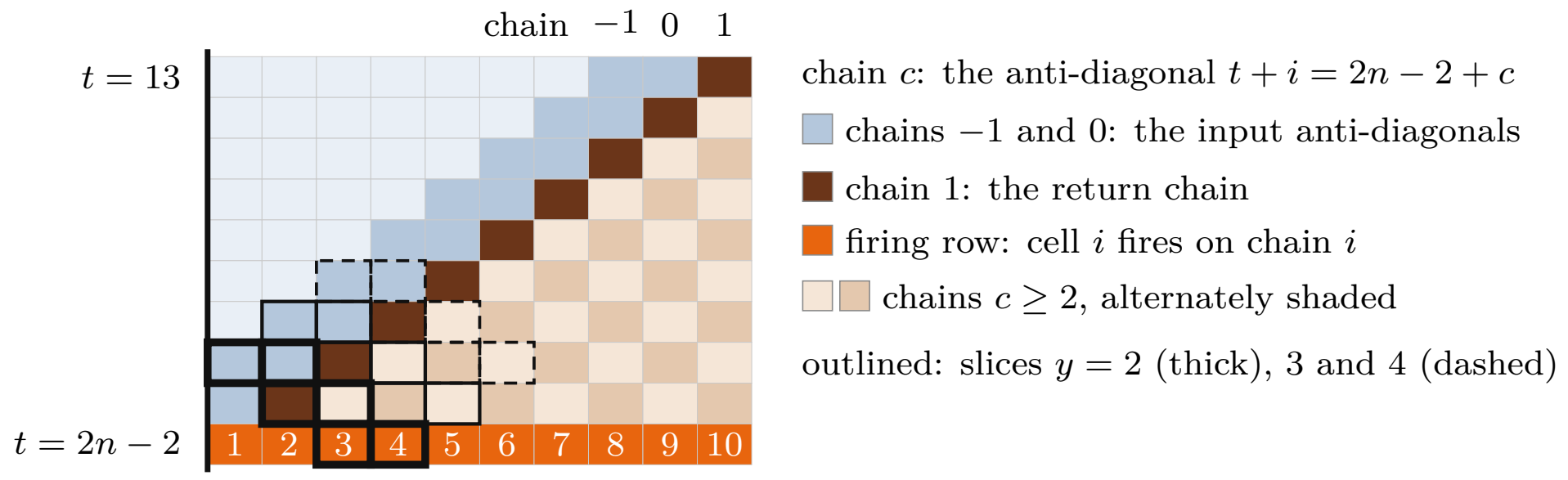
\begin{figure}[t]
\centering
%% Figure: the lower left corner of the line of length n = 12, coloured by chain, with the slices of
%% the proof of the left-border lemma for s = 2.
\begin{tikzpicture}
  \def\u{0.44}\def\v{0.33}
  \def\tz{13}
  % cell (t,i) occupies [(i-1)u, iu] x [-(t-tz+1)v, -(t-tz)v]
  \foreach \t in {13,...,22} {%
    \foreach \i in {1,...,10} {%
      \pgfmathtruncatemacro{\c}{\t+\i-22}
      \pgfmathtruncatemacro{\odd}{mod(\c+10,2)}
      \def\cc{rgAgree}%
      \ifnum\c>-2 \def\cc{rgInput}\fi
      \ifnum\c=1 \def\cc{rgReturn}\fi
      \ifnum\c>1 \ifnum\odd=1 \def\cc{rgRefl2}\else\def\cc{rgRefl}\fi\fi
      \ifnum\t=22 \def\cc{stF}\fi
      \filldraw[fill=\cc,draw=grid!45,line width=0.15pt]
        (\i*\u-\u,-\t*\v+\tz*\v) rectangle (\i*\u,-\t*\v+\tz*\v-\v);
    }}
  \draw[ink,line width=1.3pt] (0,0.06) -- (0,-10*\v-0.06);
  \def\cellbox#1#2#3{\draw[#3] (#2*\u-\u,-#1*\v+\tz*\v) rectangle (#2*\u,-#1*\v+\tz*\v-\v);}
  % slices y = 4 (dashed), y = 3 (thin) and y = s = 2 (thick), each with its two input cells
  \foreach \t/\i in {18/3,18/4,19/4,19/5,20/5,20/6} {\cellbox{\t}{\i}{ink,line width=0.7pt,densely dashed}}
  \foreach \t/\i in {19/2,19/3,20/3,20/4,21/4,21/5} {\cellbox{\t}{\i}{ink,line width=0.8pt}}
  \foreach \t/\i in {20/1,20/2,21/2,21/3,22/3,22/4} {\cellbox{\t}{\i}{ink,line width=1.8pt}}
  % on the firing row, cell i lies on chain i
  \foreach \i in {1,...,10} {\node[font=\scriptsize,text=white] at (\i*\u-0.5*\u,-9.5*\v) {\i};}
  % chains -1, 0, 1 cross the first row shown at the cells 8, 9, 10
  \node[font=\scriptsize,anchor=south] at (7.5*\u,0.03) {$-1$};
  \node[font=\scriptsize,anchor=south] at (8.5*\u,0.03) {$0$};
  \node[font=\scriptsize,anchor=south] at (9.5*\u,0.03) {$1$};
  \node[font=\scriptsize,anchor=south east] at (6.95*\u,0.03) {chain};
  \node[font=\scriptsize,anchor=east] at (-0.08,-9.5*\v) {$t=2n-2$};
  \node[font=\scriptsize,anchor=east] at (-0.08,-0.5*\v) {$t=13$};
  % legend
  \begin{scope}[font=\scriptsize]
    \node[anchor=west,inner sep=0pt] at (4.8,-0.15) {chain $c$: the anti-diagonal $t+i=2n-2+c$};
    \foreach \c/\l [count=\k from 1] in {
        rgInput/{chains $-1$ and $0$: the input anti-diagonals},
        rgReturn/{chain $1$: the return chain},
        stF/{firing row: cell $i$ fires on chain $i$}}
      \node[anchor=west,inner sep=0pt] at (4.8,-0.15-0.45*\k) {\swatch{\c}\ \l};
    \node[anchor=west,inner sep=0pt] at (4.8,-1.95) {\swatch{rgRefl}\,\swatch{rgRefl2}\ chains $c\ge2$, alternately shaded};
    \node[anchor=west,inner sep=0pt] at (4.8,-2.5) {outlined: slices $y=2$ (thick), $3$ and $4$ (dashed)};
  \end{scope}
\end{tikzpicture}
\caption{The lower left corner of the line of length $n=12$, coloured by
chain. The cell $i$ of the firing row lies on chain $i$. Each outlined group of
six cells is a slice $(u_1(y),\dots,u_4(y))$ of the band of chains
$1,\dots,4$ used in the proof of \cref{lem:left} with $s=2$, together with its
two input cells $u_{-1}(y)$ and $u_0(y)$: the slice is computed from the slice
to its right and from these two input cells. The group drawn with a thick line
is the one at $y=s$; its last cell $u_{2s}(s)$ is in the firing row.}
\label{fig:chains}
\end{figure}

Fix $n\ge2$. For $c\ge0$ and $\max(1,c)\le i\le n$, and for $c=-1$ and $1\le
i\le n-1$, put
\[
v_c(i):=C_n(2n-2+c-i,\,i),
\]
the cell at position $i$ of the \emph{chain} $t+i=2n-2+c$. Chains $-1$ and $0$
are the input anti-diagonals, and chain $1$ is the return chain
(\cref{fig:chains}). The predecessors of the cell $(2n-2+c-i,i)$ lie on the
chains $c-2$, $c-1$ and $c$, so
\begin{equation}\label{eq:chains}
v_c(i)=\delta\bigl(v_{c-2}(i-1),\,v_{c-1}(i),\,v_c(i+1)\bigr)\qquad(c\ge1),
\end{equation}
where $v_{c-2}(0)$ and $v_c(n+1)$ stand for the borders. If the line of length
$n$ is synchronized at time $2n-2$, then the cell $i$ fires on chain $i$:
$v_i(i)=\Fq$, and $v_c(i)\ne\Fq$ for $c<i$. In the skewed coordinates
$u_c(y):=v_c(y+\lfloor c/2\rfloor)$ the recursion \eqref{eq:chains} becomes
\begin{equation}\label{eq:slices}
u_c(y)=\delta\bigl(u_{c-2}(y),\,u_{c-1}(y+\varepsilon_c),\,u_c(y+1)\bigr),
\qquad \varepsilon_c=1\text{ for even }c,\ \ \varepsilon_c=0\text{ for odd }c,
\end{equation}
with $u_0(y)=C_\infty(2n-2-y,y)$ and $u_{-1}(y)=C_\infty(2n-2-y,y-1)$. Here
$u_c(y)$ is the cell $y+\lfloor c/2\rfloor$ at time $2n-2+\lceil c/2\rceil-y$;
it is $\Fq$ if $\lceil c/2\rceil=y$ and not $\Fq$ if $\lceil c/2\rceil<y$. The
\emph{slice} $y$ is the sequence $(u_1(y),u_2(y),\dots)$. Its cells form a
staircase that starts just below the input cells $u_{-1}(y)$ and $u_0(y)$, at
the cell $y$ of chain $1$, and descends to the right: from $u_c(y)$ to
$u_{c+1}(y)$ the position increases by one if $c$ is odd, and the time
increases by one if $c$ is even. By \eqref{eq:slices}, each slice is computed
from the slice $y+1$ and the two input cells $u_{-1}(y)$ and $u_0(y)$: reading
the input anti-diagonals from right to left, the slices form a finite
automaton. A word $w$ on an
integer interval $I$ is \emph{$\ell$-periodic} if $w(y+\ell)=w(y)$ whenever
$y,y+\ell\in I$.

\subsection{The Left-Border Lemma}

\begin{lemma}[left border]\label{lem:left}
Let $\delta$ have $k$ states and synchronize the line of length $n$ at time
$2n-2$. Let $s,\ell\ge1$ and $s\le Y\le n-s$, and suppose that the words
$y\mapsto C_\infty(2n-2-y,y)$ for $s\le y\le Y$ and $y\mapsto
C_\infty(2n-3-y,y)$ for $s\le y\le Y-1$ are $\ell$-periodic. Then
$Y-s\le(k-1)^{2s}\ell$.
\end{lemma}

\begin{proof}
For $s\le y\le Y$ let $\mathbf u(y):=(u_1(y),\dots,u_{2s}(y))$. These are the
cells $y+\lfloor c/2\rfloor\le n$ at the times $2n-2+\lceil c/2\rceil-y$ with
$\lceil c/2\rceil\le s\le y$, so they are defined, and none of them is $\Fq$ when
$y>s$. By \eqref{eq:slices}, $u_c(y)$ with $1\le c\le 2s$ is computed from
components of index less than $c$ of $\mathbf u(y)$, components of index at
most $c$ of $\mathbf u(y+1)$, and the input cells $u_{-1}(y)$ and $u_0(y)$.
Hence $\mathbf u(y)=\Gamma(\mathbf u(y+1),u_{-1}(y),u_0(y))$ for $s+1\le y\le
Y-1$, with $\Gamma$ depending on $\delta$ only. Unfolding \eqref{eq:slices} for
even $c$, $u_{2j}(y)=\delta(u_{2j-2}(y),u_{2j-1}(y+1),u_{2j}(y+1))$, shows that
$u_{2s}(y)=\Gamma'(\mathbf u(y+1),u_0(y))$ for $s\le y\le Y-1$. By hypothesis,
$u_0$ is $\ell$-periodic on $[s,Y]$ and $u_{-1}$ on $[s+1,Y]$.

Suppose that $Y-s>(k-1)^{2s}\ell$. The pairs $(\mathbf u(y),y\bmod\ell)$ with
$s+1\le y\le Y$ take at most $(k-1)^{2s}\ell$ values, so there are $s+1\le
y_1<y_2\le Y$ with $\mathbf u(y_1)=\mathbf u(y_2)$ and $\ell\mid h:=y_2-y_1$.
By downward induction, $\mathbf u(y)=\mathbf u(y+h)$ for $s+1\le y\le y_1$: if
this holds for $y+1\le y_1$, then
\[
\mathbf u(y)=\Gamma(\mathbf u(y+1),u_{-1}(y),u_0(y))
=\Gamma(\mathbf u(y+1+h),u_{-1}(y+h),u_0(y+h))=\mathbf u(y+h),
\]
since $y+h<y_2\le Y$. As $s+h\le Y-1$ and $u_0(s+h)=u_0(s)$,
\[
u_{2s}(s+h)=\Gamma'(\mathbf u(s+1+h),u_0(s+h))=\Gamma'(\mathbf
u(s+1),u_0(s))=u_{2s}(s)=\Fq,
\]
although $\lceil 2s/2\rceil=s<s+h$, so that $u_{2s}(s+h)$, the cell $2s+h$ at
time $2n-2-h$, is not $\Fq$. This is a contradiction.
\end{proof}

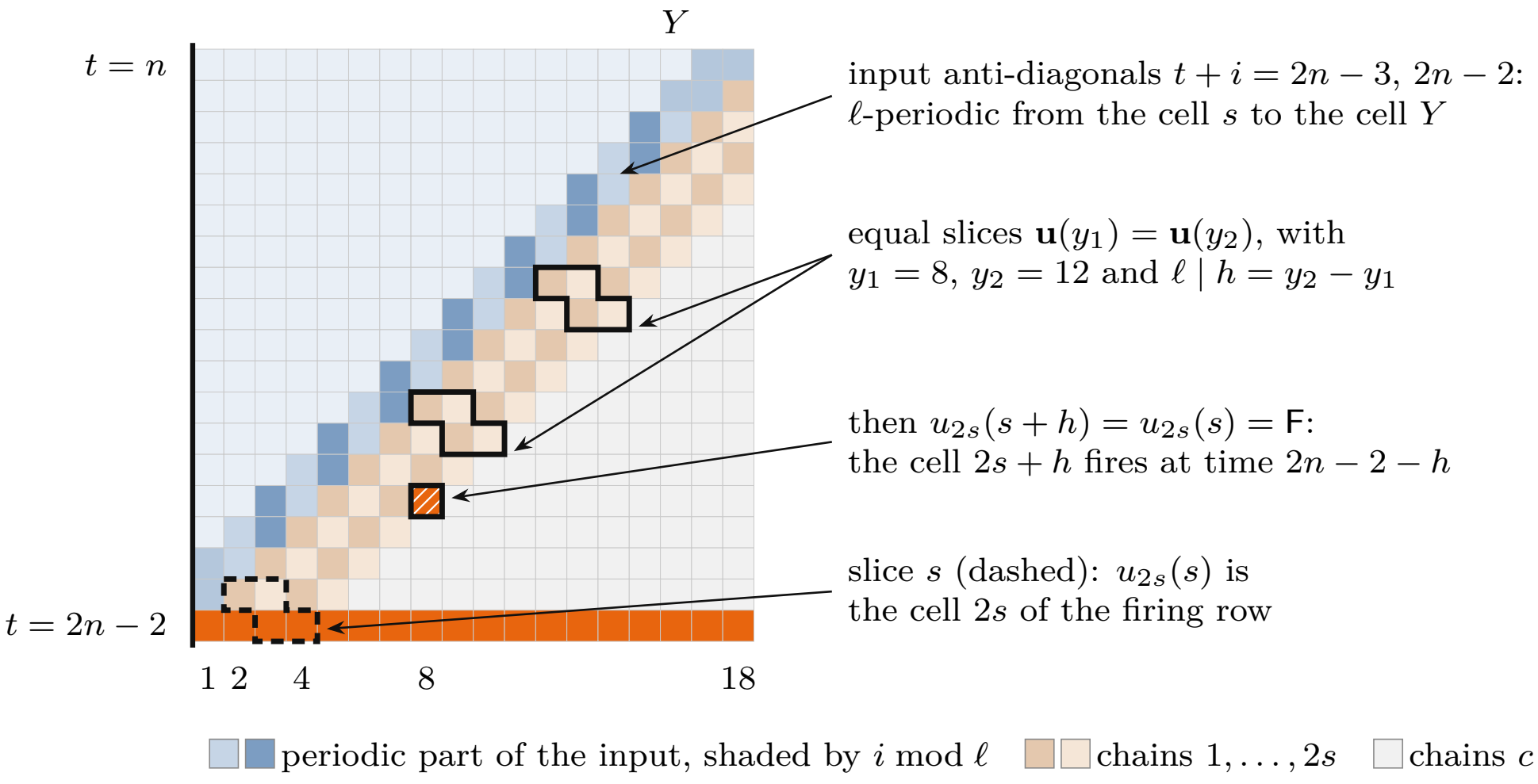
\begin{figure}[t]
\centering
%% Figure: the proof of the left-border lemma (Lemma 4.1). Schematic: the lower left corner of the
%% line of length n = 20 with s = 2, l = 2, Y = 16, y_1 = 8, y_2 = 12, h = 4. The cell (t,i), 20 <= t <= 38,
%% is [i-1,i] x [19-t,20-t]; chain c is the anti-diagonal t+i = 2n-2+c = 38+c.
\begin{tikzpicture}[x=2.6mm,y=2.6mm]
  \foreach \t in {20,...,38} {\foreach \i in {1,...,18} {%
    \pgfmathtruncatemacro{\c}{\t+\i-38}%
    \pgfmathtruncatemacro{\cls}{(\t==38) ? 1 : ((\c<=-2) ? 2 : ((\c==-1) ? (((\i>=2)&&(\i<=15)) ? 3+mod(\i,2) : 5)
      : ((\c==0) ? (((\i>=2)&&(\i<=16)) ? 3+mod(\i,2) : 5) : ((\c<=4) ? 6+mod(\c,2) : 8))))}%
    \def\cc{\ifcase\cls\or stF\or rgAgree\or stB\or stA\or rgInput\or rgRefl\or rgRefl2\or gNone\fi}%
    \filldraw[fill=\cc,draw=grid!45,line width=0.15pt] (\i-1,20-\t) rectangle (\i,19-\t);}}
  \draw[ink,line width=1.2pt] (0,0.2) -- (0,-19.2);
  % slices y_1 = 8 and y_2 = 12 (equal), slice s = 2 and the cell u_{2s}(s+h)
  \def\slice#1#2{\draw[#2] (#1-1,1-39+#1+19) -- ++(2,0) -- ++(0,-1) -- ++(1,0) -- ++(0,-1) -- ++(-2,0) -- ++(0,1) -- ++(-1,0) -- cycle;}
  \slice{8}{ink,line width=1.3pt}
  \slice{12}{ink,line width=1.3pt}
  \slice{2}{ink,line width=1.3pt,densely dashed}
  \fill[stF] (7,-14) rectangle (8,-15);
  \fill[pattern=north east lines,pattern color=white] (7,-14) rectangle (8,-15);
  \draw[ink,line width=1.3pt] (7,-14) rectangle (8,-15);
  % labels below the input anti-diagonals and of the chains
  \node[font=\scriptsize,anchor=south] at (15.5,-0.0) {$Y$};
  \node[font=\scriptsize,anchor=east] at (-0.3,-0.5) {$t=n$};
  \node[font=\scriptsize,anchor=east] at (-0.3,-18.5) {$t=2n-2$};
  \foreach \i/\l in {1/1,2/2,4/4,8/8,18/18} \node[font=\scriptsize,anchor=north] at (\i-0.5,-19.3) {$\l$};
  % annotations
  \draw[-{Stealth[length=1.5mm]},ink,line width=0.5pt] (20.5,-1.5) -- (13.7,-4.0);
  \node[font=\scriptsize,anchor=west,align=left] at (20.5,-1.5) {input anti-diagonals $t+i=2n-3,\,2n-2$:\\
    $\ell$-periodic from the cell $s$ to the cell $Y$};
  \draw[-{Stealth[length=1.5mm]},ink,line width=0.5pt] (20.5,-6.6) -- (14.3,-8.9);
  \draw[-{Stealth[length=1.5mm]},ink,line width=0.5pt] (20.5,-6.6) -- (10.3,-12.9);
  \node[font=\scriptsize,anchor=west,align=left] at (20.5,-6.6) {equal slices $\mathbf u(y_1)=\mathbf u(y_2)$, with\\
    $y_1=8$, $y_2=12$ and $\ell\mid h=y_2-y_1$};
  \draw[-{Stealth[length=1.5mm]},ink,line width=0.5pt] (20.5,-12.6) -- (8.3,-14.4);
  \node[font=\scriptsize,anchor=west,align=left] at (20.5,-12.6) {then $u_{2s}(s+h)=u_{2s}(s)=\Fq$:\\
    the cell $2s+h$ fires at time $2n-2-h$};
  \draw[-{Stealth[length=1.5mm]},ink,line width=0.5pt] (20.5,-17.4) -- (4.3,-18.6);
  \node[font=\scriptsize,anchor=west,align=left] at (20.5,-17.4) {slice $s$ (dashed): $u_{2s}(s)$ is\\
    the cell $2s$ of the firing row};
  % legend
  \node[anchor=north west,font=\scriptsize,align=left] at (0,-21.6) {%
    \swatch{stB}\,\swatch{stA}\,periodic part of the input, shaded by $i\bmod\ell$\quad
    \swatch{rgRefl2}\,\swatch{rgRefl}\,chains $1,\dots,2s$\quad \swatch{gNone}\,chains $c>2s$\quad \swatch{stF}\,$\Fq$};
\end{tikzpicture}
\caption{The proof of \cref{lem:left}, drawn for $n=20$, $s=2$, $\ell=2$ and
$Y=16$: the lower left corner of the line of length $n$, at the times
$n\le t\le2n-2$. The two input anti-diagonals $t+i=2n-3$ and $t+i=2n-2$ are
$\ell$-periodic from the cell $s$ to the cell $Y$. Each slice $\mathbf u(y)$,
a group of $2s$ cells on the chains $1,\dots,2s$ (outlined), is computed from
the slice $\mathbf u(y+1)$ and from two input cells, so the slices form a
finite automaton that reads the input anti-diagonals from right to left. If two
slices are equal, $\mathbf u(y_1)=\mathbf u(y_2)$ with $\ell\mid
h=y_2-y_1$, then $\mathbf u(y)=\mathbf u(y+h)$ for $s+1\le y\le y_1$, and the
last cell $u_{2s}(s+h)$ of the slice $s+h$ (hatched) has the state of the last
cell $u_{2s}(s)$ of the slice $s$, which is $\Fq$. It is the cell $2s+h$ at
the time $2n-2-h$, which would fire $h$ steps too early. Hence $Y-s\le
(k-1)^{2s}\ell$.}
\label{fig:leftproof}
\end{figure}

In words: if the two input anti-diagonals are $\ell$-periodic over a stretch
of more than $(k-1)^{2s}\ell$ cells that starts at the cell $s$, then the band
of the first $2s$ chains repeats a state, and the cell $2s+h$ fires $h$ steps
too early (\cref{fig:leftproof}). The bound is exponential in $s$, but for $s=1$ and $s=2$ it is small
enough to be useful in computations (\cref{prop:germs,cor:germ12}).

\subsection{No Periodic Wedge at the Left Border}

\begin{theorem}[left-border barrier]\label{thm:left}
Let $\delta$ be a minimal-time solution. There are no $s\ge1$, $\alpha>0$,
$t_0$ and $p,q\ge1$ such that $C_\infty(t+p,i)=C_\infty(t,i)$ and
$C_\infty(t,i+q)=C_\infty(t,i)$ whenever the cells involved lie in the wedge
\[
\Omega:=\{(t,i):\ s\le i\le\alpha t,\ t\ge t_0\}.
\]
\end{theorem}

\begin{figure}[t]
\centering
%% Figure: the left-border barrier (Theorem 4.2) and eventually regular half-lines (Corollary 4.3).
%% Schematic, in continuous coordinates: position i to the right, time t downwards.
\begin{tikzpicture}[x=1.6mm,y=1.6mm]
  % (a) the wedge and the input anti-diagonals of three lengths (s = 2, alpha = 0.4, t_0 = 8)
  \begin{scope}
    \fill[gNone] (0,0) -- (40,-40) -- (0,-40) -- cycle;
    \fill[stB] (2,-8) -- (3.2,-8) -- (16,-40) -- (2,-40) -- cycle;
    \draw[stC,line width=0.6pt] (2,-40) -- (2,-8) -- (3.2,-8) -- (16,-40);
    \foreach \n in {12,16,20} {
      \draw[grid,line width=0.5pt] (\n-1,-\n+1) -- (0,-2*\n+2);
      \pgfmathsetmacro{\Y}{0.4*(2*\n-3)/1.4}
      \draw[rgCone,line width=2.2pt] (2,-2*\n+4) -- (\Y,-2*\n+2+\Y);}
    \draw[stF,line width=0.9pt] (5,-25) -- (5,-22) -- (8,-22);
    \fill[ink] (5,-25) circle (0.45) (8,-22) circle (0.45);
    \draw[ink,line width=1.2pt] (0,0) -- (0,-40);
    \draw[ink,line width=0.7pt] (0,0) -- (40,-40);
    \node[font=\small] at (5.6,-35) {$\Omega$};
    \node[font=\scriptsize,anchor=north] at (2,-40.2) {$s$};
    \node[font=\scriptsize,anchor=north] at (16,-40.2) {$i=\alpha t$};
    \node[font=\scriptsize,anchor=east] at (-0.3,-8) {$t_0$};
    \node[font=\scriptsize,rotate=-45,anchor=south] at (30,-30) {front};
    \node[font=\scriptsize,anchor=west] at (8.4,-21.4) {$(t-\ell,i+\ell)$};
    \node[font=\scriptsize,anchor=north west] at (5.2,-25.1) {$(t,i)$};
    \node[font=\scriptsize,anchor=west,align=left] at (12.4,-11.6) {$t+i=2n-2$};
    \draw[-{Stealth[length=1.4mm]},ink,line width=0.45pt] (12.6,-12.3) -- (6.8,-15.0);
    \node[font=\small,anchor=south west] at (-2,0.8) {(a) the proof of \cref{thm:left}};
  \end{scope}
  % (b) an eventually regular half-line: blocks separated by signals of speeds 0, 1/4 and 1/2
  \begin{scope}[shift={(48,0)}]
    \fill[gNone] (0,0) -- (40,-40) -- (0,-40) -- cycle;
    \fill[rgInput] (0,-8) rectangle (3,-40);
    \fill[stB] (3,-8) -- (11,-40) -- (3,-40) -- cycle;
    \fill[rgRefl2] (3,-8) -- (4,-8) -- (20,-40) -- (11,-40) -- cycle;
    \fill[rgAgree] (4,-8) -- (8,-8) -- (40,-40) -- (20,-40) -- cycle;
    \draw[ink,line width=0.6pt] (3,-8) -- (3,-40) (3,-8) -- (11,-40) (4,-8) -- (20,-40);
    \draw[ink,densely dashed,line width=0.9pt] (4,-20) -- (8,-40) (4,-20) -- (4,-40);
    \draw[grid,line width=0.4pt] (0,-8) -- (8,-8);
    \draw[ink,line width=1.2pt] (0,0) -- (0,-40);
    \draw[ink,line width=0.7pt] (0,0) -- (40,-40);
    \node[font=\scriptsize,anchor=north] at (1.5,-40.2) {$w_0$};
    \node[font=\scriptsize,anchor=north] at (7.0,-40.2) {$z_1^{a_1+\nu D_1}$};
    \node[font=\scriptsize,anchor=north] at (15.5,-40.2) {$z_2^{a_2+\nu D_2}$};
    \node[font=\scriptsize,anchor=north] at (30.0,-40.2) {$z_3^{a_3+\nu D_3}$};
    \node[font=\scriptsize,anchor=east] at (-0.3,-8) {$t_1$};
    \node[font=\scriptsize,anchor=west] at (10.5,-2.5) {transient part};
    \draw[-{Stealth[length=1.4mm]},ink,line width=0.45pt] (10.6,-3.0) -- (3.0,-5.4);
    \node[font=\small,anchor=south west] at (-2,0.8) {(b) an eventually regular half-line};
  \end{scope}
  % legend
  \node[anchor=north west,font=\scriptsize,align=left] at (0,-44.6) {%
    \swatch{stB}\,in (a): the wedge $\Omega$; in (b): the first block that grows, with a wedge (dashed)\\[0.3ex]
    \swatch{rgCone}\,in (a): the stretch from the cell $s$ to the cell $Y_n$ of the input anti-diagonals
    of the lengths $n=12,16,20$};
\end{tikzpicture}
\caption{(a)~The proof of \cref{thm:left}. If the half-line were periodic in
time and in space on the wedge $\Omega=\{(t,i):s\le i\le\alpha t,\ t\ge t_0\}$,
then the input anti-diagonals $t+i=2n-3$ and $t+i=2n-2$ of the line of length
$n$ would be $\ell$-periodic, with $\ell=pq$, from the cell $s$ to the cell
$Y_n\approx2\alpha n/(1+\alpha)$ (thick), a stretch whose length grows
linearly with $n$; for large $n$, \cref{lem:left} gives a contradiction. The
path from $(t,i)$ to $(t-\ell,i+\ell)$ shows why: $\ell$ steps along an
anti-diagonal are $q$ time periods $p$ followed by $p$ space periods $q$
inside $\Omega$. (b)~An eventually regular half-line (\cref{cor:regular}):
from the time $t_1$ on, it consists of finitely many periodic blocks
separated by signals of rational speeds, here $0$, $1/4$ and $1/2$. The first
block that grows contains a wedge (dashed) on which the half-line is periodic
in time and space, which (a) excludes.}
\label{fig:wedge}
\end{figure}
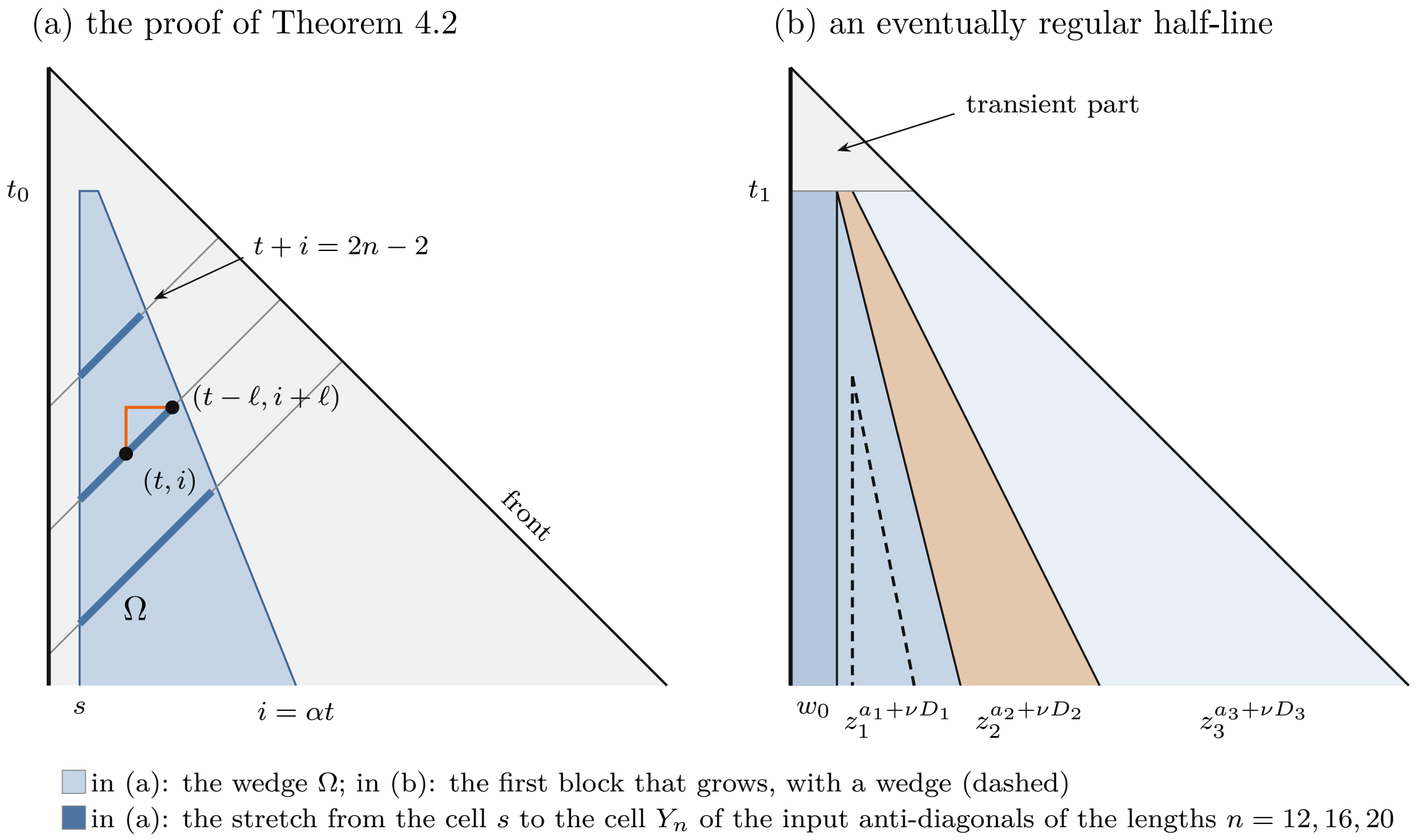

\begin{proof}
Suppose otherwise and put $\ell:=pq$. If $(t,i)$ and $(t-\ell,i+\ell)$ lie in
$\Omega$, so do the cells $(t-\ell,i+jq)$ for $0\le j\le p$ and $(t-\ell+jp,i)$
for $0\le j\le q$, hence
$C_\infty(t-\ell,i+\ell)=C_\infty(t-\ell,i)=C_\infty(t,i)$. Let
$Y_n:=\lfloor\alpha(2n-3)/(1+\alpha)\rfloor$. For $s\le y\le Y_n$ the cells
$(2n-2-y,y)$ and $(2n-3-y,y)$ satisfy $y\le\alpha(2n-3-y)$, so they lie in
$\Omega$ as soon as $2n-3-Y_n\ge t_0$, that is, for all large $n$. A step of
$\ell$ positions along an anti-diagonal is the step from $(t,i)$ to
$(t-\ell,i+\ell)$, so the words of \cref{lem:left} are $\ell$-periodic on
$[s,Y_n]$ and $[s,Y_n-1]$. As $Y_n\ge2\alpha n/(1+\alpha)-4$, for large
$n$ the number $Y:=\min(Y_n,n-s)$ satisfies $Y-s>(k-1)^{2s}\ell$, which
contradicts \cref{lem:left} (\cref{fig:wedge}(a)).
\end{proof}

The half-line of a rule is \emph{eventually regular} if there are $t_1$,
$P\ge1$ and, for each residue $r\in\{0,\dots,P-1\}$, an integer $m$, words
$w_0,\dots,w_m$, nonempty words $z_1,\dots,z_m$ over $W$ and integers
$a_j,D_j\ge0$ (all depending on $r$) such that for all $\nu\ge0$ the
configuration $C_\infty(t,1)\cdots C_\infty(t,t+1)$ at the time
$t=t_1+r+\nu P$ is $w_0z_1^{a_1+\nu D_1}w_1\cdots z_m^{a_m+\nu D_m}w_m$. In
words: from some time on, the half-line consists of finitely many periodic
regions separated by signals of rational speeds; on signals in one-dimensional
cellular automata see~\cite{MazoyerTerrier99}. The half-line of the five-state
rule $\delta_{12}$ of \cref{sec:five} looks like this on all the cells that we
simulated, those with $t+i\le1034$ (\cref{fig:halflines,fig:delta12}): the cells
between roughly the cell $t/3$ and the front are in state $\Gq$, and the cells
between the left border and roughly the cell $t/3$ alternate between $\Aq$ and
$\Lq$, both in time and in space. Whether it continues in this way is not
known, and \cref{cor:germ12} does not need it.

\begin{corollary}\label{cor:regular}
The half-line of a minimal-time solution is not eventually regular in the
sense just defined.
\end{corollary}

\begin{proof}
Suppose that it is. For each residue $r$ some $D_j$ is positive, because the
configuration grows by $P$ cells per period; let $j(r)$ be the least such
index. The blocks and the junctions to the left of $z_{j(r)}$ have constant
lengths, so the block $z_{j(r)}^{a_{j(r)}+\nu D_{j(r)}}$ starts at a fixed cell
$s_r$, and its length is at least $\beta t$ for some $\beta>0$ and all large
$t$ of the residue class $r$. Inside this block, the state of a cell depends
only on its distance to $s_r$ modulo $|z_{j(r)}|$, and it is the same at the
times $t$ and $t+P$. Let $s:=\max_r s_r$, $q:=\prod_r|z_{j(r)}|$ and
$\alpha:=\beta/2$. For $t_0$ large, every cell of the wedge $\{(t,i):s\le
i\le\alpha t,\ t\ge t_0\}$ lies in the block of index $j(r)$ of its residue
class, so $C_\infty$ is periodic with period $P$ in time and $q$ in space on
this wedge (\cref{fig:wedge}(b)). This contradicts \cref{thm:left}.
\end{proof}

Eventual regularity means exactly the definition above, and \cref{cor:regular}
excludes nothing beyond it. \Cref{thm:left}, from which it follows, excludes
every half-line that is periodic in time and space on a wedge
$\{(t,i):s\le i\le\alpha t,\ t\ge t_0\}$ with $\alpha>0$, whatever happens
elsewhere. Neither result excludes periodic regions bounded away from the left
border, periodic regions at the left border whose width grows sublinearly, or
other kinds of regular behaviour that do not produce a periodic wedge at the
left border.

%% ---------------------------------------------------------------------------
\section{The First Rows After the Reflection}\label{sec:early}

The five-state rules of \cref{prop:nearmiss}, which synchronize the lines of
lengths $2,\dots,14$, fail in the first rows of the reflected triangles of all
longer lines: some cell fires a few steps after the front has reached the right
end (\cref{fig:delta14}). The condition that this never happens concerns
infinitely many lengths. We show
that, for a given half-line, it reduces to finitely many lengths, and that a
certificate for the reduction can be computed by simulating the half-line.

\subsection{Finitely Many Early Rows}

For $T\ge0$ let $E(T):=\{(\tau,\kappa):0\le\kappa\le\tau\le T\}$ be the first
$T+1$ rows of a reflected triangle, that is, its cells at the relative times
$0,\dots,T$, during the first $T+1$ steps after the front reaches the right
end. If $n\ge T+2$, then $E(T)\subseteq R_n$,
and no cell of $E(T)$ lies in the firing row $\tau=n-1$.

\begin{lemma}[early rows]\label{lem:early}
Let $\delta$ be a rule, $T\ge0$, and $n,n'\ge T+2$ such that the partial values
$J_n(\kappa)$ and $J_{n'}(\kappa)$ coincide for $0\le\kappa\le T+1$. Then for
every $(\tau,\kappa)\in E(T)$ the values $R_n(\tau,\kappa)$ and
$R_{n'}(\tau,\kappa)$ are both undefined or both defined and equal
(\cref{fig:early}(a)).
\end{lemma}

\begin{proof}
By \cref{lem:determinacy} we may compare $\Psi(J_n,n;\tau,\kappa)$ and
$\Psi(J_{n'},n';\tau,\kappa)$. Unfolding the recursion of
\cref{lem:determinacy} from $(\tau,\kappa)\in E(T)$ visits only pairs
$(\tau'',\kappa'')$ with $0\le\kappa''\le\tau''\le T$ and reads the symbols
$x_{\kappa'}$ with $\kappa'\le\kappa''+1\le T+1$. The case $\kappa'=m$ would
require $m\le T+1$, which is excluded for $m\in\{n,n'\}$; hence the left
border is never read, and the input word is read only at positions
$\kappa'\le T+1$, where $J_n$ and $J_{n'}$ coincide. The two recursions are
therefore identical step by step, including the points where an argument is
$\Fq$ or undefined.
\end{proof}

The input word on $[0,T+1]$ consists of depth-row cells of depths at most
$2T+1$ (\cref{fig:early}(b)). By \eqref{eq:comoving}, each depth row is driven
only by itself and by the two rows of the next smaller depths, so the vector of
the first $2T+2$ depth rows evolves autonomously, and one period of it is a
certificate for all later times.

\begin{lemma}[periodicity certificate]\label{lem:certificate}
Let $\delta$ be a rule, $D\ge0$, and $V_t:=(\Phi_t(0),\dots,\Phi_t(D))$ for
$t\ge D$. There is a map $\Gamma_D$ on $W^{D+1}$, depending on $\delta$ and $D$
only, with $V_{t+1}=\Gamma_D(V_t)$ whenever $t\ge D$ and $V_t\in W^{D+1}$. If
$V_{t_0}=V_{t_0+p}$ for some $t_0\ge D$ and $p\ge1$, and $V_t\in W^{D+1}$ for
$t_0\le t<t_0+p$, then $V_t\in W^{D+1}$ and $V_{t+p}=V_t$ for all $t\ge t_0$.
\end{lemma}

\begin{proof}
For $t\ge D$ and $0\le j\le D$ we have $j\le t$, so the arguments in
\eqref{eq:comoving} are $\Phi_t(j)$, $\Phi_t(j-1)$ and $\Phi_t(j-2)$, with
indices at most $D$, and $\Lq$ for negative indices; none of them is the border
symbol $\Phi_t(t+1)=\ast$. Hence $\Gamma_D(v)_j:=\delta(v_j,v_{j-1},v_{j-2})$, with
$v_{-1}=v_{-2}=\Lq$, has the required property. For the second statement,
suppose that $V_{t'}\in W^{D+1}$ and $V_{t'+p}=V_{t'}$ for $t_0\le t'\le t$,
which is true for $t=t_0$. Then $V_{t+1+p}=\Gamma_D(V_{t+p})=\Gamma_D(V_t)=V_{t+1}$, and
$V_{t+1}\in W^{D+1}$, by hypothesis if $t+1<t_0+p$ and because
$V_{t+1}=V_{t+1-p}$ otherwise. Induction on $t$ concludes.
\end{proof}

\begin{corollary}[finitely many early rows]\label{cor:reduce}
Let $\delta$, $T\ge0$, $D:=2T+1$, $t_0\ge D$ and $p\ge1$ satisfy the hypothesis
of \cref{lem:certificate}. Then $J_n(\kappa)=J_{n+p}(\kappa)\in W^2$ for all
$0\le\kappa\le T+1$ and $n\ge t_0+3$. Hence, for every $N_0\ge T+2$, every
$R_n$ with $n\ge N_0$ coincides on $E(T)$, in the sense of \cref{lem:early},
with $R_{n'}$ for one of the finitely many lengths $N_0\le
n'<\max(N_0,t_0+3)+p$.
\end{corollary}

\begin{proof}
For $1\le\kappa\le T+1$ the components of $J_n(\kappa)$ are cells of the depth
rows $2\kappa-2$ and $2\kappa-1\le D$ at the times $n-3+\kappa$ and
$n-2+\kappa$, which are at least $n-2>t_0$; by \cref{lem:certificate} they lie
in $W$ and do not change when $n$ is replaced by $n+p$. For $\kappa=0$ the
lower component is $\Lq$, by convention or by \cref{lem:lightcone}, and the
upper one is $C_\infty(n-2,n)=\Lq$. If $n\ge\max(N_0,t_0+3)+p$, subtracting a
multiple of $p$ gives a length $n'$ in $[\max(N_0,t_0+3),\max(N_0,t_0+3)+p)$
with $J_{n'}=J_n$ on $[0,T+1]$, and \cref{lem:early} applies since
$n,n'\ge T+2$.
\end{proof}

\begin{figure}[t]
\centering
%% Figure: the early rows (Lemma 5.1, schematic) and the periodicity certificate of delta14 (Lemma 5.2,
%% Corollary 5.3; cells from figures.py). In (a) the relative cell (tau,kappa) of R_m is
%% [m-kappa-1,m-kappa] x [-tau-3,-tau-2]; here T = 3.
\begin{tikzpicture}
  % (a) two reflected triangles with the same input on [0, T+1]
  \begin{scope}[x=2.5mm,y=2.5mm]
    \foreach \m/\dx in {9/0,13/13} {
      \begin{scope}[shift={(\dx,0)}]
        \foreach \tau in {-2,...,12} {\foreach \kap in {0,...,12} {%
          \pgfmathtruncatemacro{\cls}{((\kap<=\tau)&&(\tau<=\m-1)) ? ((\tau==\m-1) ? 1 : ((\tau<=3) ? 2 : 3))
            : ((((\tau==\kap-1)||((\tau==\kap-2)&&(\kap>=1)))&&(\kap<=\m-1)) ? ((\kap<=4) ? 4 : 5) : 0)}%
          \ifnum\cls>0
            \def\cc{\ifcase\cls\or stF\or rgRefl2\or rgRefl\or rgCone\or rgInput\fi}%
            \filldraw[fill=\cc,draw=grid!45,line width=0.15pt] (\m-\kap-1,-\tau-2) rectangle (\m-\kap,-\tau-3);
          \fi}}
        \draw[ink,line width=1.2pt] (\m,0.1) -- (\m,-\m-2.1);
        \draw[ink,line width=1.2pt] (0,-\m+0.9) -- (0,-\m-2.1);
        \draw[ink,densely dashed,line width=0.7pt] (\m-4.6,-6) -- (\m+0.6,-6);
      \end{scope}}
    \node[font=\small,anchor=south] at (4.5,0.4) {$R_n$, $n=9$};
    \node[font=\small,anchor=south] at (21.0,0.4) {$R_{n'}$, $n'=13$};
    \node[font=\scriptsize,anchor=west] at (26.8,-6) {$\tau=T$};
    \draw[-{Stealth[length=1.8mm]},ink,line width=0.6pt] (8.6,-0.2) .. controls (10.6,2.7) and (15.4,2.7) .. (17.4,-0.2);
    \node[font=\scriptsize,anchor=south] at (13.0,2.25) {equal on $[0,T+1]$};
    \node[font=\scriptsize,anchor=north,align=center] at (13,-16.4) {then $R_n$ and $R_{n'}$ coincide on the first $T+1$ rows};
    \node[font=\small,anchor=south west] at (-0.5,3.6) {(a) the early rows};
  \end{scope}
  % (b) the depth rows 0..24 of the half-line of delta14 at the times 24..130
  \begin{scope}[shift={(0,-5.9cm)},x=1.22mm,y=1.22mm]
    \input{figures/cells_delta14_depth}
    \draw[grid,line width=0.3pt] (0,0) rectangle (107,-25);
    \foreach \n in {60,68} {\foreach \k in {1,...,11} {
      \draw[ink,line width=0.55pt] (\n-27+\k,-2*\k+2) rectangle ++(1,-1);
      \draw[ink,line width=0.55pt] (\n-26+\k,-2*\k+1) rectangle ++(1,-1);}}
    \draw[stF,line width=1.0pt] (27,1.0) -- (27,-26.0);
    \draw[stF,line width=1.0pt,densely dashed] (35,1.0) -- (35,-26.0);
    \draw[{Stealth[length=1.4mm]}-{Stealth[length=1.4mm]},ink,line width=0.5pt] (27,2.0) -- (35,2.0);
    \node[font=\scriptsize,anchor=south] at (31,2.1) {$p=8$};
    \node[font=\scriptsize,anchor=north] at (27,-26.2) {$t_0=51$};
    \node[font=\scriptsize,anchor=north] at (36,-26.2) {$t_0+p$};
    \node[font=\scriptsize,anchor=north] at (45,-23.4) {$J_{60}$};
    \node[font=\scriptsize,anchor=north] at (53,-23.4) {$J_{68}$};
    \foreach \x/\l in {0/24,56/80,76/100,106/130} \node[font=\scriptsize,anchor=north] at (\x+0.5,-26.2) {$\l$};
    \node[font=\scriptsize,anchor=north] at (90,-26.2) {time $t$};
    \foreach \y in {0,12,24} \node[font=\scriptsize,anchor=east] at (-0.5,-\y-0.5) {$\y$};
    \node[font=\scriptsize,rotate=90,anchor=south] at (-3.4,-12.5) {depth $j$};
    \node[font=\small,anchor=south west] at (-4,5.0) {(b) the depth rows of the half-line of $\delta_{14}$};
  \end{scope}
  % legend
  \node[anchor=north west,font=\scriptsize,align=left] at (0,-9.6) {%
    in (a):\ \swatch{rgCone}\,input cells at $\kappa\le T+1$\quad \swatch{rgRefl2}\,the first $T+1$ rows $E(T)$\quad
    \swatch{stF}\,firing row\qquad
    in (b):\ \swatch{white}\,$\Lq$\ \swatch{stG}\,$\Gq$\ \swatch{stA}\,$\Aq$\ \swatch{stB}\,$\Bq$};
\end{tikzpicture}
\caption{(a)~\Cref{lem:early}, drawn for $T=3$, $n=9$ and $n'=13$ in the
relative coordinates of \cref{fig:pumping}(a). The first $T+1$ rows $E(T)$ of a
reflected triangle are computed from the right border and the input cells at
the distances $\kappa\le T+1$ from it; if these input cells agree, so do the
first $T+1$ rows. (b)~\Cref{lem:certificate,cor:reduce} for the rule
$\delta_{14}$ of \cref{sec:five}: the depth rows $j=0,\dots,24$ (downwards) of
its half-line at the times $t=24,\dots,130$ (to the right). From $t_0=51$ on,
the vector $V_t$ of these rows is periodic with period $p=8$; the simulation
checks $V_t=V_{t+8}$ for $51\le t\le152$, and \cref{lem:certificate} extends
it to all $t\ge51$. Outlined: the input cells $J_n(\kappa)$, $1\le\kappa\le
11$, of the lengths $n=60$ and $n=68=60+p$, on the depth rows $2\kappa-2$ and
$2\kappa-1$ at times beyond $t_0$; they carry the same values, as
\cref{cor:reduce} asserts with $T=10$.}
\label{fig:early}
\end{figure}

For a minimal-time solution, the depth rows are eventually periodic
(\cref{thm:third}), so $t_0$ and $p$ as in \cref{cor:reduce} always exist.
Thus the condition that no reflected triangle $R_n$ with $n\ge N_0$ has a
firing cell in its first $T+1$ rows reduces to finitely many lengths, with a
certificate $(t_0,p)$ found by simulating the half-line.

\subsection{Early-Row Formulas}

For a search over all rules, the half-line is not known in advance, but its
first rows and the lines of small lengths can be encoded together. The
following conditions collect what a minimal-time solution must satisfy on the
complete lines of small lengths, on the early rows and return chains of the
next lengths, and on the cone of their right ends.

\begin{definition}\label{def:G}
Let $k\ge3$, $T\ge1$, $N_1\ge T+2$ and $T+1\le C\le N_1$. A $k$-state rule
$\delta$ satisfying \eqref{eq:quiescence} \emph{satisfies $\G_k(T,N_1,C)$} if
\begin{enumerate}[(G1)]
\item $\delta$ synchronizes the line of length $n$ at time $2n-2$ for $2\le
n\le T+1$;
\item for $T+2\le n\le N_1$, $1\le i\le n$ and $t\le n-1+T$: $C_n(t,i)\in W$;
\item for $T+2\le n\le N_1$ and $1\le i\le n$ with $t:=2n-1-i\le n-1+T$:
$C_n(t,i)\ne C_\infty(t,i)$;
\item $C_\infty(t,i)\in W$ whenever $t+i\le 2N_1$;
\item for $T+2\le n\le C$: $R_n(\tau,\kappa)\in W$ for all $(\tau,\kappa)\in
R_n$ with $\tau+\kappa\le n-1$ and $\tau<n-1$, $R_n(n-1,0)=\Fq$, and
$R_n(\kappa,\kappa)\ne C_\infty(n-1+\kappa,n-\kappa)$ for $2\kappa\le n-1$.
\end{enumerate}
We write $\G_k(T,N_1)$ for $\G_k(T,N_1,T+1)$, in which (G5) is void.
\end{definition}

\begin{figure}[t]
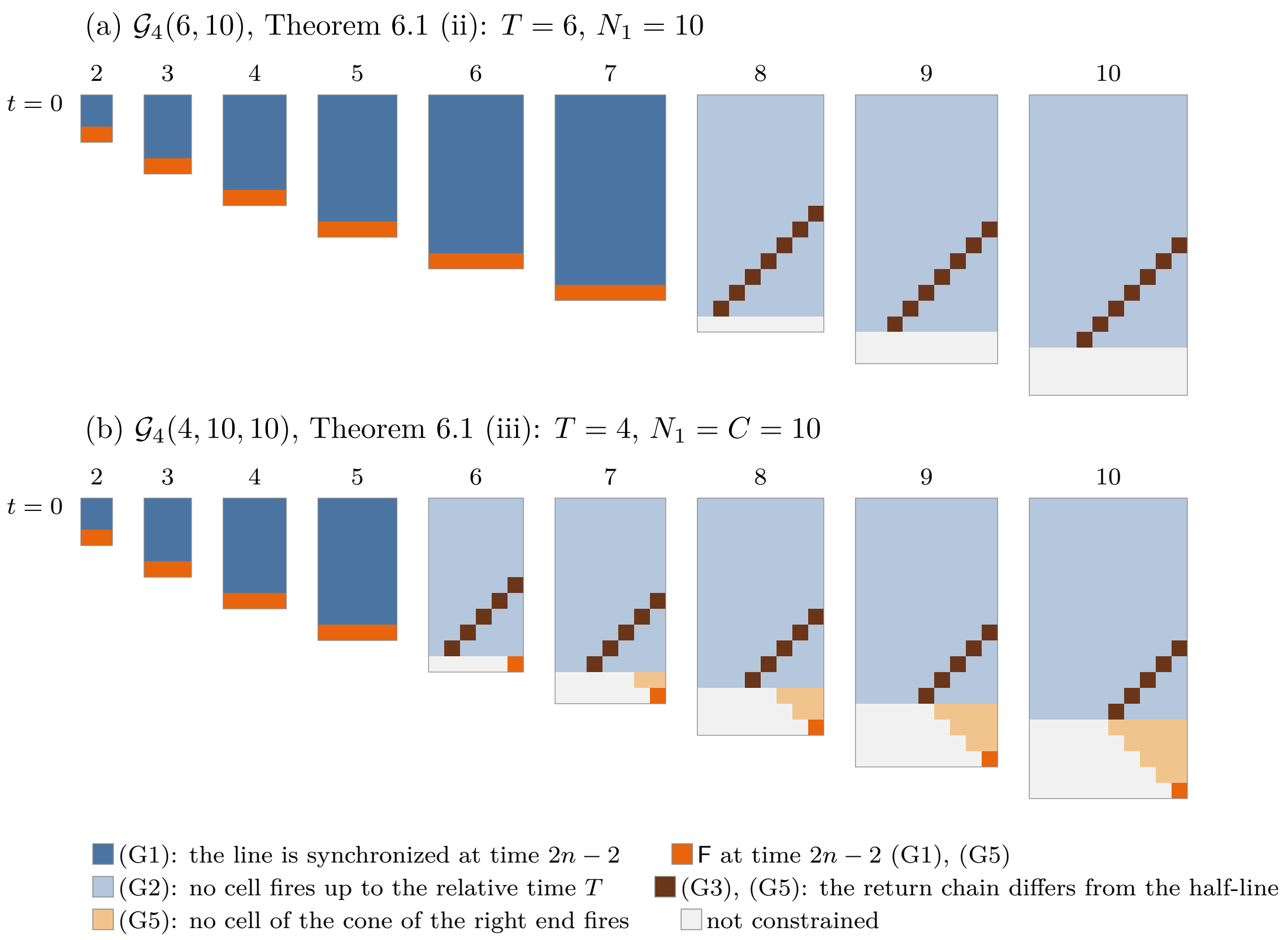

\centering
%% Figure: the cells constrained by the formulas G_4(6,10) and G_4(4,10,10) of Theorem 6.1 (ii), (iii)
%% (regions from figures.py). Each line of length n = 2,...,10 is drawn with time downwards.
\begin{tikzpicture}[x=1.85mm,y=1.85mm]
  \begin{scope}
    \input{figures/regions_G6}
    \node[font=\small,anchor=south west] at (-0.5,2.6) {(a) $\G_4(6,10)$, \cref{thm:four}~(ii): $T=6$, $N_1=10$};
  \end{scope}
  \begin{scope}[shift={(0,-25.5)}]
    \input{figures/regions_G4}
    \node[font=\small,anchor=south west] at (-0.5,2.6) {(b) $\G_4(4,10,10)$, \cref{thm:four}~(iii): $T=4$, $N_1=C=10$};
  \end{scope}
  \node[font=\scriptsize,anchor=east] at (-0.4,-0.5) {$t=0$};
  \node[font=\scriptsize,anchor=east] at (-0.4,-26.0) {$t=0$};
  \node[anchor=north west,font=\scriptsize,align=left] at (0,-46.6) {%
    \swatch{gFull}\,(G1): the line is synchronized at time $2n-2$\qquad
    \swatch{stF}\,$\Fq$ at time $2n-2$ (G1), (G5)\\[0.4ex]
    \swatch{gEarly}\,(G2): no cell fires up to the relative time $T$\qquad
    \swatch{gReturn}\,(G3), (G5): the return chain differs from the half-line\\[0.4ex]
    \swatch{gCone}\,(G5): no cell of the cone of the right end fires\qquad
    \swatch{gNone}\,not constrained};
\end{tikzpicture}
\caption{The cells of the lines of lengths $n=2,\dots,10$ (time downwards,
the line of length $n$ in a column of width $n$) that the formulas of
\cref{thm:four}~(ii) and~(iii) constrain, coloured by the condition of
\cref{def:G} that applies. Condition (G4), on the half-line, is not drawn.
(a)~In $\G_4(6,10)$ the lines of lengths $2,\dots,7$ are complete; for the
lengths $8$, $9$ and $10$, only the cells up to the relative time $\tau=6$ are
constrained. (b)~In $\G_4(4,10,10)$ only the lines of lengths $2,\dots,5$ are
complete; for the lengths $6,\dots,10$, the formula constrains the cells up to
$\tau=4$, the cone of the right end at the firing time and the return chain in
it.}
\label{fig:regions}
\end{figure}

\Cref{fig:regions} shows the cells that these conditions constrain in the
two formulas of \cref{thm:four}. Condition (G1) asks for the complete lines of
lengths up to $T+1$. Condition
(G2) says that no cell of the lines of lengths $T+2,\dots,N_1$ fires up to the
relative time $\tau=T$, and (G3) is the return chain of \cref{lem:return},
restricted to the relative times $\tau\le T$. Condition (G4) says that the
half-line does not fire on the cells with $t+i\le 2N_1$. Condition (G5)
concerns the backward dependency cone $\{(\tau,\kappa)\in R_n:\tau+\kappa\le
n-1\}$ of the firing cell $(n-1,0)$; by the local definition of $C_n$, its
cells are defined even if a cell outside the cone fires earlier.

\begin{lemma}\label{lem:G}
Every $k$-state minimal-time solution satisfies $\G_k(T,N_1,C)$ for all $T$,
$N_1$ and $C$ as in \cref{def:G}.
\end{lemma}

\begin{proof}
(G1) holds by definition. For (G2), $n\ge T+2$ gives $n-1+T<2n-2$, and every
cell of the line of length $n$ is in $W$ before time $2n-2$. (G3) is
\cref{lem:return}, which applies because the lines of lengths $n$ and $n+1$
are synchronized at times $2n-2$ and $2n$. (G4) is \eqref{eq:neverfire}. For
(G5), the cells of $R_n$ with $\tau<n-1$ are in $W$,
$R_n(n-1,0)=C_n(2n-2,n)=\Fq$, and the return chain differs from $C_\infty$ by
\cref{lem:return}.
\end{proof}

For a fixed half-line, the early rows of all lengths $n\ge N_0$ can thus be
checked on the representatives of \cref{cor:reduce}; \cref{prop:nearmiss} is
proved in this way, and the same representatives allow a formula to impose (G2)
and (G3) on all $n\ge N_0$.

%% ---------------------------------------------------------------------------
\section{Applications to Four and Five States}\label{sec:applications}

We now combine the necessary conditions with satisfiability solvers. Each
statement of this section rests on one or more of four kinds of evidence, which
we state for each result and summarize in \cref{tab:results}: proofs in the
text; exact finite computations, that is, simulations and exhaustive
enumerations that decide a finite statement completely; satisfiable formulas,
whose models we check by simulation; and unsatisfiable formulas, certified by
proofs that independent checkers verify. Statements that rest on finite
simulations alone are observations about the simulated ranges. We mark them as
such, and no proof uses them. The theorem, the propositions and the corollary
of this section are marked computer-assisted: their proofs use computations,
whereas the results of \cref{sec:lines,sec:right,sec:left,sec:early} are
proved without computation.

\subsection{Method}\label{sec:method}

\paragraph{Encodings}
The formula $\MT_k(\mathcal N;M,N')$ (MT for minimal time) has one Boolean
variable for each possible value of each entry of a $k$-state rule table, of
each cell of $C_\infty$ with $t+i\le M$, and of each cell of $R_n$ with
$n\in\mathcal N$; clauses require that exactly one value is chosen for each
entry and each cell. Its clauses state the local rule, that no
half-line cell is $\Fq$, that the line of length $n$ is synchronized at time
$2n-2$ for every $n\in\mathcal N$, the return chains of \cref{lem:return} for
$n,n+1\in\mathcal N$, and $\PUMP(n,n')$ for $4\le n<n'\le N'$; the formula
$\MT_k(\mathcal N;M)$ has no pumping clauses. The formula $\G_k(T,N_1,C)$ is
encoded in the same way, with $M=2N_1$; its regions are closed under
predecessors, so that the table determines every encoded cell.
\Cref{app:encoding} gives the clauses and shows that the encoding is equivalent
to \cref{def:G}.

\paragraph{Certificates}
Every unsatisfiability claim of \cref{thm:four,prop:germs,prop:balzer} is
certified by a proof in the LRAT format~\cite{CruzFilipeEtAl17}, which is produced by the solver
CaDiCaL~\cite{CaDiCaL2} or, for one formula, obtained from a DRAT proof of the
solver Kissat~\cite{Kissat20} with \textsf{drat-trim}~\cite{WetzlerHeuleHunt14},
and verified by the independent checker \texttt{lrat-check}. Each of these
proofs is also verified by \texttt{cake\_lpr}~\cite{TanHeuleMyreen21}, a checker
that is formally verified in the theorem prover HOL4 and compiled with the
verified compiler CakeML. The eight
unsatisfiable formulas and their proofs are published with their SHA-256
digests (\cref{app:artefacts} and the data availability statement), so that
each proof can be checked against exactly its formula. The unsatisfiable
outcomes in the checks of the encodings below, where a rule table is fixed, are
predicted by simulating that rule; they are tests, not claims of their own.

\paragraph{Checks of the encodings}
A checked proof shows that a formula is unsatisfiable, but not that the formula
expresses the intended conditions; the programs that generate the formulas are
trusted. We tested them in five independent ways.
\begin{enumerate}[(1)]
\item \emph{Differential test.} A program written for this purpose from the
definitions of this paper, sharing no code with the generators, evaluates the
conditions directly by simulating the lines and the half-line: the
synchronization of the lines, the half-line up to an anti-diagonal,
$\PUMP(n,n')$ with the cone of \cref{lem:cone}, (G1)--(G5) and (B1)--(B4) of
\cref{sec:balzer}. With a rule table fixed by assumptions, each formula must be
satisfiable exactly when the table satisfies its conditions. We compared the two
on $37$ formulas (those of \cref{thm:four}, \cref{prop:sharp}~(b),
\cref{prop:germs,prop:balzer} and \cref{tab:validation}, and variants for four,
five and six states) and $3{,}684$ tables (the known rules, models of the
satisfiable formulas, the $27$ partial rules of \cref{prop:sharp}~(a) completed
at random, single and double mutations of these, and random tables): the
answers agree on all $62{,}286$ pairs, $1{,}002$ of which are satisfiable. The
formulas of \cref{thm:four,prop:balzer} in this test are identical to the
published files (\cref{app:artefacts}), and that of \cref{prop:germs} becomes
identical to its file when the selector clauses of its $230$ germs
(\cref{app:encoding}) are added. The
clauses of (B1)--(B4) alone agree with the direct check on $7{,}280$ tables, and
the cones found by search agree with \cref{lem:cone} for $2\le n<200$. No table
satisfies the formulas of \cref{thm:four}, so for them the test can only confirm
that violated conditions are detected; the same generators with other
parameters agree with the direct evaluation on satisfiable tables as well.
\item \emph{Known rules.} With the tables of Mazoyer's solution, as extracted
from Duprat's formal proof of its correctness~\cite{Duprat97}, and of
$\delta_{14}$ fixed, the formulas have the outcomes that simulation predicts
(\cref{tab:validation}).
\item \emph{Independent methods.} The formula of \cref{thm:four}~(i) and the
exhaustive enumeration of \cref{prop:sharp}~(a), a separate program, agree that
no four-state rule synchronizes the lines of lengths $2,\dots,9$. The germ
enumeration of \cref{prop:germs} and the refutations of its $14$ remaining
germs were each repeated by an independent program.
\item \emph{Models.} Every model found by a solver was re-simulated by a checker
that shares no code with the generators.
\item \emph{Second encoder.} A second program, written from the definitions of
this paper and from \cref{app:encoding} and sharing no code with the generators,
builds the eight unsatisfiable formulas again. It encodes each condition
literally: without the structural sharing of the generators, and with the cones
of the right ends found by a search over predecessors. It agrees with the direct
evaluation of~(1) on all $36{,}648$ pairs of a formula and a table, from $21$
of its formulas and $3{,}558$ tables; $643$ of these pairs are satisfiable. Its
eight formulas are unsatisfiable, with LRAT proofs verified by
\texttt{lrat-check} and \texttt{cake\_lpr}. Each
unsatisfiability claim thus rests on two formulas written by independent
programs.
\end{enumerate}
These tests make an error in the encodings unlikely, but they do not prove its
absence: a misreading of a definition shared by both encoders and the direct
evaluation would not be detected.

\begin{table}[t]
\centering\small
\caption{Known rules fixed in the formulas (\cref{sec:method}, check~(2)): the
outcome is predicted by simulating the rule. The formula for all $n\ge15$ uses
the representative lengths of \cref{cor:reduce}.}
\label{tab:validation}
%% Table: tests of the encodings (Section 6.1); the results are rerun by check_quick.sh.
\begin{tabular}{@{}l>{\raggedright\arraybackslash}p{0.33\textwidth}l>{\raggedright\arraybackslash}p{0.3\textwidth}@{}}
\toprule
Table fixed & Conditions & Result & Prediction from simulation\\
\midrule
Mazoyer's & $\MT_6(\{3,\dots,12\};78,40)$ & satisfiable & synchronizes every line of length $n\ge3$\\
Mazoyer's & (G2), (G3) with $T=8$ for all $n\ge15$ & satisfiable & the same; $21$ representative lengths, $(t_0,p)=(42,3)$\\
$\delta_{14}$ & lines of lengths $2,\dots,14$ & satisfiable & synchronizes them\\
$\delta_{14}$ & lines of lengths $2,\dots,15$ & unsatisfiable & the line of length $15$ fires at $\tau=8$\\
$\delta_{14}$ & lines $2,\dots,14$ and (G2) for $15,\dots,30$, $T=6$ & satisfiable & no line $15\le n\le30$ fires at $\tau\le6$\\
$\delta_{14}$ & lines $2,\dots,14$ and (G2) for $15,\dots,30$, $T=8$ & unsatisfiable & some line fires at $\tau\in\{7,8\}$\\
\bottomrule
\end{tabular}

\end{table}

\paragraph{Reproducibility}
Every computational statement can be repeated with the supplementary material,
which is publicly available (see the data availability statement). It contains
the programs, the rule tables, the germ lists, the eight unsatisfiable formulas
with their SHA-256 digests, the logs, and three scripts: one writes the
formulas with the generators and compares them with the recorded digests
(about one minute); one reruns the simulations, enumerations, satisfiable
formulas and the two differential tests (about $11$ minutes); and one
regenerates every LRAT proof and checks it with \texttt{lrat-check} (and with
\texttt{cake\_lpr} if it is installed), and then
refutes the formulas of the second encoder in the same way (about two hours and
$10$\,GB). Run again with CaDiCaL~3.0.1, Kissat~4.0.4 and \textsf{drat-trim},
this script reproduced all eight recorded proofs byte for byte (its log is in
the Zenodo record of the data availability statement), so regenerated proofs
can be compared with the recorded digests.
\Cref{app:artefacts} lists each computational claim with its programs and files.

\subsection{Four States}\label{sec:four}

\begin{theorem}[computer-assisted]\label{thm:four}
No four-state rule satisfying \eqref{eq:quiescence}
\begin{enumerate}[(i)]
\item synchronizes all lines of lengths $2,\dots,9$ in minimal time;
\item satisfies $\G_4(6,10)$: the lines of lengths $2,\dots,7$ complete and,
for the lines of lengths $8$, $9$ and $10$, the first seven rows after the
reflection, with the return chains on these rows;
\item satisfies $\G_4(4,10,10)$: the lines of lengths $2,\dots,5$ complete and,
for the lines of lengths $6,\dots,10$, the first five rows after the reflection,
with the return chains on these rows, and the right ends.
\end{enumerate}
In particular, no four-state minimal-time solution exists.
\end{theorem}

\begin{proof}
(i) Such a rule yields a model of $\MT_4(\{2,\dots,9\};16)$: a half-line cell
with $t+i\le16$ is quiescent if $i\ge10$ (\cref{lem:lightcone}) and equals a
cell of the line of length $9$ at a time $t<16$ if $i\le9$ (\cref{lem:agree}),
so it is in $W$; and the return chains hold by \cref{lem:return} for $2\le
n\le8$. This formula has $753$ variables and $17{,}605$ clauses. (ii) The
encoding of $\G_4(6,10)$ has $919$ variables and $22{,}960$ clauses. (iii) The
encoding of $\G_4(4,10,10)$ has $874$ variables and $20{,}461$ clauses. CaDiCaL
refutes each of the three formulas with an LRAT proof, and \texttt{lrat-check}
and \texttt{cake\_lpr} verify each proof. The last claim follows from (i), and also from (ii) or
(iii) with \cref{lem:G}.
\end{proof}

\Cref{fig:regions} shows the cells that the formulas of parts (ii) and (iii)
constrain, and \cref{tab:four} compares them with the weaker formulas of the
next proposition.

\begin{table}[t]
\centering\small
\caption{The four-state formulas of \cref{thm:four,prop:sharp}. The columns
(G1), (G2), (G3) and (G5) list the lengths $n$ to which each condition of
\cref{def:G} is applied and, for (G2) and (G3), the relative times $\tau\le T$
up to which they are imposed; every formula $\G_4$ also contains (G4). Every
unsatisfiability is certified by an LRAT proof that \texttt{lrat-check} and
\texttt{cake\_lpr} verify, and every satisfiable formula has a model that is re-simulated. The
labels (i)--(iii) refer to the parts of \cref{thm:four}, and (a), (b) to those
of \cref{prop:sharp}; the first row is the exhaustive enumeration of part (a),
which also finds no partial rule for the lengths $2,\dots,9$.}
\label{tab:four}
\setlength{\tabcolsep}{4.5pt}
%% Table: the four-state formulas of Theorem 6.1 and Proposition 6.2.
\begin{tabular}{@{}lllll@{}}
\toprule
Formula & (G1): & (G2), (G3): early rows & (G5): & Result\\
& complete lines & and return chains & right ends &\\
\midrule
(a) enumeration & $2,\dots,8$ & --- & --- & $27$ partial rules\\
(i) $\MT_4(\{2,\dots,9\};16)$ & $2,\dots,9$ & --- & --- & unsatisfiable\\
\midrule
(ii) $\G_4(6,10)$ & $2,\dots,7$ & $8,\dots,10$; $\tau\le6$ & --- & unsatisfiable\\
(b) $\G_4(5,30)$ & $2,\dots,6$ & $7,\dots,30$; $\tau\le5$ & --- & satisfiable\\
(b) $\G_4(6,9)$ & $2,\dots,7$ & $8,9$; $\tau\le6$ & --- & satisfiable\\
(b) $\G_4(6,10)$ without (G3) & $2,\dots,7$ & $8,\dots,10$; $\tau\le6$; (G2) only & --- & satisfiable\\
\midrule
(iii) $\G_4(4,10,10)$ & $2,\dots,5$ & $6,\dots,10$; $\tau\le4$ & $6,\dots,10$ & unsatisfiable\\
(b) $\G_4(4,30,9)$ & $2,\dots,5$ & $6,\dots,30$; $\tau\le4$ & $6,\dots,9$ & satisfiable\\
(b) $\G_4(3,30,12)$ & $2,\dots,4$ & $5,\dots,30$; $\tau\le3$ & $5,\dots,12$ & satisfiable\\
(b) (G1) and (G5) only & $2,\dots,5$ & --- & $6,\dots,12$ & satisfiable\\
\bottomrule
\end{tabular}

\end{table}

Part (i) certifies the bound observed by Yun\`es~\cite{Yunes93}; the earlier
proofs~\cite{Balzer67,Yunes93,Sanders94} rest on the correctness of search
programs. Parts (ii) and (iii) show how little information excludes four
states: in (iii), only the lines of lengths up to $5$ are complete. These
bounds are sharp in the following sense.

\begin{proposition}[computer-assisted]\label{prop:sharp}
\leavevmode
\begin{enumerate}[(a)]
\item Exactly $27$ partial four-state rules (rules restricted to the
neighbourhoods they use) synchronize all lines of lengths $2,\dots,8$ in
minimal time, and no partial four-state rule synchronizes all lines of lengths
$2,\dots,9$.
\item The formulas $\G_4(5,30)$, $\G_4(6,9)$, $\G_4(4,30,9)$ and
$\G_4(3,30,12)$, the formula $\G_4(6,10)$ without (G3), and the formula that
consists of (G1) for $T=4$ and (G5) for $6\le n\le12$ are satisfiable.
\end{enumerate}
\end{proposition}

\begin{proof}
(a) is an exhaustive enumeration. Yun\`es~\cite{Yunes93} found the same number
under the extra condition $\delta(\ast,\Lq,\Lq)=\Lq$, which is vacuous here
because none of the $27$ partial rules uses the neighbourhood $(\ast,\Lq,\Lq)$.
(b) The solver finds a model of each formula, and each model is re-simulated.
\end{proof}

In words (\cref{tab:four}): the formula of part (ii) of \cref{thm:four}
becomes satisfiable if one row is removed, even when the lines up to length $30$ are added
($\G_4(5,30)$), if the line of length $10$ is removed ($\G_4(6,9)$), or if the
return chains are removed. The formula of part (iii) becomes satisfiable if
the right end of the line of length $10$ is removed, even when the early rows
of the lines up to length $30$ are added ($\G_4(4,30,9)$), or if one row is
removed, even when the right ends of the lines up to length $12$ are added
($\G_4(3,30,12)$). The right ends alone, without the early rows and the return
chains, do not suffice either.

\subsection{Five States: Rules for the Lengths Up to Fourteen}\label{sec:five}

The following five-state rules were found with satisfiability solvers. How they
were found plays no role in what follows: every statement about them is checked
on their tables, the files \texttt{uniformA\_2-12.txt} (the rule $\delta_{12}$),
\texttt{delta13.txt} and \texttt{delta14*.txt} in \texttt{results/} of the
supplementary material and in the Zenodo record. \Cref{tab:delta14} shows $\delta_{14}$, and
\cref{tab:rules} summarizes what we know about the seven rules.

\begin{table}[t]
\centering\small
\caption{The five-state rule $\delta_{14}$. The entry in row $(x,y)$ and
column $z$ is $\delta_{14}(x,y,z)$; the neighbourhoods $(\ast,y,\ast)$ are not
used (``--'').}
\label{tab:delta14}
\setlength{\tabcolsep}{4.2pt}
\begin{tabular}{@{}cc|ccccc@{\hspace{2.2em}}cc|ccccc@{}}
\toprule
$x$ & $y$ & $z=\Lq$ & $\Gq$ & $\Aq$ & $\Bq$ & $\ast$ &
$x$ & $y$ & $z=\Lq$ & $\Gq$ & $\Aq$ & $\Bq$ & $\ast$\\
\midrule
%% generated by figures.py from results/delta14.txt; do not edit
$\ast$ & $\mathsf L$ & $\mathsf F$ & $\mathsf F$ & $\mathsf L$ & $\mathsf G$ & -- & $\ast$ & $\mathsf A$ & $\mathsf G$ & $\mathsf G$ & $\mathsf L$ & $\mathsf G$ & --\\
$\mathsf L$ & $\mathsf L$ & $\mathsf L$ & $\mathsf G$ & $\mathsf G$ & $\mathsf B$ & $\mathsf L$ & $\mathsf L$ & $\mathsf A$ & $\mathsf B$ & $\mathsf G$ & $\mathsf B$ & $\mathsf A$ & $\mathsf A$\\
$\mathsf G$ & $\mathsf L$ & $\mathsf A$ & $\mathsf B$ & $\mathsf A$ & $\mathsf B$ & $\mathsf G$ & $\mathsf G$ & $\mathsf A$ & $\mathsf A$ & $\mathsf G$ & $\mathsf L$ & $\mathsf A$ & $\mathsf B$\\
$\mathsf A$ & $\mathsf L$ & $\mathsf B$ & $\mathsf G$ & $\mathsf G$ & $\mathsf G$ & $\mathsf G$ & $\mathsf A$ & $\mathsf A$ & $\mathsf L$ & $\mathsf A$ & $\mathsf G$ & $\mathsf A$ & $\mathsf B$\\
$\mathsf B$ & $\mathsf L$ & $\mathsf B$ & $\mathsf A$ & $\mathsf A$ & $\mathsf L$ & $\mathsf A$ & $\mathsf B$ & $\mathsf A$ & $\mathsf A$ & $\mathsf A$ & $\mathsf G$ & $\mathsf A$ & $\mathsf L$\\
\midrule
$\ast$ & $\mathsf G$ & $\mathsf L$ & $\mathsf F$ & $\mathsf L$ & $\mathsf L$ & -- & $\ast$ & $\mathsf B$ & $\mathsf G$ & $\mathsf G$ & $\mathsf G$ & $\mathsf G$ & --\\
$\mathsf L$ & $\mathsf G$ & $\mathsf A$ & $\mathsf F$ & $\mathsf B$ & $\mathsf A$ & $\mathsf F$ & $\mathsf L$ & $\mathsf B$ & $\mathsf B$ & $\mathsf G$ & $\mathsf A$ & $\mathsf L$ & $\mathsf G$\\
$\mathsf G$ & $\mathsf G$ & $\mathsf B$ & $\mathsf F$ & $\mathsf G$ & $\mathsf F$ & $\mathsf B$ & $\mathsf G$ & $\mathsf B$ & $\mathsf B$ & $\mathsf A$ & $\mathsf B$ & $\mathsf B$ & $\mathsf F$\\
$\mathsf A$ & $\mathsf G$ & $\mathsf A$ & $\mathsf G$ & $\mathsf G$ & $\mathsf A$ & $\mathsf B$ & $\mathsf A$ & $\mathsf B$ & $\mathsf B$ & $\mathsf B$ & $\mathsf G$ & $\mathsf B$ & $\mathsf G$\\
$\mathsf B$ & $\mathsf G$ & $\mathsf A$ & $\mathsf G$ & $\mathsf A$ & $\mathsf L$ & $\mathsf B$ & $\mathsf B$ & $\mathsf B$ & $\mathsf L$ & $\mathsf A$ & $\mathsf A$ & $\mathsf L$ & $\mathsf A$\\
\bottomrule

\end{tabular}
\end{table}

\begin{proposition}[computer-assisted]\label{prop:nearmiss}
Each of the five-state rules $\delta_{14}$, $\delta_{14}^{\mathrm b}$,
$\delta_{14}^{\mathrm c}$, $\delta_{14}^{\mathrm d}$ and $\delta_{14}^{\mathrm
e}$ synchronizes the line of length $n$ at time $2n-2$ for $2\le n\le14$, and
for no $n\ge15$. More precisely, for every $n\ge15$, some cell of the line of
length $n$ is $\Fq$ at a relative time $\tau\le\min(T,n-2)$, where
$T=10,14,22,12,14$, respectively.
\end{proposition}

\begin{figure}[t]
\centering
%% Figure: the relative time of the first F on the lines of lengths 15..60 of the five rules of
%% Proposition 6.3 (data from figures.py), with the bound min(T, n-2) of the proposition.
\begin{tikzpicture}
  \begin{groupplot}[group style={group size=5 by 1,horizontal sep=0.32cm,y descriptions at=edge left},
      scale only axis,width=2.42cm,height=2.6cm,xmin=13,xmax=62,ymin=0,ymax=25,
      xtick={15,30,45,60},ytick={0,5,10,15,20,25},tick label style={font=\scriptsize},
      label style={font=\small},title style={font=\small,yshift=-0.8ex},axis lines=left,tick align=outside,
      xlabel={length $n$}]
    \def\panel#1#2#3{%
      \nextgroupplot[title={$#2$, $T=#3$}]
      \addplot[gEarly,line width=2.4pt,domain=15:60,samples=46] {min(#3,x-2)};
      \addplot[only marks,mark=*,mark size=0.95pt,stC] table[x=n,y=tau] {figures/firstfire_#1.dat};}
    \panel{delta14}{\delta_{14}}{10}
    \panel{delta14b}{\delta_{14}^{\mathrm b}}{14}
    \panel{delta14c}{\delta_{14}^{\mathrm c}}{22}
    \panel{delta14d}{\delta_{14}^{\mathrm d}}{12}
    \panel{delta14e}{\delta_{14}^{\mathrm e}}{14}
  \end{groupplot}
  \node[font=\small,rotate=90,anchor=south] at ([xshift=-0.75cm]group c1r1.west) {$\tau$ of the first $\Fq$};
\end{tikzpicture}
\caption{The relative time $\tau=t-(n-1)$ of the first cell in state $\Fq$ on
the line of length $n$, $15\le n\le60$, for the five rules of
\cref{prop:nearmiss} (dots, from direct simulation). The thick line is the
bound $\min(T,n-2)$ of the proposition. Every dot lies on or below it, so each
of these lines fires before time $2n-2$, at a relative time $\tau\le T$. For
$n>60$ the proposition rests on the certificates of \cref{tab:rules} and on
\cref{cor:reduce}, not on this figure.}
\label{fig:firstfire}
\end{figure}
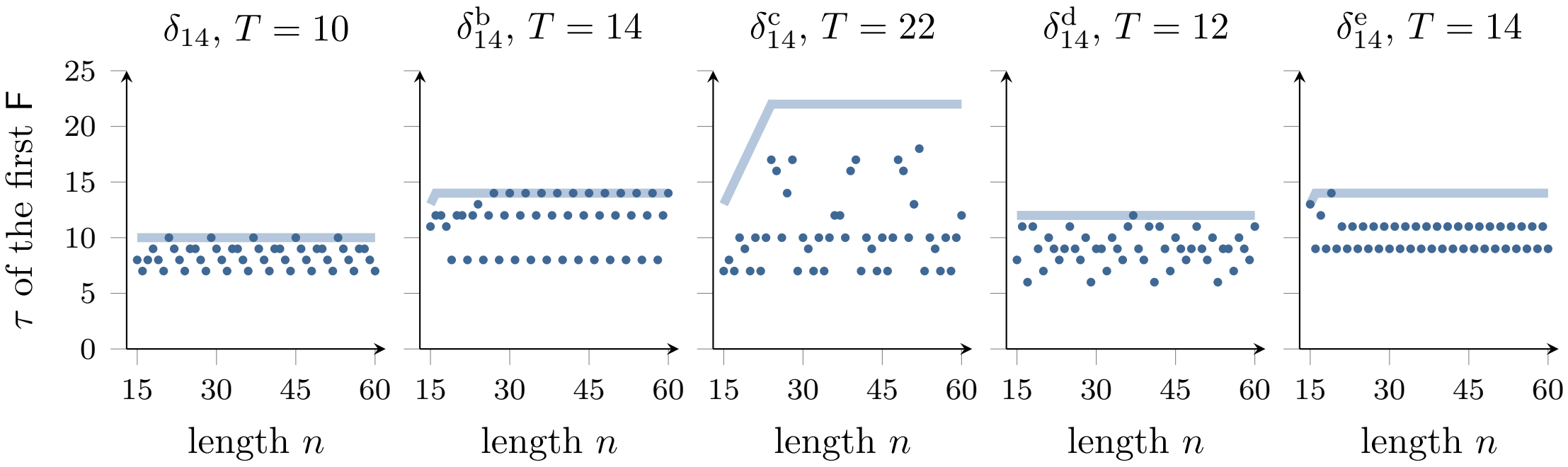

\begin{proof}
The lines of lengths $2,\dots,60$ are simulated. For $15\le n\le60$ some cell
is $\Fq$ at a relative time $\tau\le n-2$, which proves the claim for $15\le
n<T+3$. The half-line has no cell in state $\Fq$ at the times $t<3000$, so by
\cref{lem:agree} a cell of a line of length $n\le1500$ that is $\Fq$ at a time
$t\le 2n-2$ lies in $R_n$. Simulating the half-line gives $t_0$ and $p$ that
satisfy the hypothesis of \cref{lem:certificate} for $D=2T+4\ge2T+1$, namely
$(t_0,p)=(51,8)$, $(66,3)$, $(257,192)$, $(95,48)$ and $(56,2)$. For every $n$
with $\max(15,T+3)\le n<t_0+3+p\le452$, direct simulation shows that the line
of length $n$, or of a length $n'$ in this range with $J_{n'}=J_n$ on
$[0,T+1]$, has a cell in state $\Fq$ at a relative time $\tau\le T$, hence in
$E(T)$. (One length is simulated for each class of a partition that is finer
than equality of $J_n$ on $[0,T+1]$; together with the lengths below $T+3$,
these are $35$, $39$, $413$, $116$ and $28$ lengths, the largest being $49$,
$53$, $427$, $130$ and $42$.) By \cref{lem:early,cor:reduce} with
$N_0=\max(15,T+3)$, every $R_n$ with $n\ge N_0$ has a cell of $E(T)$ in state
$\Fq$, and there $\tau\le T\le n-3$ (\cref{fig:firstfire,tab:rules}).
\end{proof}

\begin{figure}[t]
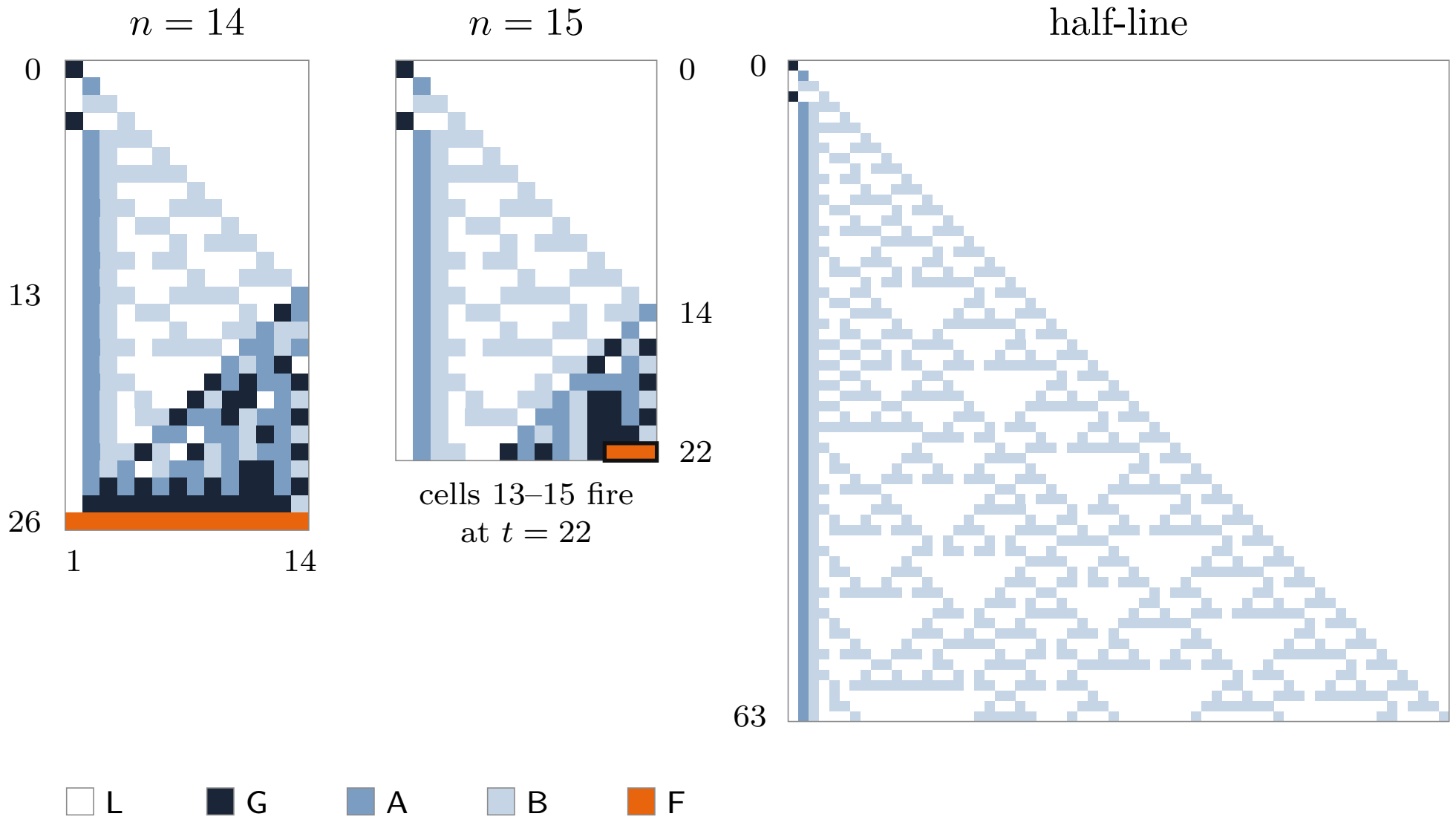

\centering
%% Figure: delta14 on the lines of lengths 14 and 15, and its half-line (cells from figures.py).
\begin{tikzpicture}
  % the line of length 14, up to its firing time 26
  \begin{scope}[x=1.55mm,y=1.55mm]
    \input{figures/cells_delta14_line14}
    \draw[grid,line width=0.3pt] (0,0) rectangle (14,-27);
    \node[font=\small,anchor=south] at (7,0.6) {$n=14$};
    \foreach \y/\l in {0/0,13/13,26/26}
      \node[font=\scriptsize,anchor=east] at (-0.5,-\y-0.5) {$\l$};
    \foreach \x/\l in {1/1,14/14}
      \node[font=\scriptsize,anchor=north] at (\x-0.5,-27.3) {$\l$};
  \end{scope}
  % the line of length 15, up to its first firing cell (time 22)
  \begin{scope}[x=1.55mm,y=1.55mm,shift={(19,0)}]
    \input{figures/cells_delta14_line15}
    \draw[grid,line width=0.3pt] (0,0) rectangle (15,-23);
    \draw[ink,line width=1.0pt] (12,-22) rectangle (15,-23);
    \node[font=\small,anchor=south] at (7.5,0.6) {$n=15$};
    \node[font=\scriptsize,anchor=north,align=center] at (7.5,-23.4) {cells $13$--$15$ fire\\ at $t=22$};
    \foreach \y/\l in {0/0,14/14,22/22}
      \node[font=\scriptsize,anchor=west] at (15.4,-\y-0.5) {$\l$};
  \end{scope}
  % the half-line, 0 <= t <= 63
  \begin{scope}[x=0.92mm,y=0.92mm,shift={(70,0)}]
    \input{figures/cells_delta14_half63}
    \draw[grid,line width=0.3pt] (0,0) rectangle (64,-64);
    \node[font=\small,anchor=south] at (32,1) {half-line};
    \foreach \y/\l in {0/0,63/63}
      \node[font=\scriptsize,anchor=east] at (-0.6,-\y-0.5) {$\l$};
  \end{scope}
  % legend
  \begin{scope}[shift={(0,-6.6)},font=\scriptsize]
    \foreach \c/\l [count=\k from 0] in {white/$\Lq$,stG/$\Gq$,stA/$\Aq$,stB/$\Bq$,stF/$\Fq$}
      \node[anchor=west,inner sep=0pt] at (\k*1.25,0) {\swatch{\c}\ \l};
  \end{scope}
\end{tikzpicture}
\caption{The rule $\delta_{14}$. Left: the line of length $14$ is synchronized
at time $26$. Middle: on the line of length $15$, the cells $13$, $14$ and $15$
fire at time $22$, that is, at the relative time $\tau=8$, six steps early.
Right: the half-line for $0\le t\le63$. In this range, and in our simulations
up to $t<1500$, the cells $i\ge3$ take only the values $\Lq$ and $\Bq$ from time
$3$ on, and there $\delta_{14}$ acts as the elementary cellular automaton $22$.}
\label{fig:delta14}
\end{figure}

\begin{table}[t]
\centering\small
\caption{The seven five-state rules of \cref{sec:five}, with their files in
\texttt{results/} of the supplementary material. Synchronized: the lengths $n$
whose line is synchronized at time $2n-2$, among $2\le n\le60$ for
$\delta_{12}$ and $\delta_{13}$ and among all $n\ge2$ for the five rules of
\cref{prop:nearmiss}. For these five rules: the bound $T$ of the proposition,
the certificate $(t_0,p)$ of \cref{lem:certificate} for $D=2T+4$, and the
number of lengths simulated in the proof, with the largest of them. $\Fq$ at
$\tau\le9$: the least length whose line has a cell in state $\Fq$ at a
relative time $\tau\le9$. Last column: the complexity $c_{78}$ of
\cref{sec:germs}.}
\label{tab:rules}
\setlength{\tabcolsep}{4.5pt}
%% Table: the seven five-state rules of Section 6.3 (rerun by check_quick.sh).
\begin{tabular}{@{}lccccccc@{}}
\toprule
Rule & File & Synchronized & $T$ & $(t_0,p)$ & Simulated lengths & $\Fq$ at $\tau\le9$ & $c_{78}$\\
\midrule
$\delta_{12}$ & \texttt{uniformA\_2-12} & $2,\dots,12$ & & & & $n=16$ & $22$\\
$\delta_{13}$ & \texttt{delta13} & $2,\dots,13$ & & & & $n=14$ & $22$\\
$\delta_{14}$ & \texttt{delta14} & $2,\dots,14$ & $10$ & $(51,8)$ & $35$, up to $49$ & $n=15$ & $16$\\
$\delta_{14}^{\mathrm b}$ & \texttt{delta14b} & $2,\dots,14$ & $14$ & $(66,3)$ & $39$, up to $53$ & $n=19$ & $16$\\
$\delta_{14}^{\mathrm c}$ & \texttt{delta14c} & $2,\dots,14$ & $22$ & $(257,192)$ & $413$, up to $427$ & $n=15$ & $15$\\
$\delta_{14}^{\mathrm d}$ & \texttt{delta14d} & $2,\dots,14$ & $12$ & $(95,48)$ & $116$, up to $130$ & $n=15$ & $15$\\
$\delta_{14}^{\mathrm e}$ & \texttt{delta14e} & $2,\dots,14$ & $14$ & $(56,2)$ & $28$, up to $42$ & $n=16$ & $17$\\
\bottomrule
\end{tabular}

\end{table}

For example, $\delta_{14}$ synchronizes the line of length $14$ at time $26$,
whereas on the line of length $15$ the cells $13$, $14$ and $15$ fire at time
$22$, eight steps after the front has reached the right end and six steps too
early (\cref{fig:delta14}). The following are observations from simulations,
valid in the stated ranges. For $3\le t<1500$ the half-line of $\delta_{14}$
takes the values $\Lq,\Gq$ at cell $1$, $\Lq,\Aq$ at cell $2$ and $\Lq,\Bq$ at
the cells $i\ge3$, and $\delta_{14}$ restricted to $\{\Lq,\Bq\}^3$ is the
elementary rule $22$~\cite{Wolfram83}. The finite conditions behind the two
barriers do not exclude this half-line in the ranges we checked: $\PUMP(n,n')$ holds for $4\le n<n'\le800$, and
\cref{lem:left} does not apply for $n\le5000$, $s\le3$ and $\ell\le24$. The
following are exact finite computations. The rules $\delta_{12}$ and
$\delta_{13}$ synchronize the lines of lengths $2,\dots,12$ and $2,\dots,13$,
respectively, and no other line of length at most $60$. Each of the seven rules
has a line of some length $14\le n\le19$ with a cell in state $\Fq$ at a
relative time $\tau\le9$, so none of them satisfies $\G_5(9,N_1)$ for
$N_1\ge19$.

\paragraph{What is new}
Synchronizing finitely many lengths is not difficult in itself: for every $k$
there is a largest $N_k\in\{1,2,\dots\}\cup\{\infty\}$ such that some $k$-state
rule synchronizes the lines of lengths $2,\dots,N_k$ in minimal time, and since
there are finitely many rules, $N_k=\infty$ exactly when a $k$-state
minimal-time solution exists. The question is how far $N_k$ reaches.
Exhaustive searches give $N_4=8$ (Yun\`es~\cite{Yunes93};
\cref{prop:sharp}~(a)), and six-state minimal-time solutions
exist~\cite{Mazoyer87}. \Cref{prop:nearmiss} gives $N_5\ge14$. We are not aware
of an earlier published bound for five states: the constructions with at most
five states that we know of synchronize restricted families of lengths, some of them in non-minimal
time~\cite{UmeoYanagihara07,Yunes08}. The rules add three things to the method
of this paper. First, a proof that five states do not suffice must use lines of
length at least $15$, or conditions other than the synchronization of the lines
of lengths $2,\dots,14$. Second, the statement that each rule fails for every
$n\ge15$ is not a finite simulation: it combines the simulation of finitely
many lengths with \cref{cor:reduce}. Third, the rules fail in the first rows
after the reflection, which is what the formulas of \cref{sec:early} target.
They are not candidates for minimal-time solutions.

\subsection{Five States: Half-Lines of Low Complexity}\label{sec:germs}

The barriers constrain the half-line, so it is natural to ask how simple the
half-line of a five-state minimal-time solution can be. We measure this by the
number of distinct neighbourhoods that it uses in a fixed region. The
\emph{germ} below $M$ of a rule is its restriction to the neighbourhoods of the
cells $(t,i)$ of its half-line with $t\ge1$ and $t+i\le M$, and its
\emph{complexity} $c_M(\delta)$ is the number of these neighbourhoods other than
$(\Lq,\Lq,\Lq)$. We use $M=78$: the half-line below the anti-diagonal $t+i=78$
contains the input anti-diagonals of all lines of lengths $n\le40$, so the germ
below $78$ determines the conditions $\PUMP(n,n')$ for $n'\le40$. For example,
$c_{78}=57$ for Mazoyer's solution, $22$ for $\delta_{12}$ and $16$ for
$\delta_{14}$ (\cref{tab:rules}).

\begin{proposition}[computer-assisted]\label{prop:germs}
Every five-state minimal-time solution $\delta$ satisfies $c_{78}(\delta)\ge15$.
\end{proposition}

\begin{proof}
Suppose that $c_{78}(\delta)\le14$. Exchanging $\Aq$ and $\Bq$ if necessary, we
may assume that $\Aq$ occurs before $\Bq$ (if at all) in the order of
anti-diagonals and, within an anti-diagonal, of time. By \eqref{eq:neverfire}
and \cref{lem:return,lem:pumping}, the germ of $\delta$ below $78$ produces no
$\Fq$, never produces $\Lq$ at the front, and satisfies $\PUMP(n,n')$ for $4\le
n<n'\le40$. A depth-first search that chooses the value of each neighbourhood
at its first occurrence enumerates all such germs, visiting about $10^8$ nodes:
there are $244$, and an independent re-implementation finds the same $244$.
For $230$ of them, the formula $\MT_5(\{2,\dots,10\};78,39)$ with the germ fixed
is unsatisfiable. A single formula covers all $230$: it has a selector variable
for each germ, clauses that make each selector fix its germ, and a clause that
requires some selector; CaDiCaL refutes it with an LRAT proof that
\texttt{lrat-check} and \texttt{cake\_lpr} verify. Each of the other $14$ germs determines
$C_\infty$ up to the anti-diagonal $3300$, and there \cref{lem:left} applies
with $s=1$: for some $n\le163$ and $\ell\le5$, the upper input anti-diagonal of
the line of length $n$ is $\ell$-periodic on $[1,Y]$ and the lower one on
$[1,Y-1]$ with $Y-1>16\ell$ (for example $n=28$, $\ell=1$ and $Y=19$,
\cref{fig:germs}(b)), as an independent simulation confirms. Hence no rule
that contains one of the $244$ germs is a minimal-time solution
(\cref{fig:germs}(a)).
\end{proof}

\begin{figure}[t]
\centering
%% Figure: the proof of Proposition 6.4 (diagram) and its example of Lemma 4.1: the half-line of germ 10
%% of results/germs/germs_K14_surv10.txt for 0 <= t <= 55 (cells from figures.py).
\begin{tikzpicture}
  % (a) the case analysis
  \begin{scope}[font=\scriptsize,every node/.style={align=flush center,inner sep=3pt,rounded corners=1.5pt}]
    \node[draw=ink,fill=gNone,text width=5.6cm] (all) at (0,0) {germs below $78$ of five-state rules with
      $c_{78}\le14$ that produce no $\Fq$, no $\Lq$ at the front, and satisfy $\PUMP(n,n')$ for
      $4\le n<n'\le40$};
    \node[draw=ink,fill=rgAgree,text width=5.6cm] (g244) at (0,-1.75) {$244$ germs, up to exchanging $\Aq$ and $\Bq$
      (depth-first search with about $10^8$ nodes;\\ an independent program finds the same $244$)};
    \node[draw=ink,fill=gEarly,text width=2.62cm] (g230) at (-1.5,-4.35) {$230$ germs: with the germ fixed,
      the formula $\MT_5(\{2,\dots,10\};$ $78,39)$ is unsatisfiable (one LRAT proof)};
    \node[draw=ink,fill=gEarly,text width=2.62cm] (g14) at (1.5,-4.35) {$14$ germs: \cref{lem:left} with $s=1$
      applies for some $n\le163$ and $\ell\le5$; see (b)};
    \node[draw=ink,fill=gNone,text width=5.6cm] (none) at (0,-6.75) {no five-state minimal-time solution has
      $c_{78}\le14$};
    \draw[-{Stealth[length=1.6mm]},ink,line width=0.5pt] (all) -- (g244);
    \draw[-{Stealth[length=1.6mm]},ink,line width=0.5pt] (g244.south) -- ++(0,-0.2) -| (g230.north);
    \draw[-{Stealth[length=1.6mm]},ink,line width=0.5pt] (g244.south) -- ++(0,-0.2) -| (g14.north);
    \draw[-{Stealth[length=1.6mm]},ink,line width=0.5pt] (g230.south) -- ++(0,-0.2) -| ([xshift=-0.4cm]none.north);
    \draw[-{Stealth[length=1.6mm]},ink,line width=0.5pt] (g14.south) -- ++(0,-0.2) -| ([xshift=0.4cm]none.north);
    \node[font=\small,anchor=south west,align=left,rounded corners=0pt] at (-3.1,0.85) {(a) the proof of \cref{prop:germs}};
  \end{scope}
  % (b) germ 10 of the 14: on t+i = 54 the cells 1..19 are G, on t+i = 53 the cells 1..18 are A
  \begin{scope}[shift={(4.1cm,0.95cm)},x=1.07mm,y=1.07mm]
    \input{figures/cells_germ10_half}
    \draw[grid,line width=0.3pt] (0,0) -- (0,-56) -- (56,-56);
    \foreach \y in {1,...,19} \draw[white,line width=0.9pt] (\y-1,-54+\y) rectangle (\y,-55+\y);
    \foreach \y in {1,...,18} \draw[white,line width=0.9pt] (\y-1,-53+\y) rectangle (\y,-54+\y);
    \foreach \y in {1,...,19} \draw[stF,line width=0.45pt] (\y-1,-54+\y) rectangle (\y,-55+\y);
    \foreach \y in {1,...,18} \draw[stF,line width=0.45pt] (\y-1,-53+\y) rectangle (\y,-54+\y);
    \foreach \t in {0,27,54} \node[font=\scriptsize,anchor=east] at (-0.5,-\t-0.5) {$\t$};
    \node[font=\scriptsize,anchor=east] at (-3.2,-40.5) {$t$};
    \foreach \i in {1,19,56} \node[font=\scriptsize,anchor=north] at (\i-0.5,-56.3) {$\i$};
    \draw[-{Stealth[length=1.4mm]},ink,line width=0.45pt] (27.2,-21.6) -- (15.2,-38.6);
    \node[font=\scriptsize,anchor=west,align=left] at (26.6,-20.6) {$t+i=54$, cells $1,\dots,19$: $\Gq$\\
      $t+i=53$, cells $1,\dots,18$: $\Aq$};
    \node[font=\small,anchor=south west] at (-3,1.4) {(b) germ $10$ of the $14$};
  \end{scope}
  % legend and the numbers of Lemma 4.1 in (b)
  \node[anchor=north west,font=\scriptsize,align=left] at (4.1cm,-5.55cm) {in (b):\ \swatch{white}\,$\Lq$\
    \swatch{stG}\,$\Gq$\ \swatch{stA}\,$\Aq$\ \swatch{stB}\,$\Bq$\\[0.4ex]
    $n=28$, $s=1$, $\ell=1$, $Y=19$: $Y-s=18>(k-1)^{2s}\ell=16$};
\end{tikzpicture}
\caption{(a)~The case analysis in the proof of \cref{prop:germs}. (b)~One of
the $14$ germs that the formula does not refute: the half-line of germ number
$10$ of the file \texttt{germs\_K14\_surv10.txt} for $0\le t\le55$. For $n=28$,
the upper input anti-diagonal $t+i=2n-2=54$ is in state $\Gq$ on the cells
$1,\dots,19$, and the lower one, $t+i=53$, is in state $\Aq$ on the cells
$1,\dots,18$ (outlined). So the hypothesis of \cref{lem:left} holds with $k=5$,
$s=1$, $\ell=1$ and $Y=19$, and $Y-s=18>16=(k-1)^{2s}\ell$: no rule that
contains this germ synchronizes the line of length $28$ in minimal time.}
\label{fig:germs}
\end{figure}

Complexity $15$ does occur for rules that synchronize many lines: the germs
below $78$ of $\delta_{14}^{\mathrm c}$ and $\delta_{14}^{\mathrm d}$ have
complexity $15$, and both rules synchronize the lines of lengths $2,\dots,14$
(\cref{prop:nearmiss}).

The rules $\delta_{12}$ and $\delta_{13}$ have the same half-line on the cells
with $t+i\le78$, where they use the $22$ neighbourhoods
\begin{center}\small
\texttt{*GL>A GLL>G *AG>B AGL>L *BL>A BLG>L LGL>B *AL>G ALB>A LBG>L BGL>G}\\
\texttt{*GA>A GAL>L ALG>B LGG>G GGL>G BGG>L GGG>G ALL>A LLG>L ALA>A LAL>L}
\end{center}
besides $(\Lq,\Lq,\Lq)$; here \texttt{xyz>d} means $\delta(x,y,z)=d$. On these
cells the half-line has the pattern described before \cref{cor:regular}
(\cref{fig:halflines}). If this pattern continued forever, \cref{thm:left}
would exclude the half-line; \cref{lem:left} gives an explicit length without
that assumption.

\begin{corollary}[computer-assisted]\label{cor:germ12}
No five-state rule that has the $22$ transitions above synchronizes the line of
length $518$ in minimal time.
\end{corollary}

\begin{figure}[t]
\centering
\input{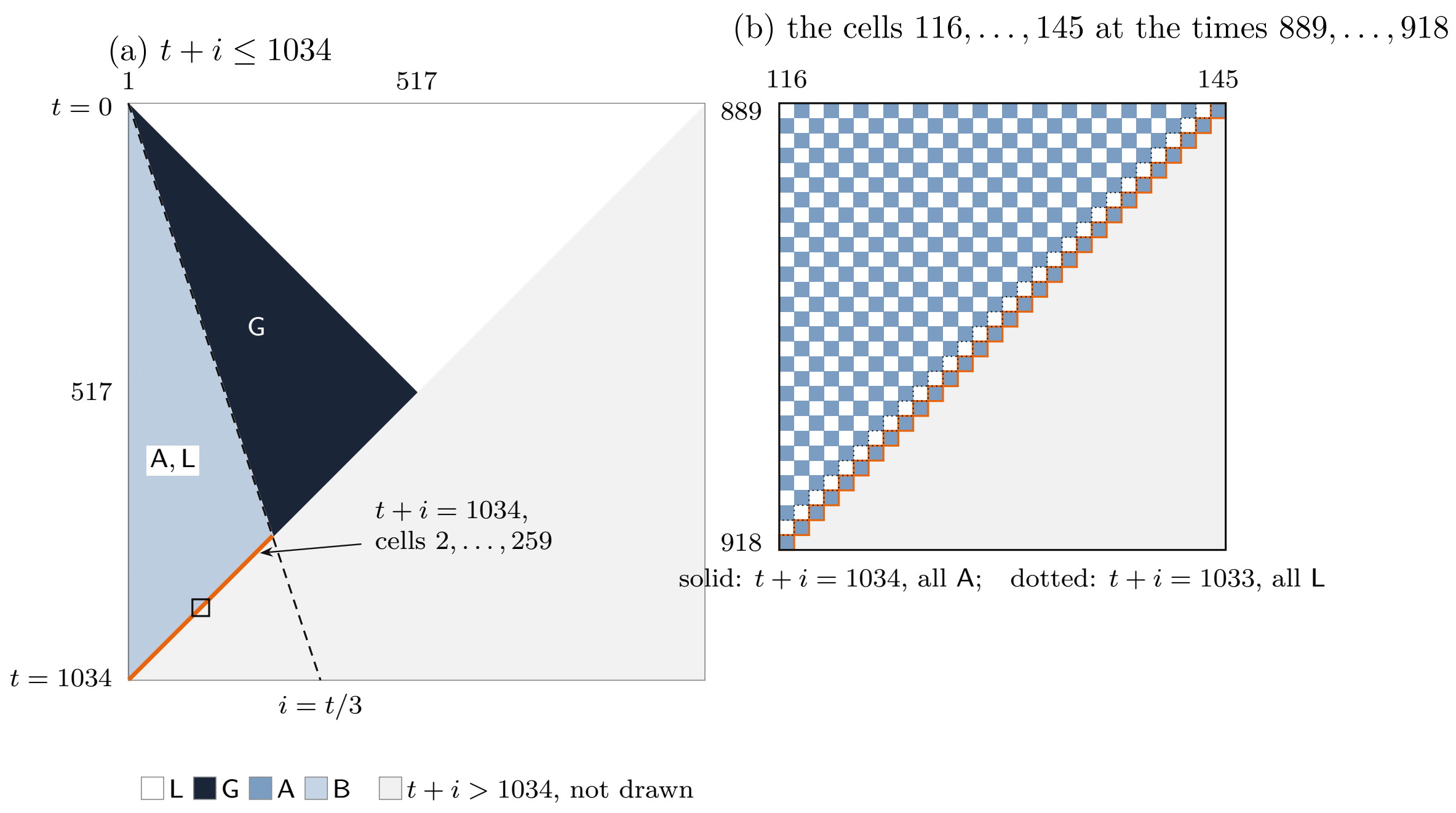}
\caption{The half-line of $\delta_{12}$ on the cells with $t+i\le1034$, which
every rule with the $22$ transitions of \cref{cor:germ12} shares. (a)~All
these cells, one pixel per cell: the region in state $\Gq$ (dark) lies between
roughly the line $i=t/3$ (dashed) and the front, and the cells between the left
border and roughly the line $i=t/3$ alternate between $\Aq$ and $\Lq$. The thick
segment marks the cells $2,\dots,259$ of the anti-diagonal $t+i=1034$.
(b)~The cells in the small square of (a). On $t+i=1034$ (solid outlines) they
are all in state $\Aq$, and on $t+i=1033$ (dotted outlines) all in state $\Lq$.
This holds for the cells $2,\dots,259$ and $2,\dots,258$, respectively, so
\cref{lem:left} applies to the line of length $518$ with $s=2$, $\ell=1$ and
$Y=259$.}
\label{fig:delta12}
\end{figure}

\begin{proof}
With these transitions and $\delta(\Lq,\Lq,\Lq)=\Lq$, no other neighbourhood
occurs on the cells with $t+i\le1034$, so every such rule has the same diagram
$C_\infty$ there. On the anti-diagonal $t+i=1034$ the cells $2,\dots,259$ are in
state $\Aq$, and on $t+i=1033$ the cells $2,\dots,258$ are in state $\Lq$.
\Cref{lem:left} with $k=5$, $n=518$, $s=2$, $\ell=1$ and $Y=259$ would give
$257\le4^4=256$ (\cref{fig:delta12}).
\end{proof}

\subsection{Balzer's Conditions}\label{sec:balzer}

Balzer~\cite{Balzer67} reported that no five-state minimal-time solution
satisfies four extra conditions. In Mazoyer's account~\cite[\S7]{Mazoyer86L},
they are the stability of~$\Gq$, the condition of an image solution, that only
the neighbourhoods $(\Gq,\Gq,\Gq)$, $(\ast,\Gq,\Gq)$ and $(\Gq,\Gq,\ast)$ give
$\Fq$, and the transitions $(\Gq,v,\Gq)\mapsto\Gq$ for $v\ne\Gq$. We state them
as follows:
\begin{enumerate}[(B1)]
\item $\delta(x,\Gq,y)=\Gq$ unless $x,y\in\{\Gq,\ast\}$;
\item $\delta(I(z),I(y),I(x))=I(\delta(x,y,z))$ for every neighbourhood
$(x,y,z)$, including those with a border, where $I$ is an involution of the
states, extended by $I(\ast)=\ast$;
\item $\delta(\Gq,\Gq,\Gq)=\delta(\ast,\Gq,\Gq)=\delta(\Gq,\Gq,\ast)=\Fq$, and no
other neighbourhood gives $\Fq$;
\item $\delta(\Gq,v,\Gq)=\Gq$ for all $v\in W\setminus\{\Gq\}$.
\end{enumerate}
We call this the \emph{strong reading}. In the \emph{weak reading}, (B1) is
required only for working states $x,y$, and (B3) without its last clause.
Because (B2) is also required at the borders, the image of a computation is a
computation: if $c$ is a configuration of the line of length $n$ in working
states and $c'(i)=I(c(n+1-i))$ for $1\le i\le n$, then the successor of $c'$ is
the image of the successor of~$c$.
Yun\`es~\cite[\S2.7]{Yunes93} states (B1) and (B3) in the weak reading, but
(B2) only for the neighbourhoods without a border, and he admits states other
than~$\Gq$ in (B1), (B3) and (B4); we keep~$\Gq$, as in Mazoyer's account, and
return to the borders after \cref{prop:balzer}.

In both readings, $I$ fixes $\Fq$ and $\Gq$. Since $\delta$ is defined only on
neighbourhoods, which contain no~$\Fq$, (B2) can hold only if $I$ maps $W$ to
itself; hence $I(\Fq)=\Fq$. Then $I(\Gq)=\Gq$: otherwise
$g:=I(\Gq)\in W\setminus\{\Gq\}$, and (B2) and (B4) give
$\delta(g,\Gq,g)=I(\delta(\Gq,g,\Gq))=I(\Gq)=g$, whereas (B1) gives
$\delta(g,\Gq,g)=\Gq$. The conclusion is the same if, as in~\cite{Yunes93},
$\delta$ is defined on all triples of states and (B1), (B2) and (B4) are
required for all of them: the same argument excludes $g=\Fq$, and then (B2) and
(B3) give
$I(\Fq)=I(\delta(\Gq,\Gq,\Gq))=\delta(I(\Gq),I(\Gq),I(\Gq))=\delta(\Gq,\Gq,\Gq)=\Fq$.
So $I$ is one of the four involutions of $\{\Lq,\Aq,\Bq\}$, and (B1)--(B4)
become unit and binary clauses. As a check of this formalization, we evaluated
(B1)--(B4), extended to six states in the obvious way, on the transitions that
Mazoyer's six-state solution uses on the lines of lengths $3,\dots,150$ (an
exact finite simulation): (B1) in both readings, (B4), and (B2) for each of the
$10$ involutions of $\{\Lq,\Aq,\Bq,\Cq\}$, even without the borders, are
violated, and (B3) holds in the strong reading, in accordance with Mazoyer's
statement that his solution satisfies only the third
condition~\cite[\S7]{Mazoyer86L}.

\begin{proposition}[computer-assisted]\label{prop:balzer}
No five-state rule that satisfies (B1)--(B4) in the weak reading synchronizes
all lines of lengths $2,\dots,12$ in minimal time; a fortiori, no rule that
satisfies them in the strong reading does. The lengths up to $11$ do not
suffice to prove this.
\end{proposition}

\begin{proof}
For each involution, the clauses of (B1)--(B4) are added to
$\MT_5(\{2,\dots,N\};2N-2)$. For $I=(\Aq\,\Bq)$ and $N=10$, and for $I=(\Lq\,\Aq)$
and $I=(\Lq\,\Bq)$ and $N=9$, CaDiCaL refutes the formula with an LRAT proof.
For $I$ the identity and $N=12$, Kissat refutes it; \textsf{drat-trim} turns its
DRAT proof into an LRAT proof. Each of the four proofs is verified by
\texttt{lrat-check} and by \texttt{cake\_lpr}. The rule \texttt{balzer\_sym11} in \texttt{results/} of the
supplementary material satisfies
(B1)--(B4) in the strong reading with $I$ the identity and synchronizes every
line of length at most $11$ (direct simulation); see \cref{tab:balzer}.
\end{proof}

The condition (B2) at the borders is needed in \cref{prop:balzer}. The rule
\texttt{balzer\_noborder12} in \texttt{results/}, found by a solver, satisfies (B1)--(B4) in the weak reading with $I$ the
identity, except that (B2) holds only for the neighbourhoods without a border,
and synchronizes every line of length at most $12$ (direct simulation;
\cref{tab:balzer}). So with (B2) as Yun\`es states it,
the lines of lengths $2,\dots,12$ do not suffice, and we do not know whether
longer lines do.

\begin{table}[t]
\centering\small
\caption{The formulas of \cref{prop:balzer}: the clauses of (B1)--(B4) for the
involution $I$ added to $\MT_5(\{2,\dots,N\};2N-2)$, where $2,\dots,N$ are the
lengths in the second column. Every proof is verified by \texttt{lrat-check} and
by \texttt{cake\_lpr}.
The last two rows are rules, found by a solver and checked by direct
simulation: the first satisfies the strong reading, hence the weak one; the
second satisfies the weak reading except (B2) at the borders.}
\label{tab:balzer}
%% Table: Balzer's conditions (Proposition 6.6); the proofs are checked by check_certificates.sh.
\begin{tabular}{@{}lcll>{\raggedright\arraybackslash}p{0.27\textwidth}@{}}
\toprule
Involution $I$ & Lengths & Reading & Result & Certificate\\
\midrule
$(\Aq\,\Bq)$ & $2,\dots,10$ & weak & unsatisfiable & LRAT proof of CaDiCaL\\
$(\Lq\,\Aq)$ & $2,\dots,9$ & weak & unsatisfiable & LRAT proof of CaDiCaL\\
$(\Lq\,\Bq)$ & $2,\dots,9$ & weak & unsatisfiable & LRAT proof of CaDiCaL\\
identity & $2,\dots,12$ & weak & unsatisfiable & DRAT proof of Kissat, turned into LRAT by \textsf{drat-trim}\\
identity & $2,\dots,11$ & strong & satisfiable & the rule \texttt{balzer\_sym11}\\
identity & $2,\dots,12$ & weak, (B2) without borders & satisfiable & the rule \texttt{balzer\_noborder12}\\
\bottomrule
\end{tabular}

\end{table}

%% ---------------------------------------------------------------------------
\section{Discussion}\label{sec:discussion}

The conditions of this paper are of two kinds. The barriers of
\cref{sec:right,sec:left} are proved for every number of states and concern the
half-line alone. The conditions of \cref{sec:early} concern the first rows after
the reflection and the right ends, and we used them through satisfiability
solvers, for four and five states.

The conditions of \cref{sec:early} do not exclude every rule. For the germ of
$\delta_{14}^{\mathrm d}$, which has complexity $15$, conditions (G1)--(G3) with
$T=10$ and all $n\ge12$ have a model, and so do (G1)--(G3) with $T=7$ and all
$n\ge9$ together with (G5) for $9\le n\le20$. In the second model, the lines of
lengths $9$, $11,\dots,17$ and $19$ have both end cells, but not all cells, in
state $\Fq$ at time $2n-2$, and no cell in state $\Fq$ before
(\cref{fig:model}(a); both models are in the supplementary material). This suggests that excluding such rules needs
conditions on the middle cells of the firing row, which are expensive to
encode: for $n=15$, the dependency cone of the middle cell of the firing row
contains $76$ of the $120$ cells of $R_{15}$ (\cref{fig:model}(b)).

\begin{figure}[!htb]
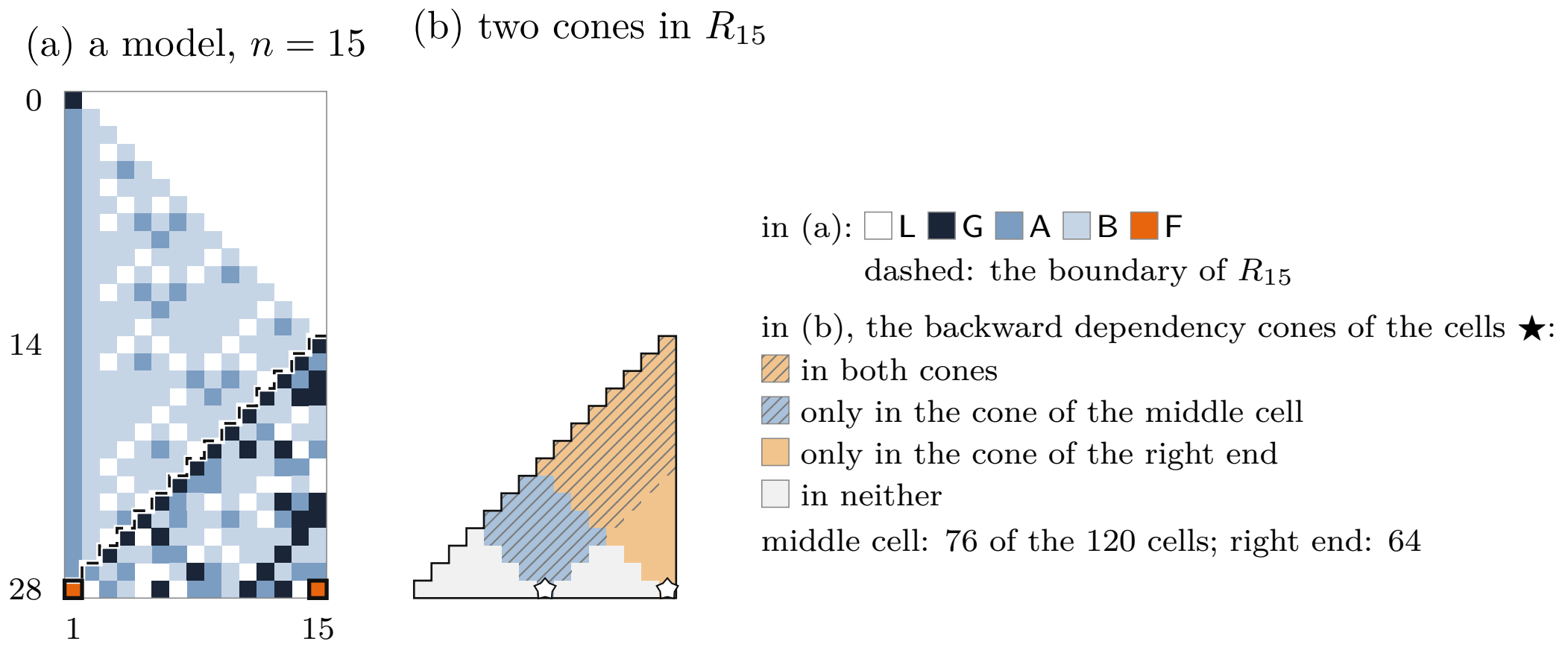

\centering
%% Figure: the second model of Section 7 on the line of length 15, and two dependency cones of R_15
%% (cells and cones from figures.py). In (b), the relative cell (tau,kappa) is [14-kappa,15-kappa] x [-tau-1,-tau],
%% shifted so that it is level with the cell (14+tau, 15-kappa) of (a).
\begin{tikzpicture}[x=1.55mm,y=1.55mm]
  \def\styleswatch#1{\tikz[baseline=-0.6ex,x=1cm,y=1cm]\filldraw[#1,draw=grid,line width=0.3pt] (0,-0.09) rectangle (0.24,0.15);}
  % (a) the line of length 15 up to time 28
  \begin{scope}
    \input{figures/cells_model_line15}
    \draw[white,line width=1.6pt] \staircase{28}{14} -- (15,-14);
    \draw[ink,densely dashed,line width=0.7pt] \staircase{28}{14} -- (15,-14);
    \draw[grid,line width=0.3pt] (0,0) rectangle (15,-29);
    \draw[ink,line width=1.0pt] (0,-28) rectangle (1,-29) (14,-28) rectangle (15,-29);
    \foreach \y/\l in {0/0,14/14,28/28} \node[font=\scriptsize,anchor=east] at (-0.4,-\y-0.5) {$\l$};
    \foreach \x/\l in {1/1,15/15} \node[font=\scriptsize,anchor=north] at (\x-0.5,-29.3) {$\l$};
    \node[font=\small,anchor=south] at (7.5,0.6) {(a) a model, $n=15$};
  \end{scope}
  % (b) the dependency cones of the middle cell and of the right end of the firing row of R_15
  \begin{scope}[shift={(20,-14)}]
    \input{figures/cones_R15}
    \draw[ink,line width=0.5pt] (15,0) -- (14,0) \foreach \k in {1,...,14} { -- (15-\k,-\k) -- (14-\k,-\k) }
      -- (0,-15) -- (15,-15) -- cycle;
    \node[star,star points=5,fill=white,draw=ink,line width=0.4pt,inner sep=0pt,minimum size=0.2cm] at (7.5,-14.5) {};
    \node[star,star points=5,fill=white,draw=ink,line width=0.4pt,inner sep=0pt,minimum size=0.2cm] at (14.5,-14.5) {};
    \node[font=\small,anchor=south west] at (-1,15.6) {(b) two cones in $R_{15}$};
  \end{scope}
  % legend
  \node[anchor=north west,font=\scriptsize,align=left] at (39,-6) {%
    in (a):\ \swatch{white}\,$\Lq$\ \swatch{stG}\,$\Gq$\ \swatch{stA}\,$\Aq$\ \swatch{stB}\,$\Bq$\
    \swatch{stF}\,$\Fq$\\[0.4ex]
    \phantom{in (a):}\ dashed: the boundary of $R_{15}$\\[1.4ex]
    in (b), the backward dependency cones of the cells $\bigstar$:\\[0.4ex]
    \styleswatch{coneBoth}\ in both cones\\[0.3ex]
    \styleswatch{coneMid}\ only in the cone of the middle cell\\[0.3ex]
    \styleswatch{coneRight}\ only in the cone of the right end\\[0.3ex]
    \styleswatch{coneNone}\ in neither\\[0.6ex]
    middle cell: $76$ of the $120$ cells; right end: $64$};
\end{tikzpicture}
\caption{(a)~The line of length $15$ under the second model above, which
contains the germ of $\delta_{14}^{\mathrm d}$ and satisfies (G1)--(G3) with
$T=7$ for all $n\ge9$ and (G5) for $9\le n\le20$. No cell is in state $\Fq$
before time $2n-2=28$, and at time $28$ only the end cells $1$ and $15$
(outlined) are. (b)~The backward dependency cones, inside $R_{15}$, of the
middle cell ($\kappa=7$) and of the right end ($\kappa=0$) of the firing row
(both marked $\bigstar$), drawn level with (a). The first covers $76$ of the
$120$ cells of $R_{15}$, the second $64$; condition (G5) constrains only the
second.}
\label{fig:model}
\end{figure}

The computational part has three limitations. Its results concern four and
five states and, apart from the reduction of \cref{cor:reduce}, finitely many
lengths. The formula generators are tested (\cref{sec:method}) and
cross-checked by a second, independent encoder, but, unlike the proof checker
\texttt{cake\_lpr}, they are not formally verified. And the behaviour of particular
rules beyond the simulated ranges, such as the patterns of the half-lines of
$\delta_{12}$ and $\delta_{14}$, is not known.

We close with four questions.
\begin{enumerate}[(1)]
\item Does a five-state minimal-time solution exist?
\item Is there a condition on the middle of the reflected triangles that is as
economical as \cref{lem:G}?
\item Can the bound $(k-1)^{2s}\ell$ of \cref{lem:left} be made polynomial in
$s$?
\item Do Balzer's conditions exclude six states? Do they exclude five states
without (B2) at the borders (\cref{sec:balzer})?
\end{enumerate}

%% ---------------------------------------------------------------------------
\appendix
\section{The Encodings}\label{app:encoding}

The formulas $\MT_k$ and $\G_k$ are built from the same pieces. Let $S$ be
the set of $k$ states and $W=S\setminus\{\Fq\}$.
\begin{enumerate}[(E1)]
\item \emph{Rule table.} A variable $x_{e,s}$ for each neighbourhood $e$ and
each $s\in S$, with the clause $\bigvee_s x_{e,s}$ and the clauses $\neg
x_{e,s}\vee\neg x_{e,s'}$ for $s\ne s'$. The entries $\delta(\Lq,\Lq,\Lq)$ and
$\delta(\Lq,\Lq,\ast)$ are fixed to $\Lq$.
\item \emph{Cells.} A variable $y_{u,s}$ for each encoded cell $u$ and each
$s\in W$, with exactly-one clauses as in (E1). So every encoded cell is in a
working state. The cells at time $0$, the quiescent cells $t<i-1$
(\cref{lem:lightcone}) and the borders are constants, and so is every cell of a
firing row, with the value $\Fq$.
\item \emph{Local rule.} For each encoded cell $u$ with predecessors
$u_1,u_2,u_3$ and each choice $a,b,d$ of their values, let $e=(a,b,d)$ and let
$P$ be the conjunction of $y_{u_1,a}$, $y_{u_2,b}$ and $y_{u_3,d}$, constants
omitted. If $u$ is a working cell, the clauses are $P\wedge x_{e,s}\to y_{u,s}$
for $s\in W$ and $P\to\neg x_{e,\Fq}$; if $u$ is in a firing row, the clause is
$P\to x_{e,\Fq}$.
\item \emph{Return chain.} For an encoded cell $u=(t,i)$ of the line of length
$n$ with $t+i=2n-1$ that is not in its firing row, and the half-line cell
$h=(t,i)$: the clauses $\neg y_{u,s}\vee\neg y_{h,s}$ for $s\in W$.
\item \emph{Pumping.} For $\PUMP(n,n')$: the clause $\bigvee_p z_p$ over the
positions $p$ of $\mathrm{cone}_n(n-1,0)$ (\cref{lem:cone}). The input cells
$v_p$ of $J_n$ and $w_p$ of $J_{n'}$ at $p$ are compared by an auxiliary
variable $z_p$ with the clauses $\neg z_p\vee\neg y_{v_p,s}\vee\neg y_{w_p,s}$
for $s\in W$; constant cells are compared directly.
\end{enumerate}

$\MT_k(\mathcal N;M,N')$ consists of (E1); the half-line cells with $t+i\le M$
and, for each $n\in\mathcal N$, the cells of $R_n$, with (E2) and (E3); (E4) for
the lengths $n$ with $n,n+1\in\mathcal N$; and (E5) for $4\le n<n'\le N'$.
$\G_k(T,N_1,C)$ consists of (E1) and, with (E2)--(E4), the half-line cells with
$t+i\le 2N_1$, the cells of $R_n$ for $2\le n\le T+1$, the cells of $R_n$ with
$\tau\le T$ for $T+2\le n\le N_1$, and the cone of the right end at the firing
time for $T+2\le n\le C$. These regions are closed under predecessors, so with
the table fixed every encoded cell is determined. The clauses (E4) for the
complete lines $2\le n\le T+1$ are not part of \cref{def:G}, but they follow from
(G1) and (G4) by the proof of \cref{lem:return}, in which (G4) gives
$C_\infty(2n-2,1)\ne\Fq$; so the formula is equivalent to \cref{def:G}. For the
formulas $\G_4$ of \cref{thm:four,prop:sharp}, the generator gives one variable
to cells that compute the same function of the table, that is, cells with the
same tree of predecessors down to the same half-line cells and borders; this
does not change the satisfying assignments of the table variables. The formula
of \cref{prop:germs} adds, for each germ $g$, a selector variable $\sigma_g$
with the clauses $\neg\sigma_g\vee x_{e,s}$ for the transitions $(e,s)$ of $g$,
and the clause $\bigvee_g\sigma_g$. The conditions (B1)--(B4) of \cref{prop:balzer} are unit and
binary clauses on the variables $x_{e,s}$.

%% ---------------------------------------------------------------------------
\section{Computational Claims and Artefacts}\label{app:artefacts}

\Cref{tab:artefacts} lists the computational claims of
\cref{sec:applications} with the programs behind them and the published files.
The digests are SHA-256 digests of the uncompressed files, abbreviated to their
first eight hexadecimal digits; the supplementary material lists them in full.

\begin{table}[ht]
\centering\footnotesize
\caption{The computational claims, their programs, and the unsatisfiable
formulas with their LRAT proofs. The programs are in \texttt{src/} and
\texttt{research/} of the supplementary material, and the formulas are the files
\texttt{formulas/*.cnf.xz}; the formulas and the proofs are also archived at
Zenodo (see the data availability statement). In the column Claim, $(\Aq\,\Bq)$ etc.\ is the
involution $I$ of \cref{prop:balzer}. The proof of the last formula was obtained
from a DRAT proof of Kissat (digest \texttt{2dd3e052}, $958$\,MB) with
\textsf{drat-trim}. Each proof is verified by \texttt{lrat-check} and by
\texttt{cake\_lpr} (\texttt{results/cake\_lpr.log}).}
\label{tab:artefacts}
\setlength{\tabcolsep}{3pt}
\begin{tabular}{@{}l>{\raggedright\arraybackslash}p{0.205\textwidth}lrrll@{}}
\toprule
Claim & Programs & Formula & Variables & Clauses & SHA-256 & LRAT proof\\
\midrule
\ref{thm:four}\,(i) & \texttt{gencnf.py} & \texttt{mt4\_9} & $753$ & $17{,}605$ & \texttt{d18d6548} & \texttt{1d511a75}, $177$\,MB\\
\ref{thm:four}\,(ii) & \texttt{gfix.py} & \texttt{k4\_G6} & $919$ & $22{,}960$ & \texttt{0e0ba286} & \texttt{7943a698}, $167$\,MB\\
\ref{thm:four}\,(iii) & \texttt{gfix.py} & \texttt{k4\_G4c} & $874$ & $20{,}461$ & \texttt{124f0954} & \texttt{7f76fe69}, $173$\,MB\\
\ref{prop:germs} & \texttt{germcert.py} & \texttt{k14\_n10} & $17{,}266$ & $559{,}558$ & \texttt{4f9306a8} & \texttt{b2324436}, $11$\,MB\\
\ref{prop:balzer}, $(\Aq\,\Bq)$ & \texttt{balzer.py} & \texttt{balzer\_AB\_10} & $1{,}496$ & $67{,}508$ & \texttt{d87457fe} & \texttt{18809db2}, $380$\,MB\\
\ref{prop:balzer}, $(\Lq\,\Aq)$ & \texttt{balzer.py} & \texttt{balzer\_LA\_9} & $1{,}244$ & $49{,}571$ & \texttt{401ef851} & \texttt{d74de977}, $188$\,MB\\
\ref{prop:balzer}, $(\Lq\,\Bq)$ & \texttt{balzer.py} & \texttt{balzer\_LB\_9} & $1{,}244$ & $49{,}571$ & \texttt{d7ba0d40} & \texttt{5784ee12}, $184$\,MB\\
\ref{prop:balzer}, id & \texttt{balzer\_k.py} & \texttt{balzer\_id\_12} & $2{,}148$ & $114{,}965$ & \texttt{023b99e2} & \texttt{e0231f2f}, $3{,}078$\,MB\\
\midrule
\ref{prop:sharp}\,(a) & \texttt{fsspdfs.c} (enumeration) & \multicolumn{5}{l}{---}\\
\ref{prop:sharp}\,(b) & \texttt{gfix.py}; models checked by \texttt{encoding\_check.py} & \multicolumn{5}{l}{satisfiable formulas}\\
\ref{prop:nearmiss} & \texttt{fsspcheck.c}, \texttt{firstfire.py}, \texttt{failall.py} & \multicolumn{5}{l}{simulations; certificates $(t_0,p)$ in \cref{tab:rules}}\\
\ref{prop:germs} & \texttt{germdfs.c}, \texttt{germdfs\_check.py}, \texttt{germinc.py}, \texttt{germext.c}, \texttt{germ\_k14\_check.py} & \multicolumn{5}{l}{enumeration of the $244$ germs; the $14$ refutations by \cref{lem:left}}\\
\ref{cor:germ12} & \texttt{antidiag.py}, \texttt{germext.c} & \multicolumn{5}{l}{simulation of the cells with $t+i\le1034$}\\
\ref{sec:balzer} & \texttt{balzer\_mazoyer.py}, \texttt{balzer\_rulecheck.py}, \texttt{fsspcheck.c} & \multicolumn{5}{l}{(B1)--(B4) on Mazoyer's solution and on the rules of \cref{tab:balzer}; simulation}\\
\ref{sec:method} & \texttt{encoding\_check.py}, \texttt{encode\_\allowbreak independent.py}, \texttt{validate\_mazoyer.py}, \texttt{gfix.py} & \multicolumn{5}{l}{checks of the encodings, the second encoder; \cref{tab:validation}}\\
\bottomrule
\end{tabular}
\end{table}

%% ---------------------------------------------------------------------------
\FloatBarrier
\section*{Acknowledgements}
This research received no specific grant from any funding agency. The author
declares no competing interests.

\paragraph{Use of generative AI}
The author used the AI system Claude (Anthropic) as an assistant. Under the
author's direction, it assisted in checking and developing the arguments, in
writing and running the programs and experiments, in drawing the figures, and in
drafting and revising the text. The author reviewed and edited all content and
takes full responsibility for it.

\paragraph{Data availability}
{\sloppy All data behind the computational statements are archived at Zenodo,
\url{https://doi.org/10.5281/zenodo.24002680}: the eight unsatisfiable formulas
of \cref{thm:four,prop:germs,prop:balzer}; their LRAT proofs, $4.4$\,GB in
total, and the DRAT proof from which one of them was obtained; the eight
formulas of the second encoder of \cref{sec:method}; the SHA-256 digests of the
formulas and the proofs; the partial rules and the models behind
\cref{prop:sharp}; the data of the figures; the logs of the runs that
regenerated the proofs and reran every other computation; and this version of
the supplementary
material, which contains the programs, the rule tables, the germ lists and the
logs behind every computational statement, the three scripts that rerun every
computation, the solver times and proof sizes, and the program that draws the
figures. The supplementary material is also in the public repository
\url{https://github.com/giamkiakeng/proj1}, directory
\texttt{fssp5/paper/supplementary}; the version described in this paper is
commit \texttt{8f3996c}, with the permanent link
\url{https://github.com/giamkiakeng/proj1/tree/8f3996cc242690d139bac913f5de652914ecb5ae/fssp5/paper/supplementary}.
The third script regenerates the proofs, checks them, and compares them with the
recorded digests (\cref{sec:method,app:artefacts}).\par}

%% ---------------------------------------------------------------------------
\section*{Nomenclature}
\begin{center}\small
\begin{tabular}{@{}>{$}l<{$}>{\raggedright\arraybackslash}p{0.62\textwidth}l@{}}
\toprule
\text{Symbol} & Meaning & Where\\
\midrule
S,\ W,\ k & the set of states, the working states $S\setminus\{\Fq\}$, and $k=|S|$ & \cref{sec:lines}\\
\Lq,\ \Gq,\ \Fq & the quiescent state, the general and the firing state & \cref{sec:lines}\\
\Aq,\ \Bq,\ \Cq & auxiliary states & \cref{sec:lines}\\
\ast & the border & \cref{sec:lines}\\
\delta & a rule, defined on the neighbourhoods $(x,y,z)$ & \cref{sec:lines}\\
C_n(t,i) & the state of cell $i$ at time $t$ in the line of length $n$ & \cref{sec:lines}\\
C_\infty(t,i) & the same in the half-line & \cref{sec:lines}\\
R_n & the reflected triangle of the line of length $n$ & \cref{sec:lines}\\
\tau,\ \kappa & relative time $t-(n-1)$ and distance $n-i$ to the right end & \cref{sec:right}\\
R_n(\tau,\kappa) & the state of $C_n$ at $(n-1+\tau,n-\kappa)$ & \cref{sec:right}\\
J_n(\kappa) & the input word: the cells of the half-line on $t+i=2n-3$ and $2n-2$ & \cref{sec:right}\\
\Psi & the map that computes $R_n$ from $J_n$ & \cref{lem:determinacy}\\
\mathrm{cone}_n(\tau,\kappa) & the input cells on which $R_n(\tau,\kappa)$ depends & \cref{sec:right}\\
\PUMP(n,n') & the conclusion of the pumping lemma & \cref{lem:pumping}\\
\Phi_t(j) & the cell of the half-line at depth $j$ behind the front at time $t$ & \cref{sec:right}\\
\theta(j) & the time from which depth row $j$ is periodic & \cref{thm:third}\\
\theta_H(j) & the same, observed in a simulation up to time $H$ & \cref{rem:division}\\
v_c(i),\ u_c(y) & the cells of chain $c$, in direct and in skewed coordinates & \cref{sec:left}\\
E(T) & the first $T+1$ rows of a reflected triangle & \cref{sec:early}\\
V_t,\ \Gamma_D & the first $D+1$ depth rows at time $t$, and the map that advances them & \cref{lem:certificate}\\
\G_k(T,N_1,C) & the early-row conditions (G1)--(G5) & \cref{def:G}\\
\MT_k(\mathcal N;M,N') & the formula for the lines in $\mathcal N$, the half-line up to $t+i=M$ and $\PUMP$ up to $N'$ & \cref{sec:method}\\
c_M(\delta) & the number of neighbourhoods besides $(\Lq,\Lq,\Lq)$ on the half-line up to $t+i=M$ & \cref{sec:germs}\\
\delta_{12},\ \delta_{13},\ \delta_{14} & five-state rules for the lengths up to $12$, $13$ and $14$ & \cref{sec:five}\\
\bottomrule
\end{tabular}
\end{center}

%% generated by make_refs.py from the citations in paper.aux; do not edit


\begin{thebibliography}{99}
\small
\bibitem{Moore64} Moore, E.~F. (1964). The firing squad synchronization problem. In: E.~F. Moore (ed.), \emph{Sequential Machines: Selected Papers}, pp.~213--214. Addison-Wesley.
\bibitem{Goto62} Goto, E. (1962). A minimal time solution of the firing squad problem. Dittoed course notes for Applied Mathematics~298, Harvard University, pp.~52--59.
\bibitem{Waksman66} Waksman, A. (1966). An optimum solution to the firing squad synchronization problem. \emph{Inform.\ and Control}, 9, 66--78.
\bibitem{Mazoyer86} Mazoyer, J. (1988). An overview of the firing squad synchronization problem. In: \emph{Automata Networks}, Lecture Notes in Comput.\ Sci. 316, pp.~82--94. Springer.
\bibitem{Umeo09} Umeo, H. (2009). Firing squad synchronization problem in cellular automata. In: R.~A. Meyers (ed.), \emph{Encyclopedia of Complexity and Systems Science}, pp.~3537--3574. Springer.
\bibitem{Correa17} Correa, T., Gustavo, B., Lemos, L., Settle, A. (2017). An overview of recent solutions to and lower bounds for the firing synchronization problem. arXiv:1701.01045.
\bibitem{Balzer67} Balzer, R. (1967). An 8-state minimal time solution to the firing squad synchronization problem. \emph{Inform.\ and Control}, 10, 22--42.
\bibitem{Gerken87} Gerken, H.-D. (1987). \emph{{\"U}ber Synchronisationsprobleme bei Zellularautomaten}. Diplomarbeit, Institut f{\"u}r Theoretische Informatik, Technische Universit{\"a}t Braunschweig.
\bibitem{Mazoyer87} Mazoyer, J. (1987). A six-state minimal time solution to the firing squad synchronization problem. \emph{Theoret.\ Comput.\ Sci.}, 50, 183--238.
\bibitem{Yunes93} Yun{\`e}s, J.-B. (1993). \emph{Synchronisation et automates cellulaires: la ligne de fusiliers}. PhD thesis, Universit{\'e} Paris~7. Available from HAL, tel-00139049.
\bibitem{Sanders94} Sanders, P. (1994). Massively parallel search for transition-tables of polyautomata. In: \emph{Parcella '94: Proceedings of the VI International Workshop on Parallel Processing by Cellular Automata and Arrays}, pp.~99--108. Akademie Verlag.
\bibitem{Clergue18} Clergue, M., Verel, S., Formenti, E. (2018). An iterated local search to find many solutions of the 6-states firing squad synchronization problem. \emph{Appl.\ Soft Comput.}, 66, 449--461.
\bibitem{NguyenMaignan20} Nguyen, T.~T., Maignan, L. (2020). Exploring millions of 6-state FSSP solutions: the formal notion of local CA simulation. In: \emph{Cellular Automata and Discrete Complex Systems (AUTOMATA 2020)}, Lecture Notes in Comput.\ Sci. 12286, pp.~1--13. Springer.
\bibitem{UmeoYanagihara07} Umeo, H., Yanagihara, T. (2007). A smallest five-state solution to the firing squad synchronization problem. In: \emph{Machines, Computations, and Universality (MCU 2007)}, Lecture Notes in Comput.\ Sci. 4664, pp.~291--302. Springer.
\bibitem{Yunes08} Yun{\`e}s, J.-B. (2008). A 4-states algebraic solution to linear cellular automata synchronization. \emph{Inform.\ Process.\ Lett.}, 107, 71--75.
\bibitem{UmeoKamikawaYunes09} Umeo, H., Kamikawa, N., Yun{\`e}s, J.-B. (2009). A family of smallest symmetrical four-state firing squad synchronization protocols for ring arrays. \emph{Parallel Process.\ Lett.}, 19, 299--313.
\bibitem{Mazoyer96} Mazoyer, J. (1996). On optimal solutions to the firing squad synchronization problem. \emph{Theoret.\ Comput.\ Sci.}, 168, 367--404.
\bibitem{YamashitaEtAl14} Yamashita, K., Nishitani, Y., Hirose, S., Okawa, S., Osato, N. (2014). The firing squad synchronization problem with sub-generals. \emph{Inform.\ Process.\ Lett.}, 114, 60--65.
\bibitem{Kobayashi19} Kobayashi, K. (2019). Nonexistence of minimal-time solutions for some variations of the firing squad synchronization problem having simple geometric configurations. \emph{J.\ Cell.\ Autom.}, 14, 351--396.
\bibitem{MazoyerTerrier99} Mazoyer, J., Terrier, V. (1999). Signals in one-dimensional cellular automata. \emph{Theoret.\ Comput.\ Sci.}, 217, 53--80.
\bibitem{BerthiaumeEtAl04} Berthiaume, A., Bittner, T., Perkovi{\'c}, L., Settle, A., Simon, J. (2004). Bounding the firing synchronization problem on a ring. \emph{Theoret.\ Comput.\ Sci.}, 320, 213--228.
\bibitem{DAntonioDelzanno04} D'Antonio, M., Delzanno, G. (2004). SAT-based analysis of cellular automata. In: \emph{Cellular Automata (ACRI 2004)}, Lecture Notes in Comput.\ Sci. 3305, pp.~745--754. Springer.
\bibitem{WetzlerHeuleHunt14} Wetzler, N., Heule, M.~J.~H., Hunt, W.~A., Jr. (2014). DRAT-trim: efficient checking and trimming using expressive clausal proofs. In: \emph{Theory and Applications of Satisfiability Testing (SAT 2014)}, Lecture Notes in Comput.\ Sci. 8561, pp.~422--429. Springer.
\bibitem{CruzFilipeEtAl17} Cruz-Filipe, L., Heule, M.~J.~H., Hunt, W.~A., Jr., Kaufmann, M., Schneider-Kamp, P. (2017). Efficient certified RAT verification. In: \emph{Automated Deduction -- CADE 26}, Lecture Notes in Comput.\ Sci. 10395, pp.~220--236. Springer.
\bibitem{HeuleKullmannMarek16} Heule, M.~J.~H., Kullmann, O., Marek, V.~W. (2016). Solving and verifying the Boolean Pythagorean triples problem via cube-and-conquer. In: \emph{Theory and Applications of Satisfiability Testing (SAT 2016)}, Lecture Notes in Comput.\ Sci. 9710, pp.~228--245. Springer.
\bibitem{Heule18} Heule, M.~J.~H. (2018). Schur number five. In: \emph{Proceedings of the Thirty-Second AAAI Conference on Artificial Intelligence (AAAI-18)}, pp.~6598--6606. AAAI Press.
\bibitem{CaDiCaL2} Biere, A., Faller, T., Fazekas, K., Fleury, M., Froleyks, N., Pollitt, F. (2024). CaDiCaL~2.0. In: \emph{Computer Aided Verification (CAV 2024)}, Lecture Notes in Comput.\ Sci. 14681, pp.~133--152. Springer.
\bibitem{Kissat20} Biere, A., Fazekas, K., Fleury, M., Heisinger, M. (2020). CaDiCaL, Kissat, Paracooba, Plingeling and Treengeling entering the SAT Competition 2020. In: \emph{Proceedings of SAT Competition 2020: Solver and Benchmark Descriptions}, Department of Computer Science Report Series~B, B-2020-1, pp.~51--53. University of Helsinki.
\bibitem{TanHeuleMyreen21} Tan, Y.~K., Heule, M.~J.~H., Myreen, M.~O. (2021). cake\_lpr: verified propagation redundancy checking in CakeML. In: \emph{Tools and Algorithms for the Construction and Analysis of Systems (TACAS 2021)}, Lecture Notes in Comput.\ Sci. 12652, pp.~223--241. Springer.
\bibitem{Duprat97} Duprat, J. (1997). Firing Squad Synchronization Problem. Coq user contribution FiringSquad (formal proof of the correctness of Mazoyer's six-state solution), LIP, ENS Lyon. Archived at \url{https://github.com/rocq-archive/firing-squad}.
\bibitem{Wolfram83} Wolfram, S. (1983). Statistical mechanics of cellular automata. \emph{Rev.\ Modern Phys.}, 55, 601--644.
\bibitem{Mazoyer86L} Mazoyer, J. (1986). A six states minimal time solution to the firing squad synchronization problem. \emph{Publ.\ D{\'e}p.\ Math.\ (Lyon)}, 1A, 1--92. Available from Numdam.
\end{thebibliography}
\end{document}